\documentclass[prx, twocolumn, superscriptaddress,
 showpacs,
 longbibliography,
 aps,
 showkeys]{revtex4-2}

\usepackage{dsfont} 
\usepackage{bm} 

\usepackage{bbold} 
\usepackage{xcolor}

\usepackage{float}
\usepackage{physics}
\usepackage{graphicx,braket}
\usepackage{hyperref}
\hypersetup{colorlinks=true,linkcolor=blue,citecolor=blue}
\usepackage{amsmath,amssymb,amsfonts}
\usepackage{wrapfig}
\usepackage[export]{adjustbox}
\usepackage{cleveref} 
\usepackage{hhline}
\usepackage{siunitx}
\usepackage{booktabs}
 
\usepackage{enumitem}   

\usepackage{cleveref}
\crefname{equation}{Eqs.}{Eqs.}
\Crefname{equation}{Equation}{Equations}
\crefrangelabelformat{equation}{(#3#1#4--#5#2#6)}
\crefmultiformat{equation}{Eqs. (#2#1#3}{, #2#1#3)}{#2#1#3}{#2#1#3}
\Crefmultiformat{equation}{Equations (#2#1#3}{, #2#1#3)}{#2#1#3}{#2#1#3}

\begin{document}
	
	\allowdisplaybreaks 
	
	\flushbottom
	\title{Dissipative stabilization of maximal entanglement between non-identical emitters via two-photon excitation }

	\author{Alejandro Vivas-Via{\~n}a}
	\affiliation{Departamento de Física Teórica de la Materia Condensada and Condensed
		Matter Physics Center (IFIMAC), Universidad Autónoma de Madrid, 28049 Madrid,
		Spain}	
	\author{Diego Martín-Cano}
	\affiliation{Departamento de Física Teórica de la Materia Condensada and Condensed
		Matter Physics Center (IFIMAC), Universidad Autónoma de Madrid, 28049 Madrid,
		Spain}
	\author{Carlos S\'anchez Mu\~noz}
	\email[]{carlossmwolff@gmail.com}
	\affiliation{Departamento de Física Teórica de la Materia Condensada and Condensed
		Matter Physics Center (IFIMAC), Universidad Autónoma de Madrid, 28049 Madrid,
		Spain}
	
	\newcommand{\down}{\op{g}{e}}
	\newcommand{\up}{\op{e}{g}}
	\newcommand{\downd}{\op{+}{-}} 
	\newcommand{\upd}{\op{+}{-}}
	\newcommand{\app}{a^\dagger}
	\newcommand{\ssp}{\sigma^\dagger}
	\newcommand*{\Resize}[2]{\resizebox{#1}{!}{$#2$}}%
	\newcommand{\conc}{\mathcal C(\rho)}

\begin{abstract}
Two non-identical quantum emitters, when placed within a cavity and coherently excited at the two-photon resonance, can reach stationary states of nearly maximal entanglement. 
In Ref.~\cite{Vivas-VianaFrequencyresolvedPurcell2023}, we introduce a frequency-resolved Purcell effect stabilizing entangled $W$ states among strongly-interacting quantum emitters embedded in a cavity. 
Here, we delve deeper into a specific configuration with a particularly rich phenomenology: two interacting quantum emitters under coherent excitation at the two-photon resonance. 
This scenario yields two resonant cavity frequencies where the combination of two-photon driving and Purcell-enhanced decay stabilizes the system into the sub- and superradiant states, respectively.
By considering the case of non-degenerate emitters and exploring the parameter space of the system, we show that this mechanism is merely one among a complex family of phenomena that can generate both stationary and metastable entanglement when driving the emitters at the two-photon resonance. We provide a global perspective of this landscape of mechanisms and contribute analytical characterizations and insights into these phenomena, establishing connections with previous reports in the literature and discussing how some of these effects can be optically detected.

\end{abstract}

\date{\today} \maketitle


\section{Introduction }
Generating, manipulating and preserving entanglement is a critical task in the development of quantum technologies such as quantum computation~\cite{NielsenQuantumComputation2012}, quantum communications~\cite{KimbleQuantumInternet2008,NarlaRobustConcurrent2016} or quantum metrology~\cite{PezzeQuantumMetrology2018,BraskImprovedQuantum2015}. However, the unavoidable interaction between the quantum system and its environment leads to the major obstacle of decoherence. One strategy to overcome this issue is reservoir engineering, whereby interactions between the system and environment are designed to dissipatively stabilize a target computation result or quantum state~\cite{PoyatosQuantumReservoir1996,VerstraeteQuantumComputation2009,LiuComparingCombining2016}.
Examples of this strategy applied to the stabilization of entangled states have been demonstrated in platforms such as superconducting circuits~\cite{LeghtasStabilizingBell2013,ShankarAutonomouslyStabilized2013,LiuComparingCombining2016}, atomic ensembles~\cite{KrauterEntanglementGenerated2011} or trapped ions~\cite{LinDissipativeProduction2013}. 

Platforms in which optical photons are interfaced with solid-state quantum emitters, such as quantum dots (QDs)~\cite{LodahlInterfacingSingle2015}, molecules~\cite{ToninelliSingleOrganic2021} or colour centres~\cite{SipahigilIntegratedDiamond2016,AwschalomQuantumTechnologies2018}, are considered particularly promising for quantum information for their potential for fast operation, on-chip integration, low-noise, and capability of transmission over long distances~\cite{OBrienPhotonicQuantum2009,WangIntegratedPhotonic2020}. In these solid-state platforms, proposals that can be considered variants of reservoir engineering have been made for the formation of steady-state entanglement between qubits through dissipative coupling  mediated by photonic nanostructures~\cite{Gonzalez-TudelaEntanglementTwo2011,Martin-CanoDissipationdrivenGeneration2011,Gonzalez-TudelaMesoscopicEntanglement2013,RamosQuantumSpin2014,PichlerQuantumOptics2015,HaakhSqueezedLight2015,ChangColloquiumQuantum2018,ReitzCooperativeQuantum2022}. There has been experimental progress in this direction, with reports of the formation of  superradiant and subradiant states---corresponding to the symmetric triplet and antisymmetric singlet Bell states---in  systems consisting of two solid state quantum emitters, such as closely spaced molecules interacting via dipole coupling~\cite{HettichNanometerResolution2002,TrebbiaTailoringSuperradiant2022}, SiV centers in a waveguide~\cite{SipahigilIntegratedDiamond2016} or semiconductor QDs coupled to cavities~\cite{ReitzensteinCoherentPhotonic2006,LauchtMutualCoupling2010} or waveguides~\cite{KimSuperRadiantEmission2018,GrimScalableOperando2019,TiranovCollectiveSuper2023}.
Despite this progress, generating entanglement between solid-state qubits coupled to a nanophotonic structure remains a challenging task. In the case of molecules and QDs, it is recognized that the main obstacle faced is inhomogeneous broadening~\cite{GrimScalableOperando2019}, i.e., the differences in the natural frequencies of the quantum emitters due to different size, strain and composition of QDs, or the different local matrix environments of the molecules, which are also difficult to position selectively~\cite{ToninelliSingleOrganic2021,TrebbiaTailoringSuperradiant2022}.

\begin{figure*}[t!]
	\includegraphics[width=0.95\textwidth]{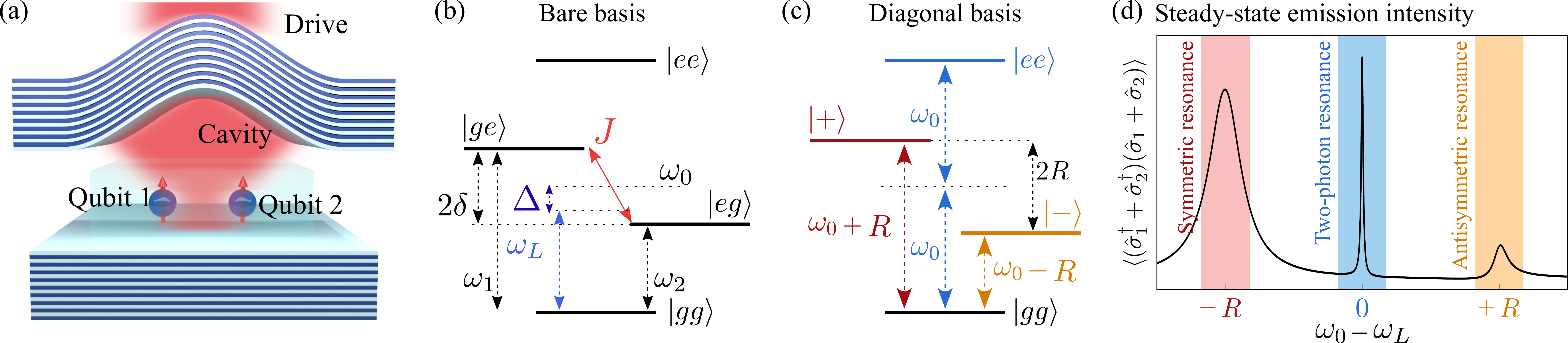}
	\caption{(a) Sketch of the system: two interacting nonidentical quantum emitters under coherent drive and coupled to a single mode cavity. (b) Energy diagram of the quantum emitters in the bare basis. The states of the one-photon manifold are detuned by an energy $2\delta$ and coupled with a coupling strength $J$. The system is driven with a laser with frequency $\omega_\mathrm L$. (c) Energy levels in the excitonic basis where the coupling between emitters has been diagonalized. Here, the system is driven at the two-photon resonance $\omega_L=\omega_0$. (d) Steady-state emission intensity as the frequency of the laser is varied.  }
	\label{fig:Fig1_Scheme}
\end{figure*}

In this work, we address this challenge by exploring the generation of steady-state entanglement between non-identical emitters weakly coupled to a lossy cavity mode and driven at the two-photon resonance, i.e., when the drive frequency is half that of the doubly-excited state~\cite{VaradaTwophotonResonance1992}. In Ref.~\cite{Vivas-VianaFrequencyresolvedPurcell2023}, we unveil a mechanism based on the cavity-enhanced decay from the highest excited state of the ensemble to an entangled state with a single de-excitation coherently shared among the degenerate quantum emitters.
In this paper, we delve further into a specific case of particular relevance: two emitters coherently driven at the two-photon resonance. In this case, the effect manifests as two resonant values of the cavity frequency in which either the entangled symmetric or antisymmetric single-excitation eigenstates are respectively stabilized.  
Notably, this scenario allows  the cavity to be optically tuned into the required resonant conditions via dressing of the quantum emitters with the external driving.

Additionally, we move away from the requisite of degenerate emitters, and consider the situation of two non-identical sources. This allows us to unveil a complex landscape of up to four different mechanisms of both stable and metastable entanglement generation, which we label as Mechanisms I-IV. We characterize the regions of the parameter phase space in which these mechanisms emerge, contributing with novel insights and analytical descriptions that connect some of them to previous reports in the literature.

 As an important example, we discuss the case  in which the two quantum emitters are strongly detuned and have negligible coherent interaction in the absence of the cavity. In this regime, as originally reported in Ref.~\cite{PichlerQuantumOptics2015}, the combination of driving at the two-photon resonance and collective decay induced by a cavity or a waveguide can stabilize the system into a very high occupation of the entangled, antisymmetric singlet state. Here, we contribute to the understanding of this effect from the perspective of recent reports of unconventional population of virtual states~\cite{Vivas-VianaUnconventionalMechanism2022}, which allows us to provide new analytical expressions for the timescale of entanglement formation, the steady-state population, and the parameter regimes where this effect is activated. These results are relevant in the context of both cavity and waveguide QED~\cite{Gonzalez-TudelaEntanglementTwo2011}.

Finally, we also propose observational schemes to detect entanglement through the optical properties (both classical and quantum) of the light emitted by the system, and test the robustness of the mechanisms of entanglement generation against extra decoherence channels---non-radiative spontaneous decay, and both local and collective pure dephasing---to prove its viability in realistic, dissipative scenarios. The fact that some of the mechanism that we discuss operate in regimes in which the quantum emitters have different frequencies is of great relevance for the experimental implementations of quantum technologies in the solid state, in which the fabrication of identical quantum emitters remains a challenging task~\cite{OBrienPhotonicQuantum2009,PatelTwophotonInterference2010,GieszCoherentManipulation2016,EvansPhotonmediatedInteractions2018,LukinIntegratedQuantum2020,LukinTwoEmitterMultimode2023}.

The paper is organized as follows. In Sec~\ref{sec:model}, we describe the model of the system and then introduce an effective description of the system dynamics when the cavity is traced out. In Sec~\ref{sec:entanglement-mechanism}, we characterize the emergent mechanisms of entanglement generation.  In Sec~\ref{sec:observation}, we propose observational schemes to detect entanglement via the quantum and classical properties of the emitted light by the system. Finally, in Sec~\ref{sec:decoherence}, we test the robustness of entanglement when extra decoherence channels are considered.

\section{Model}
\label{sec:model}

\subsection{General model}

We consider a system composed of two nonidentical quantum emitters, both coupled to a single mode cavity, and driven by a classical source of light, as depicted in Fig.~\ref{fig:Fig1_Scheme}(a). We describe the emitters as two-level systems, and therefore we will also refer to them as qubits. The state of the $i$-th emitter is written in a basis $\left\{ |g_i\rangle,   |e_i\rangle\right\}$, which sets a lowering operator given by $\hat \sigma_i\equiv|g_i\rangle \langle e_i| $.
The qubit-qubit system spans a basis $\left\{ |gg\rangle,   |ge\rangle,|eg\rangle,|ee\rangle \right\}$. 
In addition, each emitter is characterized by its position $\bm{r}_i$, natural frequency $\omega_i$, and dipole moment $\bm{\mu}_i$.
The two natural frequencies of the emitters are given by $\omega_1 \equiv \omega_0 - \delta$ and $\omega_2 \equiv \omega_0 + \delta$, so that the emitters have an average frequency of $\omega_0$ and are detuned from each other by a frequency of $2\delta$, see Fig.~\ref{fig:Fig1_Scheme}(b).
The cavity is described by a single mode cavity with frequency $\omega_a$, annihilation operator $\hat a$, and  emitter-photon coupling $g$. The classical source of light is described by a laser field of frequency $\omega_L$ driving the emitters with a Rabi frequency $\Omega$. 
In the rotating frame of the laser and under the rotating wave approximation, setting $\hbar =1$ henceforth, we can write the time-independent Hamiltonian $\hat H=\hat H_{\text{q}}+\hat H_{\text a}+\hat H_d$, $\hat H_\text{q}$ being the bare Hamiltonian of the interacting quantum emitters,
\begin{equation}
	\hat H_{\text{q}}=(\Delta -\delta)\hat \sigma_1^\dagger \hat \sigma_1 +(\Delta +\delta)\hat \sigma_2^\dagger \hat \sigma_2 + J(\hat \sigma_1^\dagger  \hat \sigma_2+\text{H.c}),
	\label{eq: H_emitters}
\end{equation}
 $\hat H_\text{a}$ being the Hamiltonian that involves the cavity
\begin{equation}
	\hat H_\text{a}=\Delta_a \hat a^\dagger \hat a+g\left[\hat a^\dagger (\hat \sigma_1 +\hat \sigma_2)+\text{H.c.}\right],
	\label{eq:H_cavity}
\end{equation}
and $\hat H_d$ being the Hamiltonian of the coherent drive,
\begin{equation}
	\hat H_d=\Omega (\hat \sigma_1+\hat \sigma_2+ \text{H.c.}),
	\label{eq: H_laser}
\end{equation}
where defined the laser-qubit detuning, $\Delta\equiv \omega_0-\omega_L$, the laser-cavity detuning, $\Delta_a\equiv \omega_a-\omega_L$, and where H.c. denotes Hermitian conjugate.

We also assume that the total system interacts with a bath that introduces incoherent processes by which the emitters are de-excited by spontaneous emission and photons leak out of the cavity.
In the Markovian regime, the dissipative dynamics of the reduced density matrix of the qubits-cavity is described by the master equation~\cite{GardinerQuantumNoise2004,FicekQuantumInterference2005,BreuerTheoryOpen2007},
\begin{equation}
	\frac{d \hat{\rho}}{dt}=-i[ \hat H, \hat{\rho}]+\sum_{i,j=1}^{2}\frac{\gamma_{ij}}{2}\mathcal{D}\left[\hat \sigma_i,\hat \sigma_j\right]  \hat{\rho}+\frac{\kappa}{2} \mathcal{D}\left[\hat a \right]  \hat{\rho},
	\label{eq:full_master_eq}
\end{equation}
where $\mathcal{D} [\hat A , \hat B ] \hat \rho  \equiv 2 \hat A \rho \hat B^\dagger -\{\hat B^\dagger \hat A ,\hat \rho \}$ and $\mathcal D [\hat A] \equiv \mathcal D [\hat A, \hat A]$. Here,  $\gamma_{ii}$ is the local decay rate of spontaneous emission of the $i$-th emitter, $\gamma_{12}=\gamma_{21}$ is the dissipative coupling rate between emitters that emerges as a consequence of collective decay, and $\kappa$ is the photon leakage rate. Incoherent excitation by thermal photons is neglected since we consider optical transitions or cryogenic temperatures in platforms such as superconducting circuits~\cite{GarciaRipollQuantumInformation2022}.

Although our model and results derived from it apply to any type of interaction between the emitters, for concreteness in what follows, we will consider in our parametrization of $J$ that the qubits are coupled by dipole-dipole interactions, i.e., the coherent coupling is mediated by the electromagnetic vacuum in free space~\cite{CarmichaelStatisticalMethods1999,FicekQuantumInterference2005,HettichNanometerResolution2002,TrebbiaTailoringSuperradiant2022}. Our results, however, are also relevant for more general scenarios, such as when this interaction is mediated by photonic structures~\cite{Gonzalez-TudelaEntanglementTwo2011,Martin-CanoDissipationdrivenGeneration2011,ChangColloquiumQuantum2018,Miguel-TorcalInversedesignedDielectric2022}. 

The local decay rates in free space depend on the natural frequencies and dipole moments of the emitters \cite{FicekQuantumInterference2005},
\begin{equation}
	\gamma_{ii}\equiv \gamma_i = \frac{\omega_i^3 \abs{\bm{\mu}_i}^2}{3\pi \epsilon_0  c^3},
	\label{eq:gamma}
\end{equation}
where $\bm{\mu}_i$ is the dipole moment of the $i$-th emitter, $\epsilon_0$ is the vacuum permittivity and $c$ is the speed of light. 
The coherent and dissipative coupling rates also depend on the relative separation between emitters \cite{FicekQuantumInterference2005,ReitzCooperativeQuantum2022}, $\mathbf{r}_{12}=\mathbf{r}_1-\mathbf{r}_2$,
\begin{align}
		J=
		&\frac{3}{4}\sqrt{\gamma_1 \gamma_2} \left\{ - \left[ 1- \left( \bm{\bar \mu}\cdot \mathbf{\bar r}_{12} \right)^2 \right] \frac{\cos\left( k r_{12} \right)}{kr_{12}} \right. \notag \\
		&\left. + \left[ 1-3 \left( \bm{ \bar\mu}\cdot \mathbf{\bar r}_{12} \right)^2 \right] \left[  \frac{\sin\left( k r_{12} \right)}{(kr_{12})^2} +\frac{\cos\left( k r_{12} \right)}{(kr_{12})^3} \right] \right\},
		\\
		\gamma_{12}=
		&\frac{3}{2} \sqrt{\gamma_1 \gamma_2} \left\{ \left[ 1- \left( \bm{\bar \mu}\cdot \mathbf{\bar r}_{12} \right)^2 \right] \frac{\sin \left( k r_{12} \right)}{kr_{12}}  \right. \notag \\
		& \left. + \left[ 1-3 \left( \bm{\bar\mu}\cdot \mathbf{ \bar r}_{12} \right)^2 \right] \left[  \frac{\cos\left( k r_{12} \right)}{(kr_{12})^2} - \frac{\sin \left( k r_{12} \right)}{(kr_{12})^3} \right] \right\},
\end{align}
where $r_{12}=|\bm r_{12}|$, $\bm{\bar \mu_1} = \bm{\bar \mu_2} = \bm{\bar \mu}$ and $\mathbf{ \bar r}_{12}$ are unit vectors, and $k=\omega_0/c$, having assumed $\omega_0 \gg (\omega_2-\omega_1)$. We consider emitters with similar spontaneous decay rate, $\gamma_{ii}\approx \gamma$, and $\Omega(\bm r_1)\approx \Omega (\bm r_2)\approx \Omega$. Also, we assume an H-aggregate configuration, in which the dipole moments are perpendicular the line that connects them, i.e.,  $\bm \mu_{i} \cdot\bm r_{12}=0$ $\forall i$ . This choice is made out of convenience without loss of generality \cite{FicekQuantumInterference2005,Vivas-VianaTwophotonResonance2021}. 
We note that, throughout this work, we use values in concordance with previous experimental implementations of two quantum emitters interacting via dipole-dipole interaction mediated by the vacuum~\cite{HettichNanometerResolution2002,TrebbiaTailoringSuperradiant2022}.

\subsection{Bare qubit system}

The Hamiltonian of the bare, undriven quantum emitters in Eq.~\eqref{eq: H_emitters} describes a dimer yielding a diagonal basis $\{ |gg\rangle , |-\rangle , |+\rangle, |ee\rangle \}$, where $\{ |-\rangle , |+\rangle  \}$ are the eigenstates of the single-excitation subspace, given by
\begin{equation}
|\pm\rangle = \frac{1}{\sqrt{2}}(\sqrt{1\mp\sin \beta }|eg \rangle \pm\sqrt{1\pm\sin \beta }|ge \rangle ), \label{eq:eigedressed}  
\end{equation}
$\beta $ denoting a mixing angle~\cite{SanchezMunozPhotonCorrelation2020,Vivas-VianaTwophotonResonance2021} defined as 
\begin{equation}
\beta\equiv \arctan (\delta/J).
\end{equation}
The energies of these single-excitation eigenstates are $E_{\pm}=\Delta \pm R$, where $R=\sqrt{J^2 +\delta^2}$ is the Rabi frequency of the dipole-dipole coupling. This energy level structure is depicted in Fig.~\ref{fig:Fig1_Scheme}(c).

The case $\beta=0$ corresponds to the limit of identical emitters ($\delta=0$) or strong dipole-dipole interactions rates ($J\gg \delta$). In that case, the eigenstates in the single-excitation subspace correspond to purely symmetric and antisymmetric superpositions $|S\rangle$ and $|A\rangle$, which are the so called superradiant and subradiant states, respectively~\cite{ReitzCooperativeQuantum2022}:
\begin{equation}
|\pm\rangle_{\beta=0} = |S/A\rangle = \frac{1}{\sqrt 2}(|eg\rangle \pm |ge\rangle).
\end{equation}

On the other hand, the limit $\beta =\pi/2$ corresponds to the case in which the interaction rate between emitters is much weaker than their detuning ($\delta\gg J$) or the dipole-dipole interaction is disabled ($J=0$). In this limit, the eigenstates tend to those of independent, non-interacting emitters, 
\begin{equation}
	|-\rangle_{\beta\approx \pi/2} \approx |eg\rangle,\quad |+ \rangle_{\beta\approx \pi/2}  \approx |ge\rangle.
\end{equation}
\begin{figure}[t]
	\includegraphics[width=0.47\textwidth]{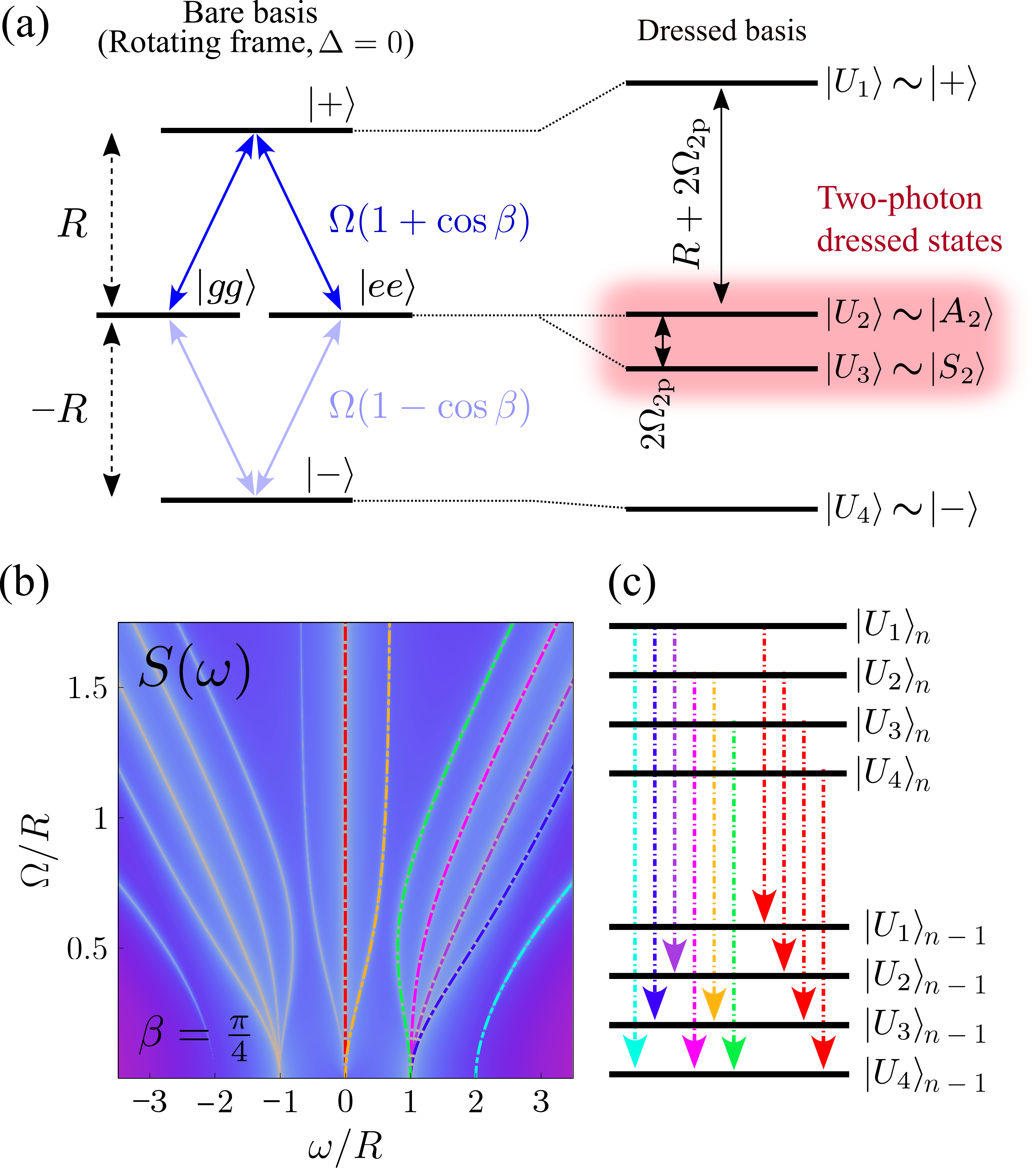}
	\caption{ (a) Schemes of the two-photon dressing process: hybridization from the bare basis (at the two-photon resonance, $\Delta=0$) to the dressed basis. (b) Two-photon resonance fluorescence spectrum of the two-qubit system.  (c) Energy manifolds, labelled by the total excitation number $n$, of the dressed laser-atom system. Arrows represent transitions between $n$ and $n-1$ excitation manifolds. Parameters (b-c): $r_{12}=2.5\ \text{nm}$, $k=2\pi/780\ \text{nm}^{-1}$, $J=9.18\times 10^4\gamma$, $\gamma_{12}=0.999\gamma$, $\delta=J$, $\Delta=0$. }
	\label{fig:Fig2_EneryDiagramSpectrum}
\end{figure}
\subsection{Dressed laser-qubit system}
\label{sec:dressed_system}
We now consider the hybridized light-matter character of the qubit-qubit system when we include the external driving laser, described by the Hamiltonian in Eq.~\eqref{eq: H_laser} (we do not consider yet the interaction with the single cavity mode). As we will justify later, we focus on the case in which 
 both emitters are excited at the two-photon resonance, $\Delta=\omega_0-\omega_\mathrm L = 0$.
 
In the perturbative regime $\Omega\ll R$, the laser is strongly detuned from all the single-photon transitions, and it can only interact with the emitters via the resonant exchange of photon pairs, which induces two-photon Rabi oscillations between the ground state $|gg\rangle$ and doubly excited state $|ee\rangle$. This two-photon driving of the system is evidenced as a resonant peak in the intensity of the radiated emission at the two-photon resonance condition, see e.g. Fig.~\ref{fig:Fig1_Scheme}(d). Crucially, this mechanism is only enabled when the emitters are interacting~\cite{VaradaTwophotonResonance1992,
HettichNanometerResolution2002,
TrebbiaTailoringSuperradiant2022}; the effective Rabi frequency of the two-photon drive is given by  $\Omega_{\text{2p}}=2\Omega^2\cos \beta/R $~\cite{HaakhSqueezedLight2015,Vivas-VianaTwophotonResonance2021}, which indeed becomes zero in the limit of non-interacting qubits, $\beta = \pi/2$.
When this Rabi frequency is larger than the spontaneous emission rate $\Omega_{\mathrm{2p}}>\gamma$, the natural way to describe the system is in terms of the dressed eigenstates resulting from the hybridization between the emitters and the driving field~\cite{DarsheshdarPhotonphotonCorrelations2021,Vivas-VianaTwophotonResonance2021}. Ordering them by decreasing energy, these eigenstates are approximately given by $\left\{|+\rangle, |A_2\rangle, |S_2\rangle , |-\rangle \right\}$, where 
\begin{equation}
|S_2/A_2\rangle=\frac{1}{\sqrt 2}(|gg\rangle \pm |ee\rangle)
\label{eq:s2a2eigenstates}
\end{equation}
are two-photon dressed states. The corresponding eigenenergies are provided in Ref.~\cite{Vivas-VianaTwophotonResonance2021}, and in the limit $\beta\rightarrow 0$ they are approximately given by
\begin{equation}
(\lambda_1, \lambda_2, \lambda_3, \lambda_4 ) \approx ( R + 2\Omega_{\mathrm{2p}},0,-2\Omega_{\mathrm{2p}},-R ).
\label{eq:eigenvalues_dressed}
\end{equation}
This configuration of states is depicted in Fig.~\ref{fig:Fig2_EneryDiagramSpectrum}(a). 

Beyond the perturbative regime, i.e., for $\Omega > R$, the eigenstates do not have a closed analytical form except for the limiting cases of $\beta = 0$, where the eigenstates are the same as in the weak-driving case except for $|S\rangle$ and $|S_2\rangle$, which get mixed (more details can be found in Ref.~\cite{Vivas-VianaTwophotonResonance2021}). Therefore, we generally label the eigenstates, for any value of $\Omega$, as $|U_i\rangle$, with $i=1,2,3,4$, in decreasing order of energy.

A clear evidence of the emergence of dressed eigenstates once the driving is included is the development of new sidebands in the fluorescence spectrum of the emitted radiation, due to the new transitions enabled in the composite qubit-laser
 system~\cite{MollowPowerSpectrum1969,Cohen-TannoudjiAtomPhotonInteractions1998,ArdeltOpticalControl2016,HargartCavityenhancedSimultaneous2016,DarsheshdarPhotonphotonCorrelations2021,Vivas-VianaTwophotonResonance2021}. A subset of these transitions is shown in Fig.~\ref{fig:Fig2_EneryDiagramSpectrum}(b-c). An example of the resulting spectrum is illustrated in Fig.~\ref{fig:Fig2_EneryDiagramSpectrum}(b), where we computed $S(\omega)=\lim_{t\rightarrow \infty} \frac{1}{\pi} \text{Re}\int_0^\infty d\tau e^{i\omega \tau} \langle \hat{E}^{(-)}(t)   \hat{E}^{(+)}(t+\tau) \rangle$, 
considering that $\hat{E}^{(+)}(\mathbf{r},t)\propto ( \hat \sigma_1 +\hat \sigma_2) (t)$ \cite{ScullyQuantumOptics1997,NovotnyPrinciplesNanoOptics2012,Vivas-VianaTwophotonResonance2021}.
We note that the transitions around $\Delta E \approx \pm R$ will be of crucial relevance for the generation of steady-state entanglement, as we will see in the next section.
%
%
%


\subsection{Adiabatic elimination of the cavity}
We now complete our description and consider the full qubits-driving-cavity system. In this work we focus on  applications of lossy cavities for the  design of dissipative phenomena; therefore, we assume the cavity to be in the weak coupling regime, restricting ourselves to the  bad-cavity limit \cite{SavageStationaryTwolevel1988,CiracInteractionTwolevel1992,ZhouDynamicsDriven1998}, $\kappa \gg g \gg \gamma$. 
In particular, we fix $g=0.1\kappa$ in all the results shown in this text.  In this regime, we can perform an adiabatic elimination of the cavity degrees of freedom and obtain an effective model for the reduced two-qubit system. Following standard adiabatic elimination techniques~ \cite{BreuerTheoryOpen2007,RivasOpenQuantum2012,GardinerQuantumNoise2004} (details can be found in Appendix~\ref{appendix:A}), one obtains a master equation for the qubits where the effect of the cavity takes the form of Bloch-Redfield equations \cite{WhitneyStayingPositive2008,JeskeBlochRedfieldEquations2015}, %
\begin{multline}
 \frac{d \hat{\rho}}{dt}=-i[\hat{H}_q+\hat{H}_d,\hat{\rho}]+\sum_{i,j=1}^2\frac{\gamma_{ij}}{2}\mathcal{L}[\hat{\sigma}_i,\hat{\sigma}_j]\hat{\rho}
\\ + \sum_{i,j,m,n=1}^4 \left( \frac{g_{ij}g_{mn}^* }{\kappa/2+i(\Delta_a-\omega_{ij})}\left[\hat \sigma_{ij}\hat \rho,\hat \sigma_{mn}^\dagger\right] + \text{H.c.} \right).
	\label{eq:Nakajima}
\end{multline}
Here,  we defined $\hat\sigma_{ij}\equiv |j\rangle \langle i |$,  $g_{ij}\equiv g\langle j| \hat \sigma_1 +\hat \sigma_2 |i\rangle$, and $\omega_{ij}=\lambda_i-\lambda_j$, where $|i\rangle$ and $\lambda_i$, $i=1,\ldots,4$, are the eigenvectors and eigenvalues of the dressed qubit-laser system.

A characteristic quantity that emerges in several limits of Eq.~\eqref{eq:Nakajima} is the effective decay rate $ \Gamma_P$ induced in the qubits by the coupling to the lossy cavity, which is given by the standard expression of the Purcell rate~\cite{PurcellResonanceAbsorption1946,KavokinMicrocavities2017},
\begin{equation}
\Gamma_P \equiv \frac{4g^2}{\kappa}.
\label{eq:purcell-rate}
\end{equation}
In order to assess to what extent the losses in the system are channeled through the cavity, one usually defines the cavity cooperativity as the ratio between the Purcell rate and the spontaneous emission rate,
\begin{equation}
C \equiv \frac{4g^2}{\kappa \gamma},
\end{equation}
so that a value $C>1$ indicates that the cavity is providing an enhancement of the decay rate. Both definitions will be used extensively throughout the text. 


\section{Steady-state entanglement generation mechanisms}
\label{sec:entanglement-mechanism}
%
%
%
\begin{figure}[t!]
	\includegraphics[width=0.45\textwidth]{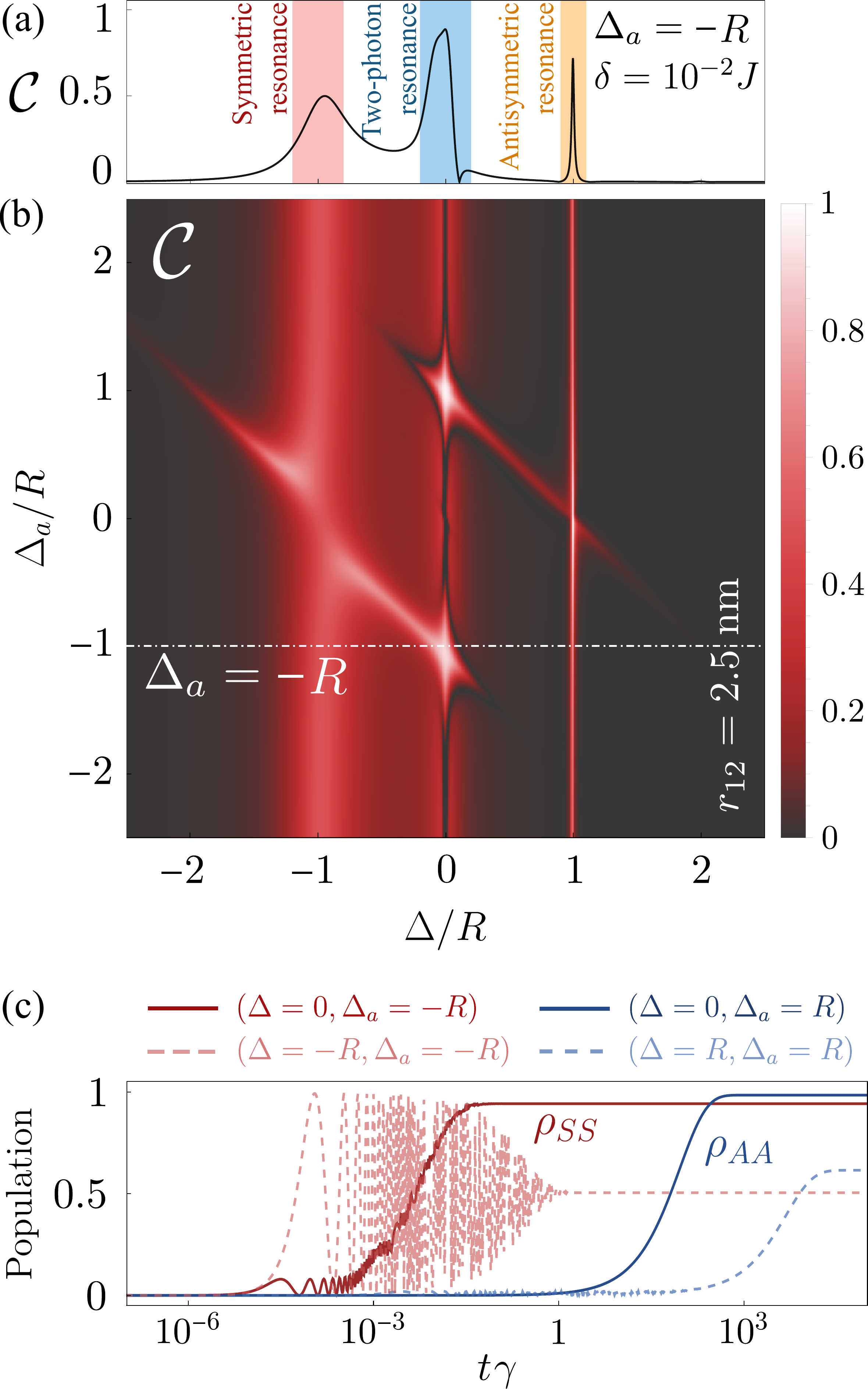}
	\caption{ Steady-state concurrence versus (a) laser-qubit detuning $\Delta$ 
	 and (b) both $\Delta$ and laser-cavity detuning $\Delta_a$ (b). In (a), the laser-cavity detuning is fixed at $\Delta_a=-R$ [dashed line in (b)]. 
	 (c) Time evolution of the symmetric ($\rho_{SS}\equiv \langle S | \hat  \rho |S\rangle$, red curves) and antisymmetric ($\rho_{AA}\equiv \langle A | \hat  \rho |A\rangle$, blue curves) populations for different qubit-laser and cavity-laser resonances. Straight lines correspond to two-photon resonance condition ($\Delta=0$), and dashed lines correspond to one-photon resonance  condition ($\Delta=\pm R$).
		Parameters (a-c): $r_{12}=2.5\ \text{nm}$, $k=2\pi/780\ \text{nm}^{-1}$, $J=9.18\times 10^4\gamma$, $\gamma_{12}=0.999\gamma$, $\delta=10^{-2}J$, $R=9.18\times 10^4\gamma$, $\Omega=10^4 \gamma$, $\kappa=10^4 \gamma$, $g=10^{-1}\kappa$. }
	\label{fig:Fig3_ConcurrenceQubitLaserDetuning}
\end{figure}
We will now discuss the different regimes of stationary entanglement emerging in this system, focusing in the condition of two-photon excitation. We will compute the stationary state of the system by solving the master equation in Eq.~\eqref{eq:full_master_eq} and quantify the degree of steady-state entanglement between the emitters by means of the concurrence $\mathcal{C}$, which is a widely used measure for detecting entanglement in bipartite systems~\cite{WoottersEntanglementFormation1998,WoottersEntanglementFormation2001, PlenioIntroductionEntanglement2007,HorodeckiQuantumEntanglement2009}.

\begin{figure*}[t]
	\begin{center}
		\includegraphics[width=1.\textwidth]{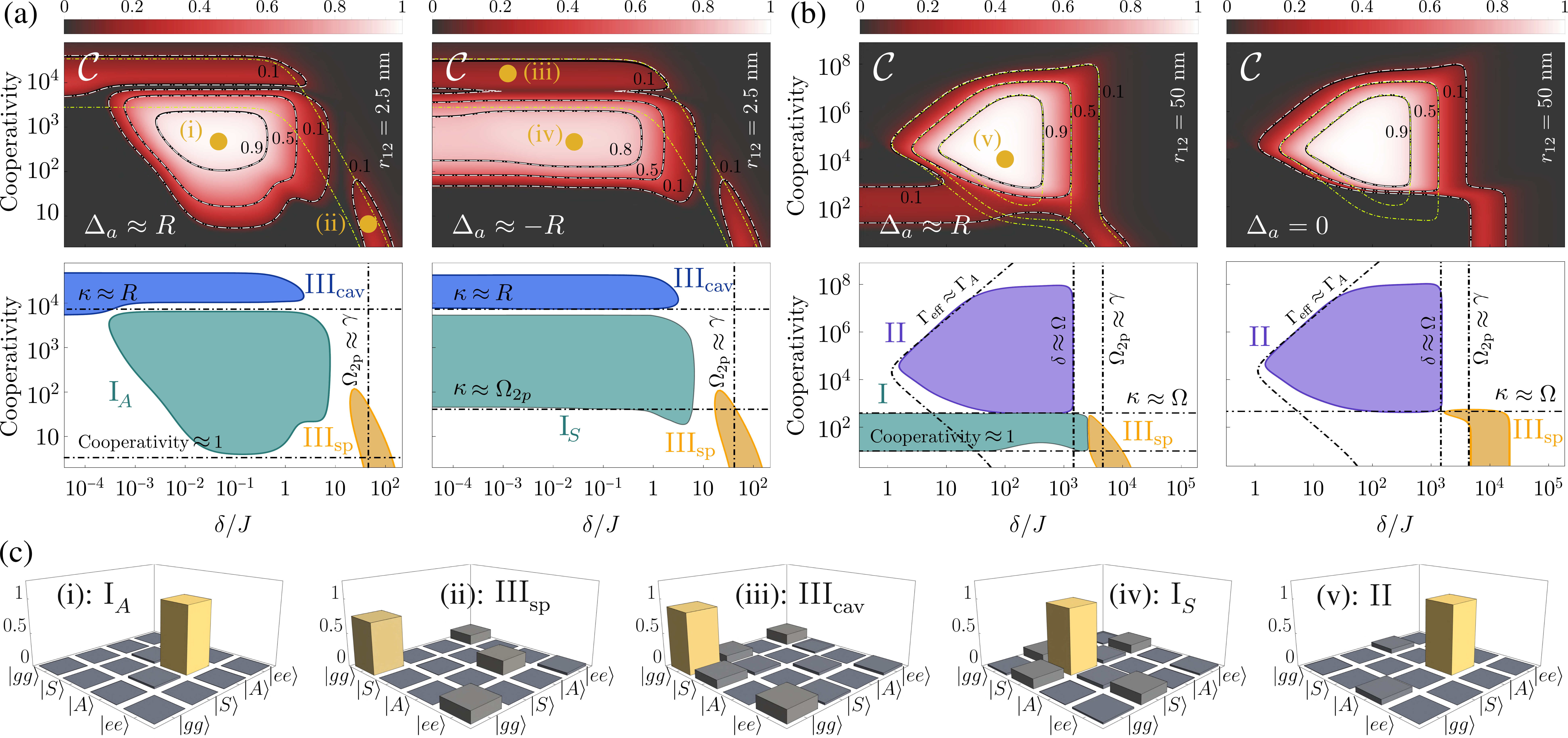}
	\end{center}
	\caption{Entanglement regimes. Steady-state concurrence (top row) and diagram of entanglement mechanisms (bottom row) versus cooperativity and qubit-qubit detuning $\delta$ for four different configurations: (a) cavity in resonance with the antisymmetric $\Delta_a=\omega_{34} \approx R$ (left) and symmetric $\Delta_a=\omega_{21}\approx -R$ (right) transitions, with $r_{12}=2.5\ \text{nm}$; (b) cavity in resonance with the antisymmetric transition $\Delta_a=\omega_{34} \approx R$ (left) and the two-photon resonance $\Delta_a=0$ (right), with $r_{12}=50\ \text{nm}$. In the concurrence plots, straight lines correspond to numerical computations, grey-dashed lines are numerical predictions from the adiabatic model in Eq.~\eqref{eq:Nakajima}, and the yellow-dashed lines are numerical predictions from the approximated collective model in Eq.~\eqref{eq:CollectivePurcell}. The diagrams of entanglement mechanisms include dot-dashed lines outlining the regions of operation of each mechanism presented in Table~\ref{tab:EntanglementTab}. (c) Absolute value of the steady-state density matrix corresponding to points (i)-(v) in panel (a).
		Parameters: (a) $r_{12}=2.5\ \text{nm}$,  $J=9.18\times 10^4\gamma$, $\gamma_{12}=0.999\gamma$; (b) $r_{12}=50\ \text{nm}$, $J=10.65 \gamma$, $\gamma_{12}=0.967\gamma$;  (a-b) $k=2\pi/780\ \text{nm}^{-1}$, $\Delta=0$, $\Omega=10^4 \gamma$, $g=10^{-1}\kappa$. }
	\label{fig:Fig9_EntanglementRegimes}
\end{figure*}

\subsection{Dependence with laser frequency}
\label{sec:laser-freq}
As a first step of our analysis, we study the impact of the choice of laser frequency in the generation of entanglement and justify our decision to focus on the two-photon excitation regime $\Delta=0$ throughout this work.

The stationary value of $\mathcal C$ versus qubit-laser detuning $\Delta$ and cavity-laser detuning $\Delta_a$ features clear resonant patterns, as we depict in Fig.~\ref{fig:Fig3_ConcurrenceQubitLaserDetuning}(a,b). In Fig.~\ref{fig:Fig3_ConcurrenceQubitLaserDetuning}(a), we plot a cut versus $\Delta$ fixing $\Delta_a=-R$ 
showing that the concurrence $\mathcal C$ versus qubit-laser detuning features three peaks centred at $\Delta=\{-R,0,R\}$. 

\emph{One-photon resonances---.} The leftmost and rightmost peaks, that we label as ``symmetric'' and ``antisymmetric'', are explained by the fact that, at laser-qubit detunings $\Delta=\pm R$, the laser is resonant with the transition between the ground state and one of the one-excitation eigenstates, which for strongly coupled qubits correspond to the superradiant and subradiant
 states $|\pm\rangle \approx |S/A\rangle$. In this case, the dynamics  gets approximately confined within a reduced subspace involving only the ground state and the symmetric state $\{|gg\rangle, |S\rangle \}$, or the ground state and the antisymmetric state $\{|gg\rangle, |A\rangle \}$. Since in the symmetric and the antisymmetric resonances the qubits are excited only by populating either the state $|S\rangle$ or $|A\rangle$, respectively, and these are maximally entangled states, the finite population of these states leads to high values of the stationary concurrence. These two resonances feature very different widths due to their respective superradiant and subradiant nature \cite{FicekQuantumInterference2005}. Here, the cavity does not play a significant role: while it slightly modifies the effective decay rates via Purcell effect, it is not responsible of the mechanism of entanglement generation.
 This type of direct resonant excitation of the subradiant and superradiant states has been reported experimentally, e.g., in Ref.~\cite{TrebbiaTailoringSuperradiant2022}.

\emph{Two-photon resonance---.} On the other hand, the highest value of the concurrence is obtained when the system is driven at the two-photon resonance, i.e., $\Delta=0$, with the concurrence reaching values close to the maximum achievable entanglement $\mathcal{C}\approx 1$ when the cavity frequency is set at one of the two resonant frequencies $\Delta_a \approx \pm R$. These two resonance conditions correspond, to the steady-state stabilization of the states $|A\rangle$ and $|S\rangle$, respectively. Notably, the stabilization time---the time the system takes to reach the steady state---is several orders of magnitude faster than the time required by exciting the corresponding one-photon resonances. This is clearly seen in Fig.~\ref{fig:Fig3_ConcurrenceQubitLaserDetuning}(c), where we show the time evolution of the populations of the symmetric and antisymmetric states when exciting the one-photon resonances $(\Delta=\pm R)$ and the two-photon resonance $(\Delta=0,\Delta_a =\pm R)$, demonstrating that in the second case, the stabilization occurs orders of magnitude faster. 
The entanglement created at the two-photon resonance is, furthermore, the only one that survives as the distance between emitters increases and the one-excitation eigenstates lose their superradiant/subradiant character $|\pm\rangle \neq |S/A\rangle$ (see Appendix~\ref{appendix:B}). All these remarkable properties justify that, in this work, we focus our attention on the regime of resonant two-photon excitation $\Delta=0$.

We remark that, in contrast to the symmetric/antisymmetric resonances, the entanglement stabilization at the two-photon resonance that we have observed is enabled by the cavity, as we discuss in following sections in detail. 
We note that cavity-enabled mechanisms of entanglement stabilization in systems under two-photon excitation have also been proposed in biexciton-exciton cascades for the case of single QDs~\cite{SeidelmannDifferentTypes2021,
SeidelmannTimedependentSwitching2021,
SeidelmannPhononinducedTransition2023}.

\subsection{Mechanisms of entanglement generation}
Based on the previous discussion, we fix our attention on the two-photon resonance ($\Delta=0$) henceforth. The example of high entanglement stabilization  that we just discussed represents only a specific point in the phase space  spanned by varying parameters such as the cavity-laser detuning $\Delta_a$, the cavity decay rate $\kappa$, the drive amplitude $\Omega$, or the strength of the emitter-emitter interaction $J$ relative to their detuning $\delta$. A systematic exploration of this parameter space reveals a complex landscape with a wealth of regions of high steady-state entanglement, which originate from different mechanisms  of entanglement generation, as we show in Fig.~\ref{fig:Fig9_EntanglementRegimes}. 
In the upper panels of this figure we plot the steady-state concurrence $\mathcal C$ as a function of the qubit-qubit detuning $\delta$ and the cavity cooperativity $C$ (while fixing $g=0.1\kappa$), for different values of the cavity detuning $\Delta_a$, and the distance between emitters $r_{12}$.  Specifically, $r_{12}$ impacts the values of $J$ and $\gamma_{12}$.
The straight contour lines correspond to numerical calculation with the full model from  Eq.~\eqref{eq:full_master_eq}, while the white-dashed contour lines correspond to numerical calculation with the adiabatic model from Eq.~\eqref{eq:Nakajima}, which confirms its validity over all the regimes of parameters considered in this work. 
In the bottom row of Fig.~\ref{fig:Fig9_EntanglementRegimes}, we show regions in the parameter phase-space corresponding to three main different mechanisms of entanglement generation. We discuss these mechanisms, as well as a metastable mechanism of entanglement generation, in the following sections.

\subsubsection*{Mechanism I: Frequency-Resolved Purcell enhancement}
The first mechanism that we present, referred to as Mechanism I, 
corresponds to the frequency-resolved Purcell effect that we present in Ref.~\cite{Vivas-VianaFrequencyresolvedPurcell2023}. It is characteristic of situations in which the emitters have a strong interaction in the absence of the cavity (e.g., via dipole-dipole interaction), forming a dimer. This occurs if $J \gg \delta$, so that $\beta \ll 1 $ and the one-excitation eigenstates are close to the perfectly superradiant and subradiant states, $|+\rangle \approx |S\rangle$ and $|-\rangle \approx |A\rangle$. This can be the case, for instance, in molecules found in close proximity~\cite{HettichNanometerResolution2002, TrebbiaTailoringSuperradiant2022} or molecular dimers assembled via insulating bridges~\cite{DiehlEmergenceCoherence2014}, which can reach distances of just a few nanometers and experience strong dipole-dipole coupling. Other possible candidates are quantum dot molecules~\cite{KrennerDirectObservation2005,ReitzensteinCoherentPhotonic2006,ArdeltOpticalControl2016}. This strong-interacting case is shown in Fig.~\ref{fig:Fig9_EntanglementRegimes}(a), where we consider two dipoles with optical resonant frequencies separated by 2.5 nm.  The first evidence of entanglement at the two-photon resonance that we presented in Fig.~\ref{fig:Fig3_ConcurrenceQubitLaserDetuning} also belongs to this type of  mechanism.

\begin{figure}[t!]
	\includegraphics[width=0.48\textwidth]{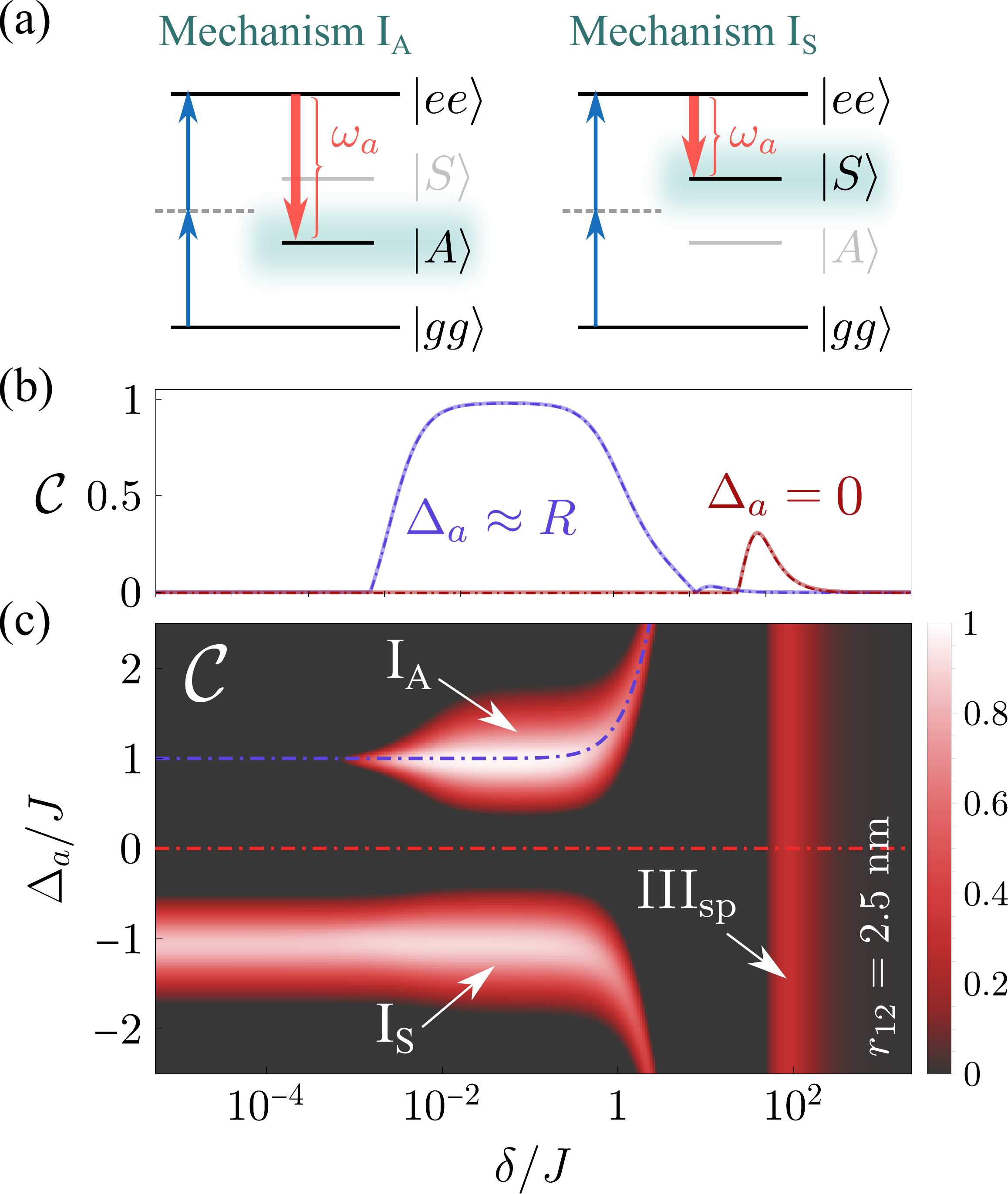}
	\caption{(a) Sketches of Mechanism I for the stabilization of the states $|A\rangle$ and $|S\rangle$. (b)
	Steady-state concurrence versus laser-cavity detuning $\Delta_a$ and qubit-qubit detuning $\delta$ when the two-qubit system is driven at the two-photon resonance $\Delta=0$. Top panel corresponds to a cut of the concurrence when the cavity is in resonance with some frequencies: antisymmetric transition (blue) and the two-photon resonance (red).
	Parameters (b-c): $r_{12}=2.5\ \text{nm}$, $k=2\pi/780\ \text{nm}^{-1}$, $J=9.18\times 10^4\gamma$, $\gamma_{12}=0.999\gamma$, $\Delta=0$, $\Omega=10^4 \gamma$, $\kappa=10^4 \gamma$, $g=10^{-1}\kappa$. }
	\label{fig:Fig4_ConcurrenceQubitCavityDetuning}
\end{figure}

Mechanism I relies on the cavity providing a Purcell enhancement of the decay from the doubly excited state $|ee\rangle$ to the superradiant state $|S\rangle$ or the subradiant state $|A\rangle$, both of which are maximally entangled states. This enhancement is activated when the cavity frequency is close to one of the two main transition energies in the dimer, $\Delta_a \approx \pm R$. Specifically, when $\Delta_a = R$, the cavity enhances resonantly the transition $|ee\rangle \rightarrow |A\rangle$, and when $\Delta_a = -R$, it enhances the transition $|ee\rangle \rightarrow |S\rangle$ [see Fig.~\ref{fig:Fig4_ConcurrenceQubitCavityDetuning}(a)]. 
By combining the direct two-photon excitation of the doubly-excited state $|ee\rangle$ and the subsequent cavity-enhanced decay towards either $|S\rangle$ or $|A\rangle$, a stationary occupation probability of these states close to one can be reached. We label each of these two possibilities as Mechanism I$_S$ and I$_A$, depending on whether the cavity enhances the decay towards $|S\rangle$ or $|A\rangle$, respectively. 
In Ref.~\cite{Vivas-VianaFrequencyresolvedPurcell2023}, we show that this mechanism can also be implemented with incoherent driving, in which case it can be extended to an arbitrary number of emitters to generate entangled $W$ states.

By utilizing the Bloch-Redfield equation Eq.~\eqref{eq:Nakajima}, we have demonstrated (see Appendix~\ref{appendix:A2}) that, when resonances activating either Mechanism I$_A$ or Mechanism I$_S$ are selected, the effective contribution of the cavity to the dynamics of the emitters is described by a single Lindblad term 
\begin{equation}
\mathcal L_\mathrm{eff}[\hat\rho] = \Gamma_P\mathcal D[\hat\xi_{A/S}]\hat\rho,
\end{equation}
with jump operators $\hat\xi_A\equiv |gg\rangle\langle + | - \frac{\beta}{2}|-\rangle \langle ee|$ for the case of Mechanism I$_A$, and $\hat\xi_S \equiv -|+\rangle \langle ee| + \frac{\beta}{2}|gg\rangle\langle -|$ for the case of Mechanism I$_S$. This expression of the effective dynamics, rigorously derived in Appendix~\ref{appendix:A2}, provides analytical insights into the process of entanglement stabilization. 

In particular, we find that the steady state population of the entangled subradiant state $|-\rangle \approx |A\rangle$ via Mechanism I$_A$ is given by
\begin{equation}
\rho_{A,\mathrm{ss}} = \frac{\Gamma_{\mathrm I,A}}{\Gamma_{\mathrm I,A}+\gamma_-},
\end{equation}
where $\Gamma_{\mathrm{I},A}\equiv \beta^2\Gamma_P/2$ is the effective decay rate from $|ee\rangle$ to $|-\rangle$ provided by the cavity, and $\gamma_- = \gamma - \gamma_{12}\cos\beta$ is the subradiant decay rate. This mechanism becomes efficient when $\Gamma_{\mathrm{I},A}> \gamma_-$, which for $\beta \ll 1$ sets the condition
\begin{equation}
\beta^2 > \frac{2}{C}\left(1-\frac{\gamma_{12}}{\gamma}\right).
\label{eq:beta_min_IA}
\end{equation}
This minimum required value of $\beta$ explains why a finite detuning $\delta$ is necessary to stabilize the state $|-\rangle$, as can be clearly seen in the bottom-left of Fig.~\ref{fig:Fig9_EntanglementRegimes}(a) and in Fig.~\ref{fig:Fig4_ConcurrenceQubitCavityDetuning}(c). This finite detuning prevents the state $|-\rangle$ from being entirely subradiant and uncoupled from the cavity. We also obtain that this state is stabilized in a timescale which, in the limit $\Gamma_{\mathrm I,A}\gg \gamma_-$ is given by 
\begin{equation}
\tau_{\mathrm I,A}\approx 2/\Gamma_{\mathrm I,A} = \frac{4}{\beta^2 \Gamma_P}.
\label{eq:tau_IA}
\end{equation}

On the other hand, we find that the steady-state population of the entangled superradiant state $|+\rangle \approx |S\rangle$ formed via Mechanism I$_S$ is given by
\begin{equation}
\rho_{S,\mathrm{ss}} = \frac{1}{1 + \gamma_+\left(P_S^{-1}+\Gamma_P^{-1}\right)},
\end{equation}
where we defined the effective pumping rate $P_S \equiv 2\Omega_\mathrm{2p}^2/\Gamma_P$ and the superradiant decay rate $\gamma_+\equiv \gamma +\gamma_{12}\cos\beta$. In this case, the mechanism will be efficient when $P_S> \gamma_+$, which sets the condition
\begin{equation}
\left(\frac{\Omega_\mathrm{2p}}{\gamma}\right)^2 > C.
\end{equation}
Here, the timescale of stabilization in the efficient regime $P_S\gg \gamma_+$ is directly given (see Appendix~\ref{appendix:A2}) by
\begin{equation}
\tau_{\mathrm I, S} =  \left(\frac{2}{\Gamma_P}\right)\frac{1}{1 - \text{Re}\sqrt{1-(2\Omega_\mathrm{2p}/\Gamma_P)^2}},
\label{eq:tau_IS}
\end{equation}
where $\text{Re}(*)$ denotes real part.
Notice the two significant limits of this equation: when $\Gamma_P < 2\Omega_\mathrm{2p}$, we have $\tau_{\mathrm I, S} =  2/\Gamma_P$, while when $\Gamma_P \gg \Omega_\mathrm{2p}$, one obtains $\tau_{\mathrm I, S}\approx \Gamma_P/\Omega_\mathrm{2p}^2$. 

There are three primary conditions for Mechanism I to occur, that we list below:
\begin{enumerate}
\item  The cavity must be able to resolve the main transitions in the dimer taking place at $\omega_0 \pm R$ in order to enhance them selectively. This requires $\kappa \ll R$.
\item To ensure that the cavity significantly enhances transition rates with respect to spontaneous emission, the cooperativity parameter should be greater than one, i.e., $C>1$.
\item  The dimer should be strongly coupled, i.e., $J\gg\delta$. 
\end{enumerate}

\begin{figure}[b]
	\includegraphics[width=0.48\textwidth]{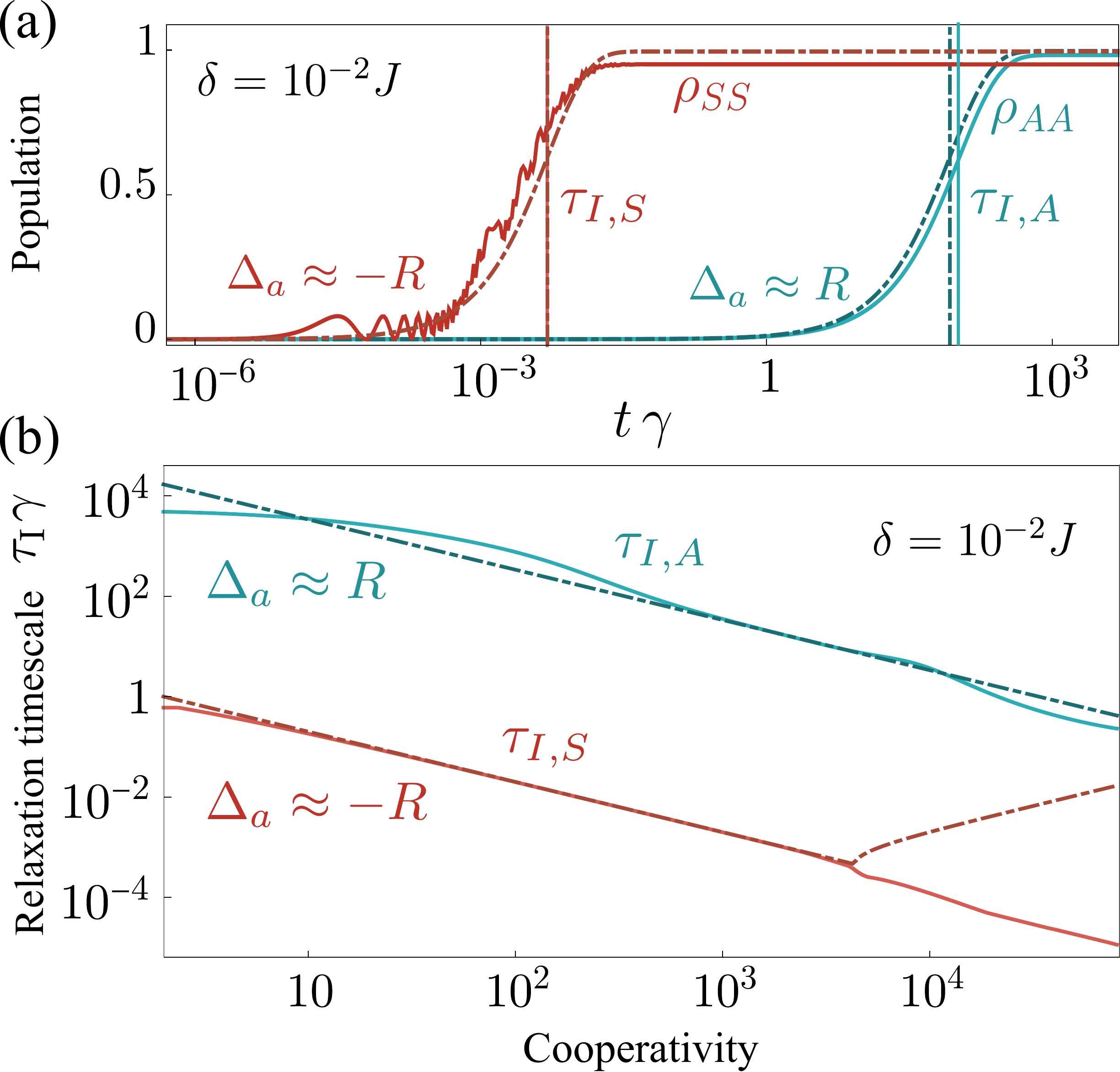}
	\caption{(a) Time evolution of the density matrix elements when the cavity is in resonance with the antisymmetric transition $\Delta_a=\omega_{34}\approx R$ in blue ($\rho_{AA}\equiv \langle A | \hat  \rho |A\rangle$, antisymmetric state) and the symmetric transition $\Delta_a=\omega_{21} \approx -R$ in red ($\rho_{SS}\equiv \langle S | \hat  \rho |S\rangle$, symmetric state). 
		(b) Relaxation time towards the entangled steady-state versus the cooperativity when the cavity is in resonance with the antisymmetric transition $\Delta_a=\omega_{34}\approx R$ (blue) and the symmetric transition $\Delta_a=\omega_{21} \approx -R$ (red).
		Parameters (a-b): $r_{12}=2.5\ \text{nm}$, $k=2\pi/780\ \text{nm}^{-1}$, $J=9.18\times 10^4\gamma$, $\gamma_{12}=0.999\gamma$, $\delta=10^{-2}J$, $\Delta=0$, $\Omega=10^4 \gamma$, $g=10^{-1}\kappa$. }
	\label{fig:Fig5_RelaxationRates}
\end{figure}

Furthermore, there are specific requirements for Mechanism I$_A$ and Mechanism I$_S$ in addition to the previously mentioned conditions. For the case of Mechanism I$_A$, we required a finite detuning $\delta$ to break perfect subradiance and fulfill the condition $\Gamma_{\mathrm I, A}>\gamma_-$, which sets the minimum for $\beta$ in Eq.~\eqref{eq:beta_min_IA}. On the other hand, for  Mechanism I$_S$, the cavity should not able to resolve the fine dressed-state structure of the system, i.e., $\kappa \gg \Omega_{\mathrm{2p}}$, to ensure that all possible transitions towards $|+\rangle$ are enhanced by the cavity and the transitions towards $|-\rangle$ are subsequently quenched, since otherwise $|-\rangle$ would acquire a non-negligible population due to its long lifetime.
This is in stark contrast to the case the stabilization of $|A\rangle$ via activation of Mechanism I$_A$, which benefits from the subradiant nature of $|-\rangle$. Even if the cavity is able to resolve the energy levels dressed by the laser, split by $\Omega_\mathrm{2p}$, the entangled state $|-\rangle \approx |A\rangle$ will be stabilized as long as $\Delta_a = \omega_{24} = R$ or $\Delta_a = \omega_{34} = R-2\Omega_\mathrm{2p}$, which correspond to the transitions $|A_2\rangle \rightarrow |A\rangle$ and $|S_2\rangle \rightarrow |A\rangle$, respectively. Further calculations of the concurrence in this limit can be found in Appendix~\ref{appendix:A2}.
The phase diagram shown in Fig.~\ref{fig:Fig9_EntanglementRegimes}(a) illustrates the upper, lower, leftmost, and rightmost limits for the green regions, which correspond to the various conditions discussed above.

Regarding the relaxation time into the entangled steady state, it is clear from the different expressions in Eqs.~\eqref{eq:tau_IA} and \eqref{eq:tau_IS} that 
the choice between the symmetric or antisymmetric transition has a strong impact on the timescale of stabilization, a difference that stems from the superradiant and subradiant nature of the corresponding target states. This is explicitly shown in Fig.~\ref{fig:Fig5_RelaxationRates}(a). In particular, the ratio between both quantities $r_\tau \equiv \tau_{\mathrm I,S}/\tau_{\mathrm I,A}$ will be given by 
\begin{equation}
r_\tau = 
\frac{\beta^2/2}{1 - \text{Re} \sqrt{1-(2\Omega_\mathrm{2p}/\Gamma_P)^2}},
\end{equation}
 which, for the particular set of parameters chosen in Fig.~\ref{fig:Fig5_RelaxationRates}, yields $r_{\tau} = \beta^2 \sim 10^{-4}$. This implies that 
the relaxation towards the symmetric state is four orders of magnitude faster than the relaxation towards the antisymmetric state. This observation is further supported by Fig.~\ref{fig:Fig5_RelaxationRates}(b), where we evaluate the relaxation timescales $\tau_\mathrm{I, A} $ and $\tau_\mathrm{I,S}$  with respect to the cooperativity $C$, confirming a ratio $\propto 10^{-4}$ between them. Our analytical calculations show a good agreement with the corresponding timescales computed numerically through direct diagonalization of the Liouvillian. We only observe a discrepancy for $\tau_{\mathrm{I},S}$ at large values of $C>10^3$, where, nevertheless, this mechanism is no longer activated, since $\kappa > R$.

\subsubsection*{Mechanism II. Collective Purcell enhancement}
The next mechanism that we consider, that we refer to as Mechanism II, occurs when the cavity is not able to resolve any of the transitions taking place within the dressed dimer, i.e., when $\kappa \gg R, \Omega$ (this includes possible dressed-state transitions when $\Omega\gg R$). Importantly, this mechanism does not require any coherent coupling between the quantum emitters; in other words, even if $J\ll \delta$ so that $R\approx \delta$, the process will be activated provided that $\delta < \kappa$. Hence, this mechanism is particularly relevant in scenarios involving weakly interacting non-identical emitters, which commonly arise in solid-state quantum systems such as quantum dots or molecules. In these systems, precise control over the emitter positioning is often challenging, resulting in emitters with different frequencies that are too separated to present any significant interaction. In this limit, the main effect brought in by the cavity is a Purcell enhancement of the collective decay along the symmetric state $|S\rangle$ which, in combination with the drive, results in a stabilization of the entangled antisymmetric state $|A\rangle$ via a mechanism that we will discuss in detail below.
We note that this effect has been previously described in the context of waveguide QED~\cite{PichlerQuantumOptics2015} and cavity QED~\cite{OliveiraSteadystateEntanglement2023}. Here, we offer a comprehensive explanation of the underlying mechanism, providing analytical expressions of the regions in parameter phase space where it occurs, the predicted steady-state populations, and the timescale of formation of entanglement.

The generation of entanglement via this mechanism is evidenced in Fig.~\ref{fig:Fig9_EntanglementRegimes}(b), where the steady-state concurrence $\mathcal C$ is evaluated in the parameter phase space spanned by the cavity cooperativity $C$ and $\delta$ for two emitters spatially separated by 50 nm. This separation is reflected in a small dipole-dipole coupling $J$, meaning that most of the phase diagram corresponds to the situation of non-interacting emitters, with $\delta \gg J$. We observe ample regions in this parameter phase space where the system reaches very high values of the concurrence, close to the maximum possible value $\mathcal  C = 1$. These regions are shaded in purple in the phase diagrams at the bottom of  Fig.~\ref{fig:Fig9_EntanglementRegimes}(b), and labeled as II. The high values of the concurrence correspond to a high population of the entangled state $|A\rangle$.

\begin{figure}[t]
	\includegraphics[width=0.45\textwidth]{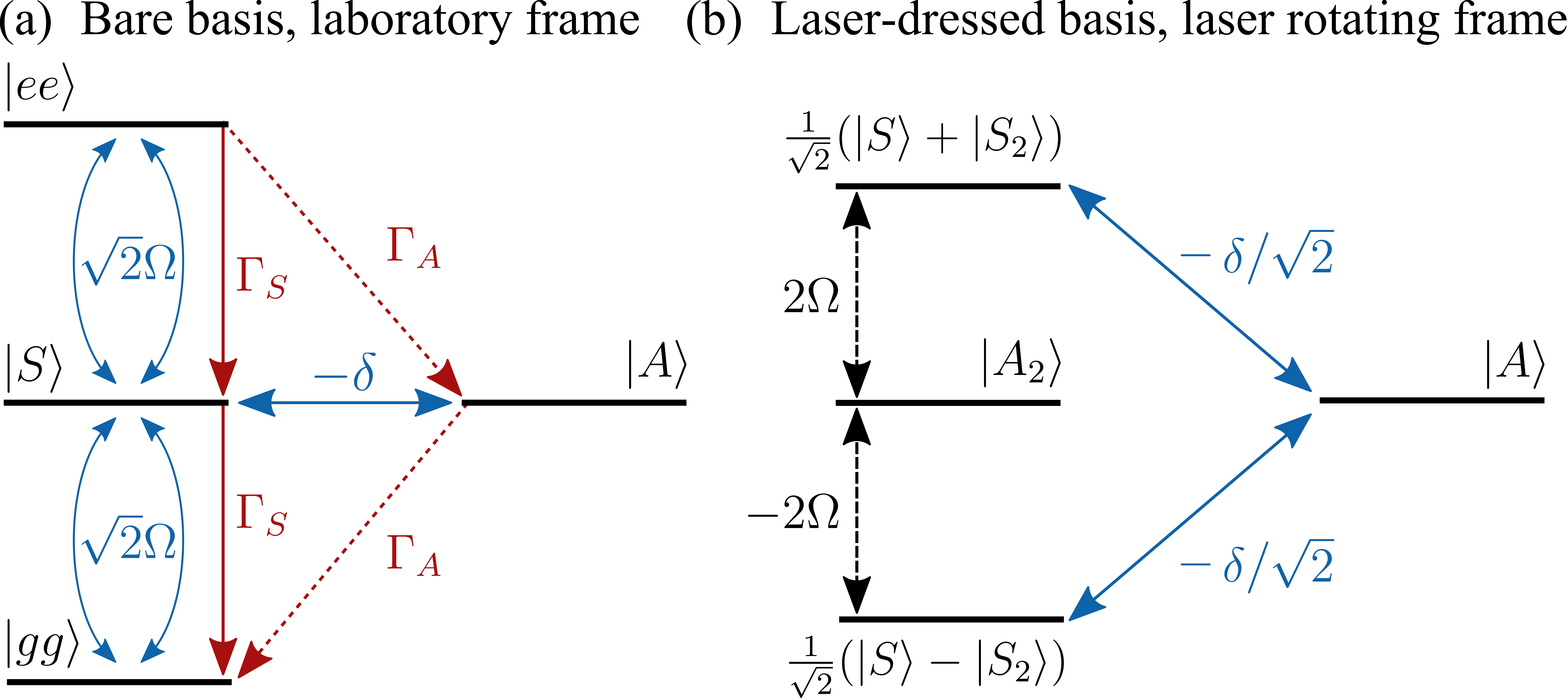}
	\caption{(a) Scheme of the qubit-laser system in the regime where Mechanism II is activated ($J\ll \delta$). Blue and red lines correspond to coherent and incoherent interactions, respectively. The symmetric channel $|gg\rangle \leftrightarrow |S\rangle \leftrightarrow |ee\rangle$ is coherently driven by the laser and dissipatively connected with an enhanced decay rate $\Gamma_S=2\Gamma_P+\gamma+\gamma_{12}$. The antisymmetric state is only coupled coherently via the qubit-qubit detuning $\delta$ and dissipatively connected with reduced decay rate $\Gamma_A\ll \Gamma_S$. (b) In the basis dressed by the laser, it becomes clear that $|A\rangle$ is detuned from the states to which it is coupled by an energy difference $2\Omega$, contributing only perturbatively to the Hamiltonian dynamics of those states as a virtual state, provided that $\Omega \gg \delta$. Incoherent processes are no longer shown. }
	\label{fig:Fig_EnergyDiagramPurcell}
\end{figure}

The origin of this mechanism resides in the collective decay induced by the cavity, which can be easily obtained by analyzing the Bloch-Redfield master equation Eq.~\eqref{eq:Nakajima}. One can see that, in the limit
 $\kappa \gg R$, all the denominators in the last line of that equation are dominated by $\kappa$, resulting in an effective master equation with a term of collective decay (details are provided in Appendix~\ref{appendix:A}):
\
\begin{equation}
	\frac{d \hat{\rho}}{dt}=-i[\hat{H}_q+\hat{H}_d,\hat{\rho}]+\sum_{i,j=1}^2\frac{\gamma_{ij}}{2}\mathcal{D}[\hat{\sigma}_i,\hat{\sigma}_j]\hat{\rho}
	 +\frac{\Gamma_P}{2}\mathcal{D}[\hat \sigma_1+\hat \sigma_2 ]\hat{\rho} ,
	\label{eq:CollectivePurcell}
\end{equation}
where $\Gamma_P$ is the standard Purcell rate, defined in Eq.~\eqref{eq:purcell-rate}.  The yellow-dashed contour lines in Fig.~\ref{fig:Fig9_EntanglementRegimes}(b) correspond to numerical calculation with the reduced effective model from Eq.~\eqref{eq:CollectivePurcell}, giving a good sense of the regimes of validity of that equation.
Writing this master equation in the completely symmetric and antisymmetric basis $\{|gg\rangle, |S\rangle,|A\rangle,|ee\rangle\}$, we find that the spontaneous decay is expressed in terms of a decay through the antisymmetric channel 
$|ee\rangle \rightarrow |A\rangle \rightarrow |gg\rangle$ with a rate $\Gamma_A\equiv \gamma -\gamma_{12}$, plus a decay through the symmetric channel $|ee\rangle \rightarrow |S\rangle \rightarrow |gg\rangle$ with a rate
\begin{equation}
\Gamma_S\equiv \gamma +\gamma_{12}+2\Gamma_P,
\label{eq:Gamma_S}
\end{equation}
which, crucially, can be greatly enhanced by the Purcell rate $\Gamma_P$ brought in by the cavity. In this basis, the Hamiltonian (with the driving at the two-photon resonance) is written in the rotating frame of the drive as $\hat H = \sqrt 2\Omega(|ee\rangle \langle S| + |S\rangle \langle gg| ) - \delta |S\rangle \langle A| + \text{H.c.}$. The resulting scheme of energy levels with the interactions and incoherent processes associated to them are depicted in Fig.~\ref{fig:Fig_EnergyDiagramPurcell}(a).

To understand this mechanism, it is useful to diagonalize the driving component proportional to $\Omega$ in the Hamiltonian. The driving couples the states $|gg\rangle$, $|ee\rangle$, and $|S\rangle$; the resulting eigenstates can be expressed in terms of the two-photon dressed-states of Eq.~\eqref{eq:s2a2eigenstates} as $|\phi_\pm\rangle = (|S_2\rangle \pm |S\rangle)/\sqrt 2$ and $|\phi_0\rangle=|A_2\rangle$. These eigenstates possess eigenenergies of $\lambda_\pm=\pm 2\Omega$ and $\lambda_0=0$, respectively. The resulting energy level structure is depicted in Fig.~\ref{fig:Fig_EnergyDiagramPurcell}(b), and it presents several characteristics that combined, leads to the stabilization of state $|A\rangle$:
\begin{enumerate}
\item $|A\rangle$ is coherently coupled with a rate $\delta/\sqrt 2$ to the states $|\phi_\pm\rangle$, which are detuned from $|A\rangle$ by an energy difference of $\pm 2\Omega$ due to the dressing by the drive. In the regime where $\Omega\gg \delta$, this coupling is substantially off-resonant, and $|A\rangle$ only affects $|\phi_\pm\rangle$ perturbatively, acting as a virtual state~\cite{Cohen-TannoudjiAtomPhotonInteractions1998}. 
\item Losses within the system predominantly induce transitions within the manifold ${ |\phi_\pm\rangle, |\phi_0\rangle }$, characterized by rates given by $\Gamma_S$, which are significantly larger than the characteristic rate $\Gamma_A$ associated with the incoherent coupling between this manifold and the state $|A\rangle$.
\end{enumerate}   

In Ref.~\cite{Vivas-VianaUnconventionalMechanism2022}, some of the present authors demonstrated  that a state satisfying these two conditions (being strongly off-resonant and having negligible dissipation) will develop a sizable population in the long-time limit. We termed this effect ``unconventional population of virtual states'', since it arguably challenges the common intuition that a virtual state---which is both out of resonance from the states involved in the dynamics and uncoupled to any incoherent channel---would remain unpopulated at all times. In the current system, this is the effect behind the formation of entanglement via the Mechanism II.
In Ref.~\cite{Vivas-VianaUnconventionalMechanism2022}, it was shown that the virtual state will slowly accumulate population through non-Hermitian evolution between quantum jumps, due to the fact that, the longer a period without quantum jumps, the more likely it is that the state is to be found in $|A\rangle$. This information provided by the absence of jumps reflects back on the state, and this effect accumulates over time eventually leading to a sizable population of the ``virtual'' state. 

Under these conditions, the system becomes metastable and develops two characteristic, dissipative timescales: a fast one, given by $\Gamma_S^{-1}$, and a slow one, given by the inverse of an effective rate $\Gamma_\mathrm{eff}^{-1}$. 
This hierarchy of timescales allowed us to develop a hierarchical adiabatic elimination (HAE) method based on the successive application of a series of adiabatic elimination procedures~\cite{Vivas-VianaUnconventionalMechanism2022}. The application of our method allows us to obtain (see Appendix~\ref{appendix:D} for details) the expression of the relaxation rate towards the antisymmetric state, given by
\begin{equation}
	\Gamma_{\text{eff}}=\frac{4\Gamma_S \delta^2}{\Gamma_S^2+24\Omega^2},
	\label{eq:Gamma_eff}
\end{equation}
which was obtained under the assumptions $\Omega \gg \delta \gg J$ and $\Gamma_A \approx 0$. The corresponding population of the state $|A\rangle$ that develops under these conditions is given by (see Appendix~\ref{appendix:D})
\begin{equation}
		\rho _{A,\mathrm{ss}}=\frac{2\Omega^2}{\delta^2+2\Omega^2}.
\end{equation}

For this mechanism to be efficient, we need the timescale of formation of the entanglement to be much faster than the decay of the state $|A\rangle$; in other words, we require $\Gamma_\mathrm{eff}\gg \Gamma_A $, which in the limit $\Omega \gg \Gamma_S$ can be rewritten as 
\begin{equation}
\Gamma_S \gg 6\Gamma_A (\Omega/\delta)^2.
\label{eq:conditionMechanismII}
\end{equation}
The essential role of the cavity in this mechanism is precisely to enable this condition; since given that we already required that $\Omega/\delta\gg 1$, it is unlikely that the condition is met without a significant Purcell enhancement of $\Gamma_S$ given by $\Gamma_P$ in Eq.~\eqref{eq:Gamma_S}.

These analytical results allow us to unambiguously establish the regimes of parameters in which Mechanism II is efficient. To summarize, we required $\kappa \gg R\approx \delta$ and $\kappa \gg \Omega$ for the cavity to provide a collective decay. We also required $\Omega \gg \delta$ to turn $|A\rangle$ into a virtual state almost decoupled from the rest of the system. This means that the strongest condition for the cavity decay rate is simply $\kappa \gg \Omega$. The final condition is that $\Gamma_\mathrm{eff}\gg \Gamma_A$. Inspecting Eq.~\eqref{eq:Gamma_eff} we see that this will impose both a lower and an upper limit to $\Gamma_S$ and, consequently, to the cavity Purcell rate $\Gamma_P$, since for values of the cooperativity $C>1$, we have $\Gamma_S\approx 2\Gamma_P$. 
Given a fixed $\Gamma_S$, the condition $\Gamma_\mathrm{eff}\gg \Gamma_A$ also poses lower and upper limits to $\delta$ and $\Omega$, respectively. In particular, it is important to notice that this condition makes it necessary to have a finite detuning between the quantum emitters, since $\delta \approx 0$ would imply $\Gamma_\mathrm{eff}\approx 0$, making the timescale of formation of entanglement divergently long. All these three main conditions are depicted as lines in the phase diagrams in the bottom of Fig.~\ref{fig:Fig9_EntanglementRegimes}(b). These lines effectively outline the frontier of the region where Mechanism II is observed, supporting our analytical results.

We have demonstrated that the effective model of collective decay, as described by Eq.~\eqref{eq:CollectivePurcell}, successfully captures the essence of Mechanism II. This model not only provides an accurate representation of quantum emitters coupled to a common cavity, but it is also applicable to other structures, including waveguides, which also lead to collective decay and Purcell enhancement phenomena~\cite{Gonzalez-TudelaEntanglementTwo2011,TiranovCollectiveSuper2023,SheremetWaveguideQuantum2023}. Indeed, waveguide QED is the context in which this mechanism was originally observed~\cite{PichlerQuantumOptics2015}. Therefore, we can conclude that the scope of our results regarding Mechanism II extend beyond purely cavity-based configurations.

\begin{figure}[t]
	\includegraphics[width=0.475\textwidth]{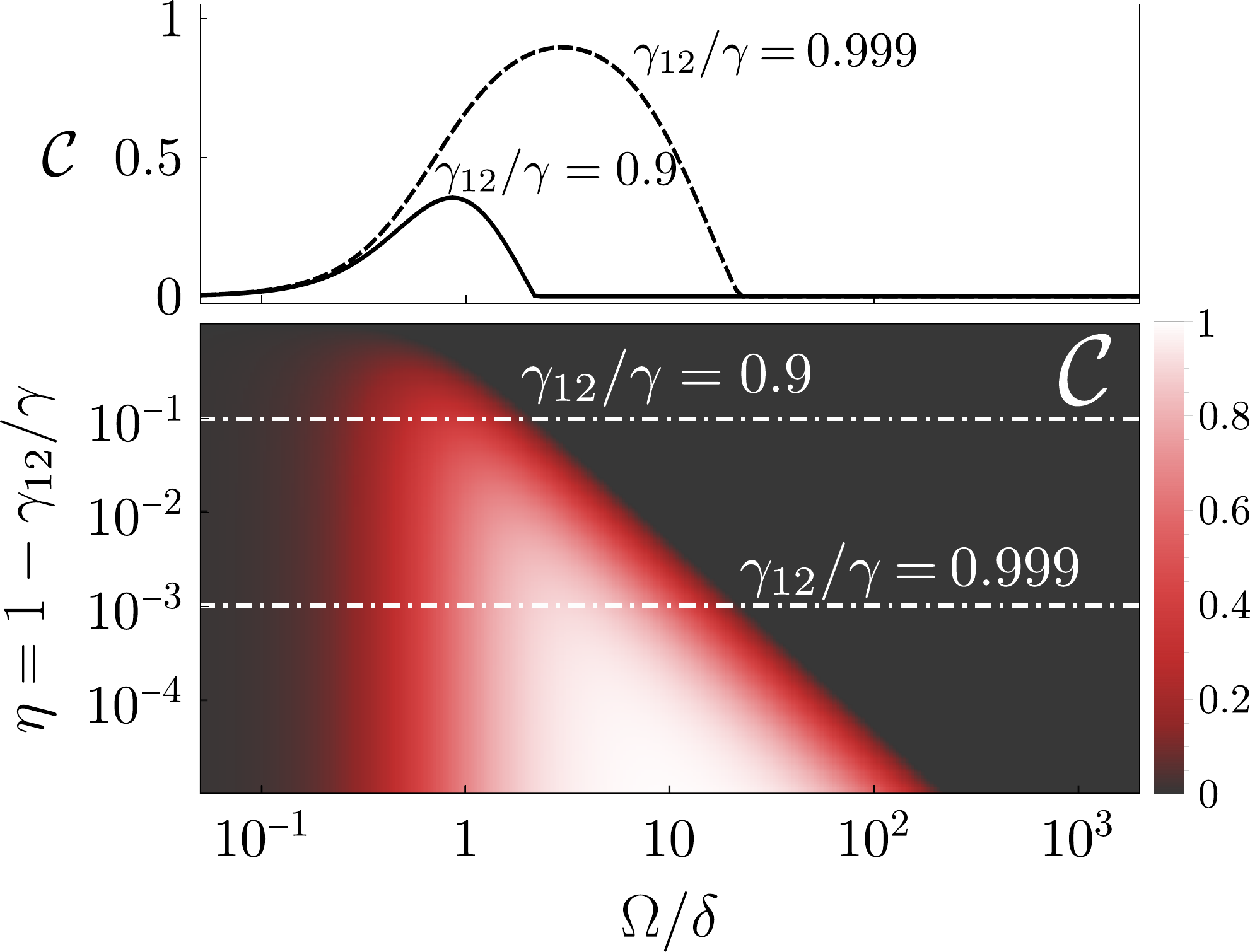}
	\caption{ Stationary concurrence for a general system of two non-identical emitters with a collective decay parametrized by $\gamma_{12}$.  The dipole-dipole coupling is disabled in this configuration and thus any dependence on the emitter-emitter distance $r_{12}$.	Parameters: $J = 0$, $\Delta=0$ , $\delta = 10^2\gamma$.}
	\label{fig:Fig_Waveguide}
\end{figure}

To explore further the possibilities of the mechanism in general platforms, we consider a simpler parametrization of the model in which $\Gamma_A = \gamma-\gamma_{12}$ and $\Gamma_S = \gamma+\gamma_{12}$, so that the ratio between $\Gamma_S$ and $\Gamma_A$ is purely controlled by the parameter $\gamma_{12}$, which can encode the dissipative coupling enabled by a structure such as a waveguide. For instance, Ref.~\cite{Gonzalez-TudelaEntanglementTwo2011} establishes the relation $\gamma_{12}/\gamma = \tilde\beta \exp[-d/(2L)]$, where $\tilde\beta$ is the standard factor that measures the fraction of the emitted radiation captured by the waveguide, $d$ is the distance between emitters, and $L$ is the propagation length of the waveguide mode.  We notice that values of $\tilde\beta > 0.9$ are within reach in state-of-the-art platforms~\cite{Gonzalez-TudelaEntanglementTwo2011,ArcariNearUnityCoupling2014}.
In Fig.~\ref{fig:Fig_Waveguide} we compute the cooperativity using this parametrization, versus  $\Omega$ and $\gamma_{12}$, for detuning between emitters set at $\delta = 10^2\gamma$, and ignoring any coherent coupling between qubits by setting $J=0$. 
These results shows that a value $\gamma_{12} = 0.9$---which would be achieved provided $\tilde\beta=0.9$ and $d\ll L$---can stabilize states with a significant steady state concurrence, reaching maximum values $\mathcal C_\mathrm{max}\approx 0.35$. Notice that this is comparable to the results of stationary concurrence reported in Ref.~\cite{Gonzalez-TudelaEntanglementTwo2011}, with the added advantage that the approach considered here enables a detuning between the emitters hundreds of times larger than their linewidths. This  holds significant implications for stabilizing entangled states among spatially separated emitters possessing distinct emission energies, without the need of tuning them in resonance by external means such as induced Stark shifts, which poses significant challenges~\cite{KoongCoherenceCooperative2022}.

\subsubsection*{Mechanism III. Two-photon resonance fluorescence}
In this section, we present an alternative mechanism (referred to as Mechanism III) for stabilizing entanglement, which targets a different type of entangled state than the ones discussed so far. Specifically, this process generates states exhibiting a coherent superposition between the ground state $|gg\rangle$ and the doubly excited state $|ee\rangle$. 
We note that this process may be present in two variants, which we label as Mechanism III$_\mathrm{sp}$ and Mechanism III$_\mathrm{cav}$, depending on whether the main decay channel is spontaneous emission or cavity-enhanced decay. 
Examples of entanglement achieved via Mechanism III can be observed in Fig.~\ref{fig:Fig4_ConcurrenceQubitCavityDetuning}, where a finite degree of entanglement is observed at large qubit-qubit detuning $\delta > J\approx 10^6\gamma$. Moreover, this mechanism is also depicted in the diagrams of the bottom row of Fig.~\ref{fig:Fig9_EntanglementRegimes}. Mechanism III results in lower levels of entanglement and reduced purity compared to the practically pure states achieved through Mechanisms I and II.  
Regarding previous reports of this mechanism in the literature, we note that the observation of stationary entanglement between non-identical emitters enabled by a plasmonic structure reported in Ref.~\cite{HaakhSqueezedLight2015} is of a similar nature as Mechanism III. However, the specific nanophotonic environment considered in that work lead to a parametrization of the decay rates  in the effective master equation of the emitters slightly different from the one obtained in this work, preventing us from drawing a full parallelism between both situations.

Regarding the underlying physics, Mechanism III is a two-photon analogue of the formation of stationary coherences of the form $\langle \hat\sigma\rangle\neq 0$ in resonance fluorescence, i.e., in a single two-level system excited by a coherent drive~\cite{SanchezMunozSymmetriesConservation2019}. In this analogue case of a single two-level system, one would find that $\langle \hat\sigma\rangle = -2i \Omega\gamma/(\gamma^2+8\Omega^2) $, which is zero for both $\Omega \ll\gamma$ and $\Omega \gg \gamma$, but non negligible when $\Omega\sim\gamma$, reaching its maximum at $\Omega_\mathrm{max} = \gamma/(2\sqrt 2)$. In Mechanism III, a similar interplay between the coherent two-photon drive of the transition $|gg\rangle \leftrightarrow |ee\rangle$, characterized by the Rabi frequency $\Omega_\mathrm{2p}$, and losses, with a characteristic rate $\gamma$, 
can enable the formation of stationary coherences between $|gg\rangle$ and $|ee\rangle$ when $\Omega_\mathrm{2p}\approx \gamma$. The resulting state takes the form:
\begin{equation}
\hat\rho \approx  (1-\epsilon)\hat\rho_\mathrm{mix} + \epsilon  |\psi_{\text{Bell}}\rangle \langle \psi_{\text{Bell}}|,
\end{equation}
where $\hat\rho_\mathrm{mix}$ is a mixed density matrix, and $ |\psi_{\text{Bell}}\rangle \equiv(|gg\rangle+ i |ee\rangle)/\sqrt{2}$. 

The first variant of this effect, Mechanism III$_\mathrm{sp}$, is solely enabled by the two-photon drive and spontaneous emission. On the other hand, the second variant, Mechanism III$_\mathrm{cav}$, requires stimulated, collective emission provided by the cavity.

\emph{Mechanism III$_\mathrm{sp}$}. This process is  inherent of the qubit-laser system and does not involve the cavity. It is, therefore, independent of the cavity-laser detuning $\Delta_a$. This mechanism is highlighted in yellow in the diagrams at the bottom of Fig.~\ref{fig:Fig9_EntanglementRegimes}, and can also be observed at large detuning in Fig.~\ref{fig:Fig4_ConcurrenceQubitCavityDetuning}(c).  
These results can be explained with a model that does not include the cavity; this simplification allows us to 
effectively describe the dynamics by a reduced system of just three energy levels, 
which yields analytical expressions for the density matrix elements and concurrence, as was recently shown in Ref.~\cite{Vivas-VianaTwophotonResonance2021}. These calculations are presented in detail in  Appendix~\ref{appendix:C}, and they align well with our numerical calculations. Our analytical treatment allows us to obtain the exact expression of the optimum detuning that maximizes the concurrence
\begin{equation}
\delta_{\text{max}}=\Omega\sqrt{2(1+\sqrt{5})J/\gamma}.
\end{equation}
This expression closely resembles the expression $\delta\approx \Omega\sqrt{2J/\gamma}$ that is directly derived from the condition $\Omega_\mathrm{2p}\approx \gamma$.  While we emphasize that this mechanism is not influenced by the cavity and thus remains mostly independent of the cavity detuning $\Delta_a$, it is important to note that the cavity can indeed interfere with it and degrade the mechanism in the case in which its frequency matches the antisymmetric resonance (as depicted by the blue line in Fig.~\ref{fig:Fig4_ConcurrenceQubitCavityDetuning}, which does not present this feature as high detuning).

\emph{Mechanism III$_\mathrm{cav}$.}  
Contrary to Mechanism III$_\mathrm{sp}$, this variant of Mechanism III, which is highlighted in blue in the diagrams at the bottom of Fig.~\ref{fig:Fig9_EntanglementRegimes}(a), is enabled by the cavity. The role of the cavity here is to enhance the spontaneous decay rates, thereby altering the conditions under which the interplay of two-photon pumping and losses can stabilize coherences between $|gg\rangle$ and $|ee\rangle$.

 Providing a systematic and analytical description of the parameter regimes where Mechanism III$_\mathrm{cav}$ is activated is a complex task beyond the scope of this text, as this feature can appear in non-perturbative limits $\Omega\sim R$ where analytical expressions of the eigenstates are unavailable. We can, nevertheless, provide general guidelines based on numerical evidence. We find that a cavity-induced collective decay---like the one responsible for Mechanism II---is necessary, i.e., $\kappa \gg (\Omega, \delta, J)$. This observation is supported by the dashed-yellow contours contours in Fig.~\ref{fig:Fig9_EntanglementRegimes}, which show that this feature is approximately described by master equation with collective decay Eq.~\eqref{eq:CollectivePurcell}. 	
Also, when $\Omega \gtrsim J$, we find that this mechanism emerges when the cavity-enabled Purcell rate is larger than the drive amplitude, $\Gamma_P > \Omega$. However, it is important to note that once $R$ becomes perturbative with respect to $\Omega$, i.e., $\Omega \gg R$,  Mechanism III$_\mathrm{cav}$ is no longer enabled, with Mechanism II being activated instead.

Some regions in parameter space may fulfil simultaneously the conditions for Mechanism III$_\mathrm{sp}$ and III$_\mathrm{cav}$. In this case, the result is a competition between processes in which the overall formation of entanglement is inhibited.

\begin{figure}[b]
	\includegraphics[width=0.45\textwidth]{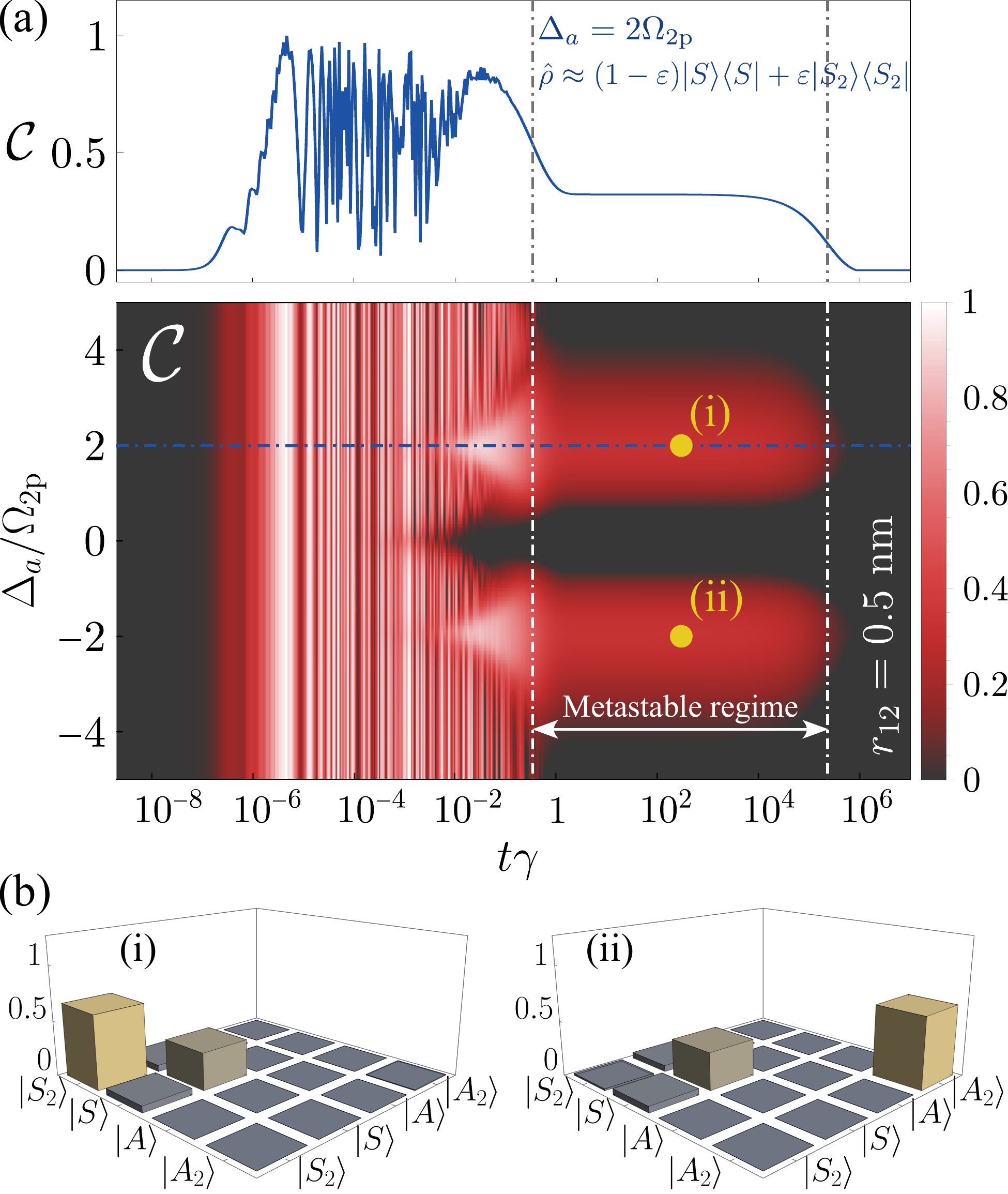}
	\caption{Time evolution of steady-state concurrence versus laser-cavity detuning $\Delta_a$. The system reaches a metastable entangled state when the cavity is in resonance with two-photon dressed energy transitions, i.e., $\Delta_a=\pm\omega_{23}=\pm 2\Omega_\mathrm{2p}$. Top panel corresponds to a cut of the concurrence when the cavity is in resonance with a dressed energy transition (blue curve). (b) Absolute value of the metastable density matrix corresponding to points (i) and (ii) in panel (a).
		Parameters: $r_{12}=0.5\ \text{nm}$, $k=2\pi/780\ \text{nm}^{-1}$, $J=1.15\times 10^7 \gamma$, $\gamma_{12}=0.999\gamma$, $\delta=10^{-4}J$, $\Delta=0$, $\Omega=10^6 \gamma$, $\Omega_{\mathrm{2p}}=1.74\times 10^5 \gamma$, $\kappa=10^5 \gamma$, $g=10^{-1}\kappa$. }
	\label{fig:Fig8_TwoPhotonEntanglement}
\end{figure}
\begin{table*}[t]
	\centering
	\caption{Summary of mechanisms of entanglement generation, conditions and entangled states. We note that the two-photon resonance condition ($\Delta=0$) is assumed in every mechanism.}
	\label{tab:EntanglementTab}
	\resizebox{\textwidth}{!}{%
		\begin{tabular}{lllll}
			\toprule
			\toprule
			\multicolumn{2}{l}{\textbf{Mechanism}}  & \multicolumn{2}{l}{\textbf{Conditions}}  & \textbf{Entangled states} \\
			\midrule
			\multicolumn{2}{l}{{I. Frequency-Resolved Purcell enhancement}} & \multicolumn{2}{l}{(i) $\text{Cooperativity}>1$}						      & $\hat  \rho_{\text{I}_A}\approx |A\rangle \langle A|$ \\
			\multicolumn{2}{l}{}                               				& \multicolumn{2}{l}{(ii) $R\gg \kappa, \delta,\Omega$ }							  &   $ \hat  \rho_{\text{I}_S}\approx |S\rangle \langle S|$ \\
			\multicolumn{2}{l}{}                               				& \multicolumn{2}{l}{(iii) $\Omega_{\text{2p}} \lesssim \kappa \lesssim J$ }  &  \\	  
			\multicolumn{2}{l}{}                                			& \multicolumn{2}{l}{(iv) $\Delta_a\approx R$, $\delta \neq 0$, and $\Gamma_{I,A}>\gamma_-$ (Mechanism $\text{I}_A$)}						  &  \\	     
			\multicolumn{2}{l}{}                                			& \multicolumn{2}{l}{(iv) $\Delta_a\approx -R$, $\kappa\gg \Omega_{\text{2p}}$, $P_S> \gamma_{S}$ (Mechanism $\text{I}_S$)}		          &  \\	    
			\midrule
			\multicolumn{2}{l}{{II. Collective Purcell enhancement}}         & \multicolumn{2}{l}{(i)  $\text{Cooperativity}>1$}   &  $\hat  \rho_{\text{II}}\approx  |A\rangle \langle A|$    \\
			\multicolumn{2}{l}{}                                             & \multicolumn{2}{l}{(ii) $\kappa \gg R, \Omega $} &             						        \\
			\multicolumn{2}{l}{}                                             & \multicolumn{2}{l}{(iii) $\Omega \gg \delta$} &              							    \\
			\multicolumn{2}{l}{}                                             & \multicolumn{2}{l}{(iv) $\Gamma_{\text{eff}} \gg \Gamma_A$} &              				    \\ 	
			\midrule
			\multicolumn{2}{l}{{III. Two-photon resonance fluorescence}}              & \multicolumn{2}{l}{Mechanism $\text{III}_{\text{sp}}$} &     $\hat  \rho_{\text{III}_{\text{sp}}}\approx (1-\varepsilon) \hat  \rho_{\text{mix}}+\varepsilon |\psi_{\text{Bell}} \rangle \langle \psi_{\text{Bell}}|$              \\ 			
			\multicolumn{2}{l}{{}}  												  & \multicolumn{2}{l}{(i)  $\delta \gtrsim J$}  					 &  where $|\psi_{\text{Bell}} \rangle \equiv 1/\sqrt{2}(|gg\rangle +i |ee\rangle)$   \\
			\multicolumn{2}{l}{}                                           			  & \multicolumn{2}{l}{(ii) $\kappa < J$}   						 &            \\
			\multicolumn{2}{l}{}                                           			  & \multicolumn{2}{l}{(iii) $\Omega_{\text{2p}} \approx \gamma$}   &              \\
			\multicolumn{2}{l}{}                                            		  & \multicolumn{2}{l}{Mechanism $\text{III}_{\text{cav}}$} &      $ \hat \rho_{\text{III}_{\text{cav}}}\approx (1-\varepsilon) \hat \rho_{\text{mix}}+\varepsilon |\psi_{\text{Bell}} \rangle \langle \psi_{\text{Bell}}|$              \\
			\multicolumn{2}{l}{{}}    												  & \multicolumn{2}{l}{(i)   $\text{Cooperativity}>1$}   &    where $|\psi_{\text{Bell}} \rangle \equiv 1/\sqrt{2}(|gg\rangle +i |ee\rangle)$   \\
			\multicolumn{2}{l}{}                                           			  & \multicolumn{2}{l}{(ii)  $\kappa\gtrsim J,\delta, \Omega$} &                \\
			\multicolumn{2}{l}{}                                            		  & \multicolumn{2}{l}{(iii) $\Gamma_P >\Omega $ when $\Omega \gtrsim J$} &                 \\
			\midrule
			\multicolumn{2}{l}{{IV. Metastable two-photon entanglement}}        & \multicolumn{2}{l}{(i)  $\text{Cooperativity}>1$  }   							& $\hat \rho_{\text{IV}}\approx (1-\varepsilon) |S\rangle\langle S| +\varepsilon  |S_2/A_2\rangle\langle S_2/A_2|$    \\
			\multicolumn{2}{l}{}                                         	    & \multicolumn{2}{l}{(ii)  $ \kappa \lesssim J, \Omega_{\text{2p}}$ }    &         when $\Delta_a\approx \pm 2\Omega_{\text{2p}} $    \\
			\multicolumn{2}{l}{}                                         	    & \multicolumn{2}{l}{(iii)  $ R\gg \Omega$ }    &           \\
			\bottomrule
			\bottomrule
		\end{tabular}%
	}
\end{table*}
\subsubsection*{Mechanism IV. Metastable two-photon entanglement.} 
We conclude our picture of the entanglement generation processes in this system by describing a metastable mechanism that arises when the cavity frequency is resonant with the direct two-photon dressed-state transitions, i.e., when $\Delta_a$ is equal to $\omega_{23}=2\Omega_\mathrm{2p}$, corresponding to the transition $|A_2\rangle \rightarrow |S_2\rangle$, and $\omega_{32}=-\omega_{23}$, corresponding to the opposite transition. In these resonant conditions, the cavity enhances said transitions and drives the system into a metastable entangled state where the final state of the transition, either $|A_2\rangle$ or $|S_2\rangle$, is highly populated. We remind the reader that these states correspond to coherent superpositions of the ground state and the doubly-excited state, see Eq.~\eqref{eq:s2a2eigenstates}.
For instance, when $\Delta_a = \omega_{23}$, a metastable state of the form $\hat  \rho \approx (1-\varepsilon)|S\rangle\langle S|  +\varepsilon |S_2\rangle\langle S_2|$ is generated. We stress that $\epsilon$ can reach sizable magnitudes, as evidenced by the high values of the concurrence shown in Fig.~\ref{fig:Fig8_TwoPhotonEntanglement}, which for that choice of parameters corresponds to $\epsilon \approx 0.66$. We observe that the lifetime of this metastable state can reach very long values of $\Delta t\sim 10^6/\gamma$.

To provide a complete and compact overview of all the mechanisms present in the space of parameters, Table~\ref{tab:EntanglementTab} summarizes the various mechanisms of entanglement generation discussed here, including their corresponding conditions and the resulting entangled steady-states.

%
%

\section{Optical tunability and detection of entanglement }
\label{sec:observation}
In this section, we discuss possible measurement schemes of the optical properties of the light emitted by the system that signal the formation of entanglement. Furthermore, we discuss how changing the amplitude of the driving field can be used to optically tune the system in and out of the resonant entanglement condition necessary for the activation of Mechanism I. 

A first element to discuss before describing experimental measurements is the impact of the coherent fraction of light generated by the 
strong driving laser, which may hinder the measurements of interest. We note that our model describes a configuration in which the two qubits are coherently driven, but the cavity is not. This situation may be challenging to realize in an experimental configuration in which the emitters are embedded into the cavity, which forces a simultaneous excitation of both systems. However, we note that the simultaneous coherent driving of both the cavity and the emitters can also be described by our model, after applying a displacement transformation into the cavity that removes its coherent fraction. In other words, our model will always describe the quantum fluctuations of the cavity field, on top of any coherent fraction that may be developed because of the driving, which we disregarded here. In the language of input-output theory~\cite{GardinerQuantumNoise2004}, the output operator will be given by
\begin{equation}
 \hat a_{\text{out}}(t) =\sqrt{\kappa}  \hat a(t) + \alpha e^{-i\omega_\mathrm L t},
\end{equation}
where $\alpha$ corresponds to the classical, coherent fraction of the output field coming from the input laser plus the coherent contribution of the cavity field. Since the protocols we proposed require rather strong driving fields, we will normally find that $|\alpha|\gg \sqrt{\langle\hat a^\dagger \hat a \rangle}$, which means that this bright coherent signal will hinder the measurement of quantum fluctuations unless it is rejected.
This is a common problem in the study of resonance fluorescence in solid state quantum optics, where the challenge has been successfully addressed via, e.g., background-suppression through orthogonal excitation-detection geometries~\cite{FlaggResonantlyDriven2009},
cross-polarization schemes~\cite{SomaschiNearoptimalSinglephoton2016} or self-homodyning of the signal, where the coherent component is eliminated through the interference with the input field~\cite{FischerSelfhomodyneMeasurement2016,FischerOnChipArchitecture2017,HanschkeOriginAntibunching2020}.
In the following, we assume that one of these approaches has been performed, and describe the direct measurement of quantum fluctuations given by an output operator without a coherent fraction, $\hat a_\mathrm{out} = \sqrt\kappa \hat a$.

\begin{figure*}[t]
	\begin{center}
		\includegraphics[width=1\textwidth]{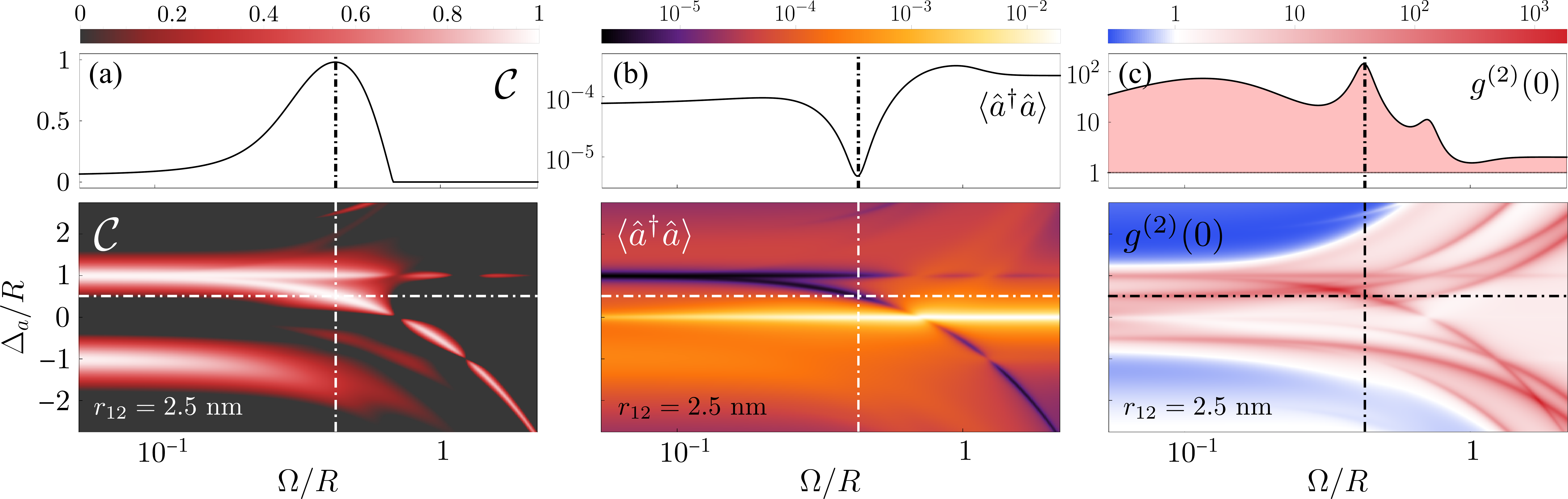}
	\end{center}
	\caption{Entanglement detection and control for Mechanism I: stationary concurrence and optical properties of the emitted light by the cavity.
		(a), (b), (c) Steady-state concurrence, transmission intensity and photon correlations versus the Rabi frequency of the laser $\Omega$ and the laser-cavity detuning $\Delta_a$. Top panels show a cut along $\Omega$, marked by the horizontal white-dashed lines in the density plots. Parameters:  $r=2.5$ nm, $k=2\pi/780\ \text{nm}^{-1}$, $J=9.18\times 10^4\gamma$, $\gamma_{12}=0.999\gamma$, $\delta=10^{-2}J$, $R=9.18\times 10^4\gamma$, $\Delta=0$, $\kappa =10^4 \gamma $, $g=10^{-1} \kappa$.}
	\label{fig:Fig10_Observability}
\end{figure*}
Figure~\ref{fig:Fig10_Observability} shows plots of the concurrence as a function of cavity detuning $\Delta_a$ and driving amplitude $\Omega$ and the corresponding properties of the radiated field, in a case in agreement with Mechanism I, in which entanglement is created when the cavity matches certain resonant conditions. The properties of the emission that we show are the emission intensity, given by $I\propto \langle \hat a^\dagger \hat a \rangle$, and photon statistics, through the zero-delay second-order correlation function $g^{(2)}(0) \equiv \langle {\hat a^\dagger}{\hat a^\dagger} \hat a \hat a\rangle/\langle \hat a^\dagger \hat a \rangle^2$, which, in this context,  has been used in the study of cooperative emission in pairs of quantum dots~\cite{SipahigilIntegratedDiamond2016,KimSuperRadiantEmission2018,KoongCoherenceCooperative2022,CygorekSignaturesCooperative2023}.
We observe that the resonances giving rise to the stabilization of entanglement correlate with strong features in the properties of the emission. For the antisymmetric transition, $\Delta_a\approx R$, a high value of concurrence corresponds to a dip in the emission intensity and bunching in the photon statistics. Here, the photon emission is suppressed due to the subradiant nature of the stabilized state $|-\rangle \approx |A\rangle$~\cite{FicekQuantumInterference2005,Vivas-VianaTwophotonResonance2021}. Consequently, the small amount of emission observed comes mostly from two-photon processes, yielding a high probability of detecting two photons simultaneously. In the symmetric transition, $\Delta_a\approx -R$, the stabilization of a superradiant state $|+\rangle \approx |S\rangle$ correlates with the observation of antibunching in the photon correlations. We note that, given the super/subradiant nature of these states, other methods such as emission dynamics measurements of the lifetime~\cite{TiranovCollectiveSuper2023} could also provide reliable evidence of their stabilization.

This figure also gives valuable insights about the optical tunability of the entanglement. In particular, we see that these effect persists even in the non-perturbative regime $\Omega \gg R$, in which the resonant conditions $\Delta_a \approx \pm R$ do not longer hold given that the eigenstates are now strongly dressed by the drive and their frequencies are subsequently shifted~\cite{Vivas-VianaTwophotonResonance2021}. Given that the strong hybridization with the drive does not preclude the generation of entanglement, as demonstrated by these results, this driving  provides a mechanism to optically tune the cavity in or out of resonance with the energy transitions associated with these dressed states by changing the driving intensity $\Omega$ and keeping the cavity frequency $\Delta_a$ fixed. This can clearly be seen in the top panels of Fig.~\ref{fig:Fig10_Observability}, where a cut of the concurrence is depicted versus the drive amplitude $\Omega$ for a fixed $\Delta_a$. The concurrence features a maximum when the corresponding dressed-state transition is placed in resonance with the cavity, correlating perfectly with a dip in the emission and a peak in the statistics.
\begin{figure}[b]
	\includegraphics[width=0.45\textwidth]{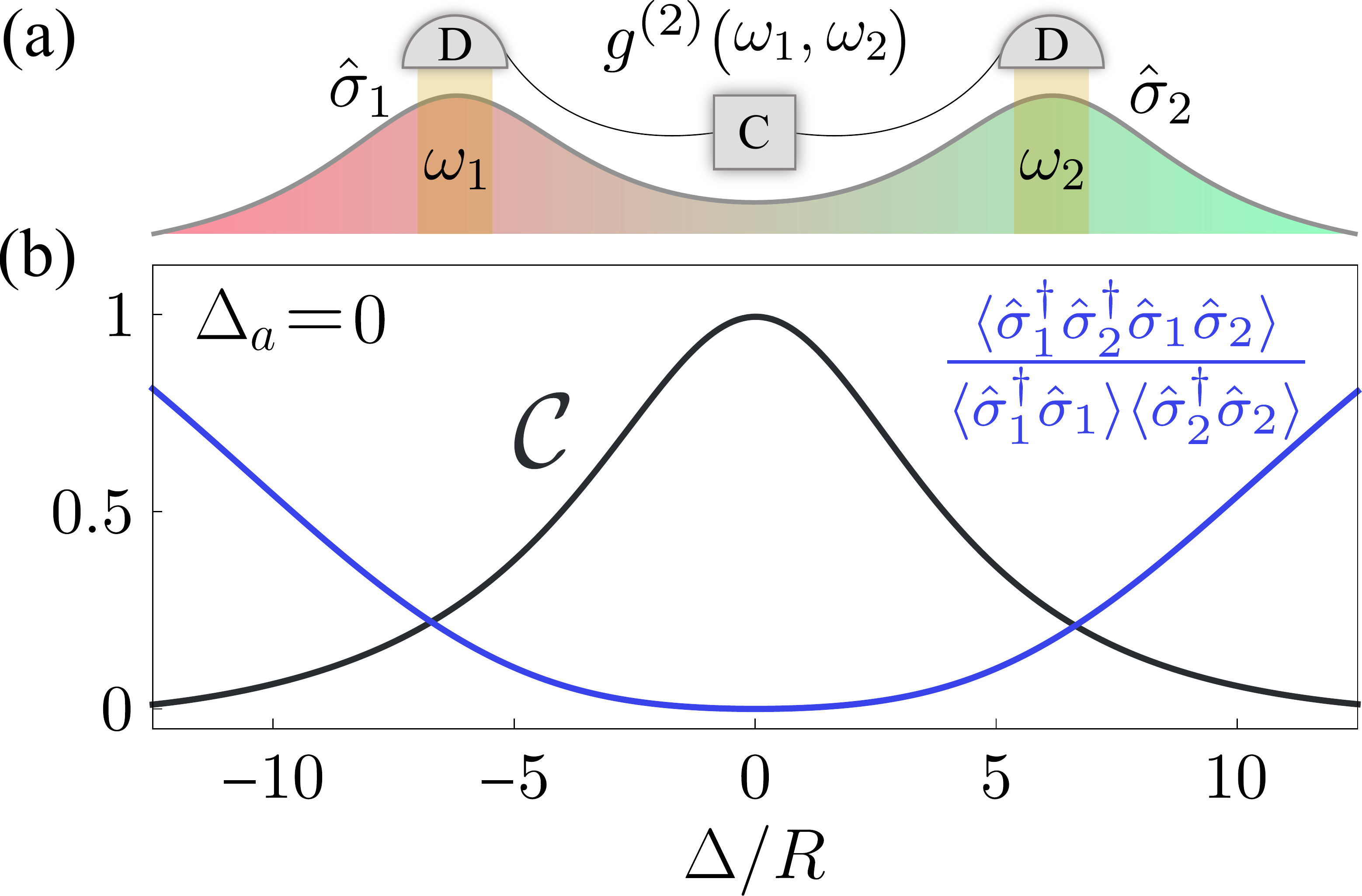}
	\caption{(a) Sketch of the use of frequency resolved correlations to evidence the formation of entanglement. (b) Correspondence between the concurrence and the frequency-resolved correlations between the two  emission peaks, each of which is associated to the individual operators of the non-degenerate emitters.  Parameters: $r_{12}=50\ \text{nm}$, $k=2\pi/780\ \text{nm}^{-1}$, $J=10.65\gamma$, $\gamma_{12}=0.967\gamma$, $\delta=10^2 J$, $R=10^3\gamma$, $\Omega=10^4\gamma$, $\kappa=10^6\gamma$, $g=10^{-1}\kappa$}
	\label{fig:FreqResolved}
\end{figure}

Considering now Mechanism II, Appendix~\ref{appendix:E} shows that the same set of measurements reveals no features that one could clearly correlate with the formation of entanglement. In order to find a measurement that could signal the formation of entanglement, we note that the main characteristic of this regime is that the detuning between emitters is much larger than their natural coupling, $\delta \gg J$. 
This means that the natural frequencies of the emitters are well resolvable, so that the detection of photons emitted at frequencies $\omega_1$ and $\omega_2$ are well described by operators $\hat\sigma_1$ and $\hat\sigma_2$, respectively. Thanks to this spectral resolution, measurement of frequency-resolved correlations in emission could  give access to the observable $\langle \hat\sigma_1^\dagger \hat\sigma_2^\dagger \hat\sigma_1\hat\sigma_2\rangle = \langle ee|\hat\rho|ee\rangle $ [see Fig.~\ref{fig:FreqResolved} (a)]. Note that a measurement during CW excitation would alter the emission frequencies due to the dressing with the drive. In order to avoid dealing with dressed-state energy levels characteristic of the CW regime, the measurement should be done on the radiation emitted after the excitation is switched off and the system is relaxing towards the ground state. In a separable, non-entangled state, this second-order correlation will be factorized $\langle \hat\sigma_1^\dagger \hat\sigma_2^\dagger \hat\sigma_1\hat\sigma_2\rangle = \langle \hat\sigma_1^\dagger \hat\sigma_1\rangle \langle \hat\sigma_2^\dagger \hat\sigma_2\rangle$. This means that the zero-delay frequency resolved correlation function $g^{(2)}(\omega_1,\omega_2)$~\cite{DelValleTheoryFrequencyFiltered2012,Gonzalez-TudelaTwophotonSpectra2013,PeirisTwocolorPhoton2015,SilvaColoredHanbury2016} would yield:
\begin{equation}
g^{(2)}(\omega_1,\omega_2) \equiv  \frac{\langle \hat\sigma_1^\dagger \hat\sigma_2^\dagger \hat\sigma_1\hat\sigma_2\rangle }{\langle \hat\sigma_1^\dagger \hat\sigma_1\rangle \langle \hat\sigma_2^\dagger \hat\sigma_2\rangle} = 1.
\end{equation}
A deviation of this value signals entanglement in the system. This is depicted in Fig.~\ref{fig:FreqResolved}(b), where the stabilization of the state $|A\rangle$ leads to a zero occupation of the state $|ee\rangle$, which for a state in which both emitters are excited is impossible unless the state is non separable, i.e., entangled. This entanglement would translate into a clear anticorrelation between the emission peaks at $\omega_1$ and $\omega_2$ corresponding to each of the emitters, $g^{(2)}(\omega_1,\omega_2)\approx 0$. 

%
%
%
%
%
%
%
%
%
\section{Effect of additional decoherence channels}
\label{sec:decoherence}
So far we have considered a general situation in which the system undergoes decoherence due to its interaction with a vacuum electromagnetic bath, giving rise to both local and collective spontaneous decay and cavity leakage~\cite{FicekQuantumInterference2005,CarmichaelStatisticalMethods1999}. 
However, considerations of more specific platforms, such as molecules or semiconductor quantum dots, may require to account for additional decoherence channels~\cite{ShammahOpenQuantum2018}. 

In order to test the robustness of the mechanisms of entanglement generation proposed here, we independently consider three distinct scenarios, each involving a different type of decoherence: (i) modified spontaneous decay, (ii) local pure dephasing, and (iii) collective pure dephasing.  Fig.~\ref{fig:Fig11_EntanglementProtection}(a-b) depicts the concurrence as a function of both the cooperativity and the additional decoherence rate for these scenarios. We will account for these effects by adding an extra term $\mathcal L_\mathrm{extra}[\hat\rho]$ to the master equation Eq.~\eqref{eq:full_master_eq}.
In order to study the effect of decoherence in all the mechanisms discussed in this text, results are provided for the same two different emitter-emitter distances,  $r_{12}=2.5\ \text{nm}$ and $r_{12}=50\ \text{nm}$, which we used in Fig.~\ref{fig:Fig9_EntanglementRegimes}, 
\subsection{Local spontaneous decay}
The presence of additional decay channels can make the value of the local spontaneous decay rate $\gamma$ to be larger than the value predicted by the parametrization used in this text, given by Eq.~\eqref{eq:gamma}. This may occur, e.g.,  due to the coupling to phonon baths or to specific nanophotonic geometries that cannot be captured by the coupling to a single, lossy bosonic mode described here~\cite{MlynekObservationDicke2014, HaakhSqueezedLight2015}. 
We take this effect into account by adding an extra Lindblad term to the master equation,
\
\begin{equation}
\mathcal L_\mathrm{extra}[\hat\rho] = \frac{\Gamma}{2} \left(\mathcal{D}[\hat\sigma_1]\hat{\rho}+\mathcal{D}[\hat\sigma_2]\hat{\rho} \right),
	\label{eq:MasterEqSpont}
\end{equation}
where $\Gamma$ is additional spontaneous decay rate. This term increases similarly the symmetric and antisymmetric spontaneous decay rates, so that $\gamma_\pm=(\gamma+\Gamma)\pm\gamma_{12}\cos \beta $. 
\begin{figure*}[t]
	\begin{center}
		\includegraphics[width=1.\textwidth]{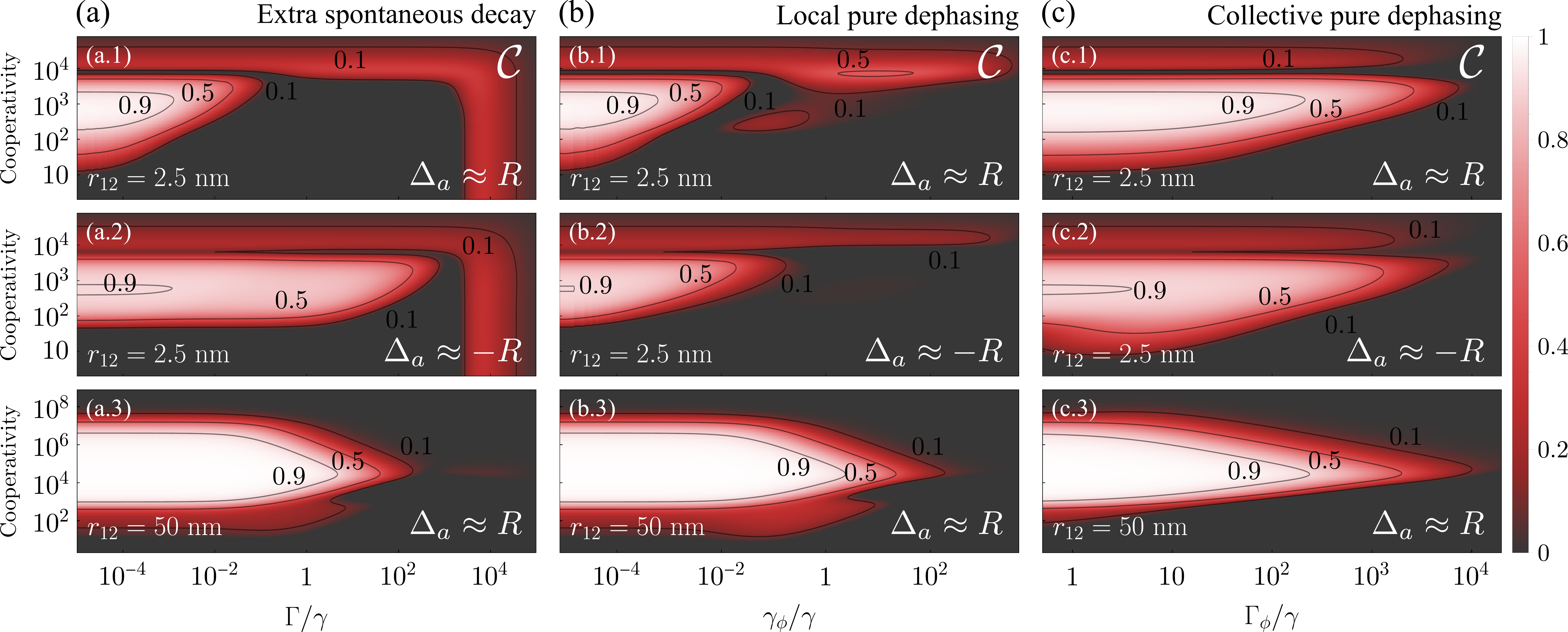}
	\end{center}
	\caption{Robustness of stationary entanglement when additional channels of dissipation are considered: (a) extra spontaneous decay rate; (b) local pure dephasing; and (c) collective pure dephasing. Each panel depicts the concurrence as a function of the cooperativity and the respective extra decoherence rate. From top to bottom: $r_{12}=2.5\ \text{nm}$ at $\Delta_a=\omega_{34}\approx R$ and $\Delta_a=\omega_{21}\approx -R$, respectively; and $r_{12}=50\ \text{nm}$ at $\Delta_a=\omega_{34}\approx R$.
Parameters (a-c): $k=2\pi/780\ \text{nm}^{-1}$, $J(r_{12}=2.5\ \text{nm})= 9.18\times 10^4 \gamma$, $\gamma_{12}(r_{12}=2.5\ \text{nm})=0.999\gamma$, $J(r_{12}=50\ \text{nm})= 10.65 \gamma$, $\gamma_{12}(r_{12}=50\ \text{nm})=0.967\gamma$,  $\delta=10^3 \gamma$, $\Delta=0$, $\Omega=10^4 \gamma$, $g=10^{-1}\kappa$.
}
	\label{fig:Fig11_EntanglementProtection}
\end{figure*}
The consequences in the concurrence are depicted in Fig.~\ref{fig:Fig11_EntanglementProtection}(a). 
When the emitters are placed at $r_{12}=2.5\ \text{nm}$, the formation of entanglement is dominated mainly by Mechanism I, leading the system to high values of entanglement, and Mechanism III, which manifests as a feature at larger values of the cooperativity than for Mechanism I, $C\sim 10^4$. The impact of the losses for Mechanism I depend strongly on the type of resonance selected. When the cavity is placed in the antisymmetric resonance, $\Delta_a=R$,  the concurrence is significantly degraded, since the condition for $\Gamma_{\mathrm I, A}$ described in Sec.~\ref{sec:entanglement-mechanism} is quickly compromised. This is due to the fact that $\gamma_-$ is now increased by $\Gamma$, making this state no longer subradiant, while it remains still weakly coupled to the cavity, with a rate which is proportional to $\beta \ll 1$. The enhancement provided by the cavity is then lost.
On the other hand, for the symmetric resonance, $\Delta_a=-R$, the mechanism is much more robust, since the coupling to the cavity is not reduced by $\beta$ in this case, and the subsequent enhancement leading to entanglement remains much larger than the spontaneous decay rate even when this is increased to values $\Gamma\approx10^2\gamma$. 

Notably, Mechanism III$_\mathrm{cav}$, which occurs at high values of the cooperativity, is very robust to increases in the local decay rate of the emitters. This is expected since the dissipative dynamics is completely dominated by the losses through the cavity channel, which is four orders of magnitude larger than the spontaneous decay rate of the emitters in the regions in which this mechanism is relevant. We notice that the extra local decay can also activate Mechanism III$_\mathrm{sp}$. This is observed as a feature that is independent of the cooperativity and that emerges when $\Gamma$ is large enough. This activation occurs since the extra decay can help fulfilling the condition $\Omega_\mathrm{2p} \approx \gamma$ necessary for Mechanism III to take place.

When the emitter-emitter distance is $r_{12}=50\ \text{nm}$, the generation of entanglement is dominated by Mechanism II. Here, the generation occurs in a characteristic timescale $\Gamma_\mathrm{eff}$ defined in Eq.~\eqref{eq:Gamma_eff}. This parameter depends on the Purcell rate $\Gamma_P$ and, therefore, the robustness of the mechanism against decoherence is largely dependent on the cooperativity. For an optimum cooperativity $C\sim 10^4$, we see that $\Gamma_\mathrm{eff}$ can be large enough to make the concurrence robust even for values of the extra decay rate almost two order of magnitudes larger than considered in the main text,  $\Gamma\sim 10^2\gamma$.

\subsection{Local pure dephasing}
Local pure dephasing, i.e., dissipative processes that destroy coherence terms in the density matrix of each qubit independently, arises in numerous systems. For instance, this effect is prevalent in semiconductor quantum dots~\cite{KrummheuerTheoryPure2002} or in
 molecular system where intra-molecular vibrations and crystal motion are taken into account~\cite{delPinoQuantumTheory2015,ReitzMoleculephotonInteractions2020}.
This effect is described by a term
\begin{equation}
\mathcal L_\mathrm{extra}[\hat\rho] = \frac{\gamma_\phi}{2}\left(\mathcal{D}[\hat\sigma_{z,1}]\hat{\rho}+\mathcal{D}[\hat\sigma_{z,2}]\hat{\rho}\right),
	\label{eq:MasterEqLocalDeph}
\end{equation}
where $\gamma_\phi$ is the local dephasing rate, and $\hat\sigma_{z,i}\equiv 2\hat \sigma_i^\dagger \hat \sigma_i -\mathbb{1}$ is the $z$-component of the Pauli matrices. The effects observed in the concurrence are very similar to the case of additional local decay just discussed, with the main difference that the stabilization of the state $|S\rangle$ is much less robust. This can be explained since the local pure dephasing tends to mix the states $|S\rangle$ and $|A\rangle$ by transferring population from one to the other, as can be seen in the corresponding rate equations
\begin{eqnarray}
		\dot{\rho}_{S,S}&\propto-\left[\Gamma_S+2\gamma_\phi \right]\rho_{S,S}+ 2\gamma_\phi \rho_{A,A}+ \Gamma_S\rho_{e,e}, \\
		\dot{\rho}_{A,A}&\propto-\left[\Gamma_A+2\gamma_\phi \right]\rho_{A,A}+ 2\gamma_\phi \rho_{S,S}+ \Gamma_A\rho_{e,e}.
	\label{eq:rateLocalDeph}
\end{eqnarray}
where one can observed that $\rho_{S,S}$ is fed by the population $\rho_{A,A}$ with a rate $\propto \gamma_\phi$, and vice versa. This implies that, if the timescale of entanglement stabilization is much shorter than $\gamma_\phi^{-1}$, we can still observe a metastable generation of entanglement that it is only degraded in the longer timescale in which state $|A\rangle$ becomes populated.

\subsection{Collective pure dephasing} Finally, we consider the effect of a collective dephasing, that may arise in different scenarios, such as color centers devices interacting with photonic baths~\cite{PrasannaVenkateshCooperativeEffects2018}, or in several qubits interacting with common dissipative bath~\cite{WangDissipationDecoherence2015}. This is described by an extra term in the master equation of the form
\begin{equation}
	\mathcal L_\mathrm{extra}[\hat\rho] = \frac{\Gamma_\phi}{2}\mathcal{D}[\hat\sigma_{z,1}+\hat\sigma_{z,2}]\hat{\rho},
	\label{eq:MasterEqColDeph}
\end{equation}
where $\Gamma_\phi$ is the collective dephasing rate. 
Notably, analysis of the rate equations associated to the resulting master equation reveals that collective dephasing only affects the coherence between the two-photon states $|S_2/A_2\rangle$. As a consequence, entanglement is maintained up to very high values of this collective decoherence term ($\Gamma_\phi \gtrsim 10^3 \gamma$) independently of the mechanism involved and the cavity resonance $\Delta_a\pm R$, as it can be seen in Fig.~\ref{fig:Fig11_EntanglementProtection}(c). 
%
%
%
\section{CONCLUSION}
We have shown that a system of two nonidentical quantum emitters placed within a lossy cavity and driven coherently at the two-photon resonance displays several mechanisms of stabilization of entangled states. 
The first mechanism, which is the focus of Ref.~\cite{Vivas-VianaFrequencyresolvedPurcell2023}, emerges in the case in which the emitters form a dimer structure. It is based on the frequency-resolved Purcell enhancement of specific transitions within the dimer, in combination with a coherent drive of the emitters at the two-photon resonance.
A systematic study in the parameter phase space demonstrates that this effect belongs to a larger landscape of several mechanisms of entanglement generation. We have presented a comprehensive description of each of these mechanisms,  providing analytical insights that define the regimes of parameters where these are activated and their associated timescales, and also describing how the formation of entanglement correlates with specific measurable properties of the light radiated by the cavity.

The stabilization of entanglement can take place in situations in which the emitters may be spatially separated, and therefore weakly interacting, as well as energetically detuned. This offers particularly relevant prospects for experimental generation of entanglement in solid-state  platforms, such as QDs~\cite{LodahlInterfacingSingle2015}, molecules~\cite{ToninelliSingleOrganic2021} or colour centres~\cite{SipahigilIntegratedDiamond2016,AwschalomQuantumTechnologies2018}, since in many of these scenarios the distance between the emitters or their relative frequency is challenging to control.

We have also proven that significant entanglement can be stabilized for a wide range of values of the cooperativity. The  minimum common requirement is a cooperativity $C>1$ that ensures that the cavity provides a Purcell enhancement of the decay. Mechanisms I$_S$ and II have more restrictive limits set by the conditions $\kappa > \Omega_\mathrm{2p}$ and $\Omega$ respectively. Nevertheless, values of natural detuning between emitters already as big as $\delta \sim 10^2\gamma $ can be small enough to allow setting the minimum required values of cooperativity in the range $C_\mathrm{min} \sim 10-100$. This makes our observations compatible with cavity QED platforms based on quantum emitters coupled to single-mode cavities, where values of the cooperativity in the range of $C\sim 1-40$ have been reported for color centres~\cite{EvansPhotonmediatedInteractions2018,LukinTwoEmitterMultimode2023}, quantum dots~\cite{SomaschiNearoptimalSinglephoton2016,TommBrightFast2021}
and molecules~\cite{	WangCoherentCoupling2017,WangTurningMolecule2019,NguyenQuantumNetwork2019}. Quantum emitters coupled to plasmonic nanoantennas, where enhancements of the emission rates of $\sim 10^3-10^6$ have been reported
~\cite{AkselrodProbingMechanisms2014, ChikkaraddySinglemoleculeStrong2016,BaumbergExtremeNanophotonics2019}, can also represent a potentially viable platform. 
Our results call for further investigations in which the dissipative stabilization of entanglement via two-photon excitation is studied while taking into account the specific characteristics of these particular systems, e.g. through a more accurate modeling of the photonic reservoir or decoherence mechanisms. Following the principles of inverse design, our findings can also guide the development of novel photonic environments for the generation of steady-state entanglement in solid state platforms~\cite{MoleskyInverseDesign2018,Miguel-TorcalInversedesignedDielectric2022}.

\acknowledgments
We acknowledge financial
support from the Proyecto Sin\'ergico CAM 2020 Y2020/TCS-
6545 (NanoQuCo-CM), and MCINN projects PID2021-126964OB-I00 (QENIGMA) and TED2021-130552B-C21 (ADIQUNANO). 
C. S. M. and D. M. C.  acknowledge the support of a fellowship from la Caixa Foundation (ID 100010434), from the European Union's Horizon 2020 Research and Innovation Programme under the Marie Sklodowska-Curie Grant Agreement No. 847648, with fellowship codes  LCF/BQ/PI20/11760026 and LCF/BQ/PI20/11760018. D. M. C. also acknowledges support from the Ramon y Cajal program (RYC2020-029730-I).

\appendix

\section{ADIABATIC ELIMINATION}
\label{appendix:A}

Here, we provide a step-by-step demonstration of the adiabatic elimination for the cavity degrees of freedom to obtain the Bloch-Redfield Master Equation for the quantum emitters presented in Eq.\eqref{eq:Nakajima}. In addition, we illustrate how this equation reduces to the Collective Purcell master equation in Eq.\eqref{eq:CollectivePurcell} when the cavity linewidth surpasses the maximum value of the dressed energy transitions.
In order to proceed, we make use of the projective adiabatic elimination method \cite{GardinerQuantumNoise2004,BreuerTheoryOpen2007,RivasOpenQuantum2012}. 

\def\trace#1#2{\mathinner{\text{Tr}_{#2}\left[{#1}\right]}}
\subsection{Projector adiabatic elimination method}
The basic idea of the projector adiabatic elimination approach~\cite{GardinerQuantumNoise2004, BreuerTheoryOpen2007,RivasOpenQuantum2012} is to separate the total system into two subsystems, that we call the system and the reservoir, and obtain the effective dynamics of the system. This is done by defining a projector superoperator $\mathcal{P}$ that projects the total density matrix onto the subspace of the system. Correspondingly, the complementary projector superoperator is defined as $\mathcal{Q}=\mathcal{I}-\mathcal{P}$. 
Assuming that a general quantum system is described by a density matrix $\hat  \rho$, whose dynamics is governed by a Liouville-von Neumann equation $d \hat  \rho(t)/dt=\mathcal{L} \hat  \rho(t)$, we can obtain an equation for the system by introducing the projector superoperators, resulting in the so-called Nakajima-Zwanzig equation
\begin{multline}
	\partial_t[\mathcal{P}\hat  \rho(t)]=\mathcal{P}\mathcal{L}\mathcal{P} \hat \rho(t)\\
	+\mathcal{P}\mathcal{L}\int_{0}^{t} d\tau e^{\mathcal{Q}\mathcal{L}\tau}\mathcal{Q}\mathcal{L}\mathcal{P} \hat \rho(t-\tau),
\end{multline}
where the Liouvillian $\mathcal{L}$ is assumed to be time-independent. However, this expression is still generally unsolvable and perturbative expansions need to be applied.

In order to obtain an effective master equation for the system density matrix we consider the following assumptions: 
(i) we consider a bipartite scenario consisting on a system (the emitters) and an environment (the lossy cavity) with Hilbert spaces $\mathcal{H}=\mathcal{H}_S\otimes\mathcal{H}_E$, where $\mathcal{H}_S$ is the subspace of interest;
(ii) we define the projector superoperator $\mathcal{P}(*)=\text{Tr}_E[(*)]\otimes \hat  \rho_E$, and decompose the total Liouvillian as $\mathcal{L}=\mathcal{L}_S+\mathcal{L}_E+\varepsilon \mathcal{L}_{\text{int}}$, where $\varepsilon \ll 1$ is a small parameter quantifying the weak interaction between the system and the environment; 
and (iii)  we assume a polynomial expansion in terms of products of system and environment operators in the interactive Liouvillian as $\mathcal{L}_{\text{int}}[(*)]=-i[\hat H_{\text{int}},(*)]=-i\sum_{m=1}^M[g_m \hat S_m\otimes \hat E_m,(*)]$, where $\hat S_m/\hat E_m \in \mathcal{H}_{S/E}$. 
In consequence, upon expanding over the small parameter $\epsilon$ we obtain the effective master equation~\cite{Navarrete-BenllochIntroductionQuantum2022}
\begin{widetext}
	\begin{equation}
		\partial_t \hat  \rho_S(t)\approx \mathcal{L}_S[\hat  \rho_S]+\left(
		\sum_{m,n=1}^M g_m g_n \int_0^{\infty} d\tau\left[  
		C_{nm}(\tau) \hat S_m \hat  \rho_S(t)\tilde S_n(\tau)-K_{mn}(\tau)\hat S_m\tilde S_n (\tau) \hat  \rho_S(t)
		\right]
		+\text{H.c.}\right),
		\label{eq:EffectiveMasterEq}
	\end{equation}
\end{widetext}
where we have defined the asymptotic two-time correlators for the environment operators
\begin{align}
	C_{nm}(\tau)&\equiv \lim_{t\rightarrow \infty} \langle \hat E_n(t)\hat E_m(t+\tau) \rangle_E , \\
	K_{mn}(\tau)&\equiv \lim_{t\rightarrow \infty} \langle \hat E_m(t+\tau)\hat E_n(t) \rangle_E.
\end{align}
In addition, we assumed that the correlation functions $C_{nm}(\tau)$ and $K_{mn}(\tau)$ are either zero or vanish at much faster rate than any other process within the system by means of the Markov approximation. Then, the system dynamics is governed by its coherent evolution
\begin{equation}
\hat 	\rho_S(t-\tau)\approx e^{i\hat H_S \tau}\hat  \rho_S(t)e^{i\hat H_S \tau},
\end{equation}
so that
\begin{equation}
e^{\mathcal{L}_S \tau}[\hat S]\approx e^{-i\hat H_S\tau}\hat S e^{i\hat H_S\tau}\equiv \tilde  S(\tau) .
\end{equation}

\subsection{Adiabatic elimination of a dressed two-qubit system coupled to a lossy bosonic mode}
Now, we apply this method in our case, consisting of a system of two interacting nonidentical quantum emitters that are both driven by a coherent field and coupled to a single mode cavity. Here, our starting point is the master equation in Eq.~\eqref{eq:full_master_eq}, which we can expand as~\cite{Navarrete-BenllochIntroductionQuantum2022}:
\begin{widetext}
	\begin{equation}	
		\partial_t \hat  \rho(t)=	
		\underbrace{-i[\hat H_\mathrm q+\hat H_\mathrm d,\hat  \rho(t)]+\sum_{i,j=1}^2 \frac{\gamma_{ij}}{2}\mathcal{D}[\hat \sigma_i,\hat \sigma_j]\hat  \rho(t)}_{\mathcal{L}_S[\hat  \rho]}
		\underbrace{\vphantom{\sum_{i,j=1}^2}-[\Delta_a \hat a^\dagger \hat a,\hat \rho(t)]+ \frac{\kappa}{2} \mathcal{D}[\hat a]\hat \rho(t)}_{\mathcal{L}_E[\hat \rho]}
		\underbrace{\vphantom{\sum_{i,j=1}^2}-i[g[\hat a^\dagger\otimes(\sigma_1+\sigma_2)+\hat a\otimes(\hat \sigma_1^\dagger+\hat \sigma_2^\dagger)],\hat \rho]}_{\mathcal{L}_{\text{int}}[\hat \rho]}.
	\end{equation} 
\end{widetext}
Because the system is assumed to be dressed, it is desirable to change to the eigenstate basis of the qubit-laser system~\cite{Vivas-VianaTwophotonResonance2021}:
\begin{multline}
	\hat H_{\text{int}}=g[\hat a^\dagger\otimes(\hat \sigma_1+\hat \sigma_2)+\text{H.c.}]\\
	=\sum_{i,j} g_{ij} \hat a^\dagger \otimes \hat \sigma_{ij} +\text{H.c.}=\hat a^\dagger \otimes \hat \xi+\text{H.c.},
\end{multline}
where we have defined $\sigma_{ij}\equiv |j\rangle \langle i|$, $g_{ij}\equiv \langle j | \hat \sigma_1+\hat \sigma_2 |i\rangle$, and $\hat \xi\equiv \sum_{i,j} g_{ij} \hat \sigma_{ij}$ ($|i\rangle$, $i=1,\ldots,4$, denote the eigenstates of the qubit-laser system). In comparison with the general formula for the interaction Hamiltonian, $\hat H_{\text{int}}=\sum_m g_m \hat S_m\otimes \hat E_m$, we identify: $g_1=g_2=1$, $\hat E_1=\hat a^\dagger=\hat E_2^\dagger$, and $\hat S_1=\hat \xi=\hat S_2^\dagger$ (note that $\hat E_1$ is already the Hermitian conjugate of the bosonic operator). 
When $1/\kappa$ is much shorter than any other timescale in the problem, we can assume that no reversible dynamics occur, and then the cavity remains in the vacuum state $\hat  \rho_E\approx |0\rangle \langle 0|$. Therefore, in this scenario we can adiabatically eliminate the cavity and get an effective master equation for the two quantum emitters of the same structure as the one in Eq.~\eqref{eq:EffectiveMasterEq}.

Firstly, we need to compute the two-time correlation functions $C_{nm}$ and $K_{mn}$. We do this by neglecting any effect of the system into the environment. To make it clearer, we include a subscript $E$ (denoting \textit{environment}) under the expected value. Then, the bosonic operator forms a closed set,
\begin{equation}
	\frac{d\langle \hat a (\tau)\rangle_E}{d\tau}=\trace{\hat a \mathcal{L}_E[\hat  \rho(t)]}{E}=-(\kappa/2+i\Delta_a)\langle \hat a (\tau)\rangle_E,
\end{equation}
that can be easily integrated in time
\begin{equation}
	\langle \hat a (\tau)\rangle_E=\langle \hat a (0)\rangle_Ee^{-(\kappa/2+i\Delta_a)\tau}=\hat a e^{-(\kappa/2+i\Delta_a)\tau}.
\end{equation}
Invoking the Quantum Regression Theorem~\cite{CarmichaelStatisticalMethods1999}, we can compute all the possible correlators $C_{nm}$ and $K_{mn}$,
\begin{align}
	\langle \hat A(t) \hat B (t+\tau )\rangle &=\trace{\hat B e^{\mathcal{L}\tau }[\hat \rho(t)\hat A]}{}, \\
	\langle \hat A(t+\tau) \hat B (t)\rangle &=\trace{\hat A e^{\mathcal{L}\tau }[\hat B \hat \rho(t)]}{}.
\end{align}
Taking into account that $\hat E_1=\hat a^\dagger=\hat E_2^\dagger$, we obtain 
\begin{align}
	C_{11}(\tau)&=\lim_{t\rightarrow \infty}\langle \hat a^\dagger(t) \hat a^\dagger(t+\tau)\rangle_E=0, \\
	C_{22}(\tau)&=\lim_{t\rightarrow \infty}\langle \hat a(t) \hat a(t+\tau)\rangle_E=0,\\
	C_{12}(\tau)&=\lim_{t\rightarrow \infty}\langle \hat a^\dagger(t) \hat a(t+\tau)\rangle_E=0, \\
	C_{21}(\tau)&=\lim_{t\rightarrow \infty}\langle \hat a(t) \hat a^\dagger(t+\tau)\rangle_E=e^{-(\kappa/2-i\Delta_a)\tau},
\end{align}
where we used the canonical commutation relation for bosonic operators $[\hat a,\hat a^\dagger]=1$.  Analogously for the $K_{mn}(\tau)$ we get
\begin{align}
	K_{11}(\tau)&=\lim_{t\rightarrow \infty}\langle \hat a^\dagger(t+\tau) \hat a^\dagger(t)\rangle_E=0, \\
	K_{22}(\tau)&=\lim_{t\rightarrow \infty}\langle \hat a(t+\tau) \hat a(t)\rangle_E=0, \\
	K_{12}(\tau)&=\lim_{t\rightarrow \infty}\langle \hat a^\dagger(t+\tau) \hat a(t)\rangle_E=0,\\
	K_{21}(\tau)&=\lim_{t\rightarrow \infty}\langle \hat a(t+\tau) \hat a^\dagger(t)\rangle_E=e^{-(\kappa/2+i\Delta_a)\tau}.
\end{align}
In consequence, we obtain the following general expressions for the two-time correlators
\begin{align}
	C_{nm}(\tau)=e^{-(\kappa/2-i\Delta_a)\tau}\delta_{n,2}\delta_{m,1}, \\
	K_{mn}(\tau)=e^{-(\kappa/2+i\Delta_a)\tau}\delta_{m,2}\delta_{n,1}.
\end{align}
We note that the indices between $C_{nm}$ and $K_{mn}$ are interchanged, which will be relevant later.

We recall that we are assuming that the correlation functions are either zero or decay at a much faster rate than any other dissipative process affecting the system. This means that $1/\kappa$ is the shortest dissipative timescale, and that in that timescale the only relevant dynamics of the system is given by its coherent evolution.
Then, by defining the action of the qubit-laser Hamiltonian $\hat H_q+\hat H_d$ onto the eigenstate basis as $(\hat H_q+\hat H_d) |i\rangle =\lambda_i |i\rangle $, we get
\begin{align}
	\hat \xi(\tau)&\approx e^{-i(\hat H_q+\hat H_d) \tau }\hat \xi e^{i(\hat H_q+\hat H_d) \tau } \notag\\
	&= \sum_{i,j} g_{ij}e^{-i(\hat H_q+\hat H_d) \tau } |j\rangle \langle i| e^{i(\hat H_q+\hat H_d) \tau }\notag\\
	&=\sum_{i,j} g_{ij}e^{i (\lambda_i-\lambda_j) \tau } |j\rangle \langle i| =\sum_{i,j} g_{ij}e^{i \omega_{ij} \tau } \hat \sigma_{ij},
\end{align}
where we have defined $\omega_{ij}\equiv \lambda_i- \lambda_j$. 

With these expressions, we can formally integrate Eq.~\eqref{eq:EffectiveMasterEq} and obtain an effective master equation for the dressed-dimer system
\begin{equation}
	\partial_t \hat \rho_S(t)\approx \mathcal{L}_S[\hat \rho_S]	+ \mathcal L_\mathrm{eff}[\hat \rho_S] 
\end{equation}
where
\begin{equation}
	\mathcal L_\mathrm{eff}[\hat \rho_S] =  \sum_{i,j,m,n} \frac{g_{ij} g_{mn}^*}{\kappa/2+i(\Delta_a-\omega_{ij})} [\hat \sigma_{ij} \hat \rho_S(t),\hat \sigma_{mn}^\dagger] + \text{H.c.}  
	\label{eq:EffectiveMasterEqsol}
\end{equation}
We note that Eq.~\eqref{eq:EffectiveMasterEqsol} has the form of a Redfield equation instead of a Lindblad form. Indeed, labeling the index pairs with greek letters $\alpha\equiv(i,j)$, $\nu\equiv (m,n)$, we see that Eq.~\eqref{eq:EffectiveMasterEqsol} has the form:
\begin{multline}
	\mathcal L_\mathrm{eff}[\hat \rho_S] =  \sum_{\alpha,\nu} \frac{a_{\alpha \nu}}{2}\left(\hat\sigma_\alpha \hat \rho \hat\sigma_\nu^\dagger - \hat\sigma_\nu^\dagger\hat\sigma_\alpha \rho \right) 
	\\
	+ \frac{a_{\nu \alpha}^*}{2}\left( \hat\sigma_\alpha \hat \rho \hat\sigma_\nu^\dagger - \hat \rho \hat\sigma_\nu^\dagger   \hat\sigma_\alpha \right),
	\label{eq:Redfield_greek}
\end{multline}
where
\begin{equation}
a_{\alpha\nu}\equiv \frac{2 g_\alpha g_\nu^*}{\kappa/2 + i(\Delta_a - \omega_\alpha)}.
\label{eq:a_alpha_beta}
\end{equation}
This Redfield equation only acquires a Lindblad form provided that
\begin{equation}
a_{\alpha\nu} = a^*_{\nu\alpha},
\label{eq:condition_lindlad}
\end{equation}
in which case we obtain
\begin{equation}
	\mathcal L_\mathrm{eff}[\hat \rho_S] =  \sum_{\alpha,\nu} a_{\alpha\nu}\left(\hat\sigma_\alpha \hat  \rho \hat\sigma_\nu^\dagger  - \frac{1}{2}\left\{\hat\sigma_\nu^\dagger\hat\sigma_\alpha, \hat  \rho \right\}\right),
\end{equation}
which can be written in a Lindblad form by diagonalizing the matrix $a_{\alpha \nu}$~\cite{BreuerTheoryOpen2007}. However, the condition in Eq.~\eqref{eq:condition_lindlad} will only be approximately true for certain choices of the system parameters.

For example, let us consider that the cavity leakage rate is much greater than any dressed energy transition, i.e., $\kappa\gg \omega_{ij}$. In this scenario the cavity is not able anymore to resolve the internal energy levels of the laser-dressed dimer, and thus we can approximate the denominators as $\kappa/2 \pm i(\Delta_a-\omega_{ij})\approx \kappa/2$. 
Defining the \textit{Purcell rate} as $\Gamma_P\equiv 4g^2/\kappa$, we recover the collective Purcell master equation described in Eq.\eqref{eq:CollectivePurcell}, which in this case  has a Lindblad structure,
\begin{equation}
	\mathcal L_\mathrm{eff}[\hat  \rho_S] \approx \frac{\Gamma_P}{2} \mathcal{D}[(\hat \sigma_1+\hat\sigma_2)]\hat  \rho_S(t).
\end{equation}

\section{DERIVATION OF A MASTER EQUATION IN LINDBLAD FORM IN THE FREQUENCY-RESOLVED PURCELL REGIME (MECHANISM I)}
\label{appendix:A2}

The frequency-resolved Purcell regime, where mechanism I is activated, is set by the condition $R\gg \kappa$, which means the cavity is able to resolve the different sets of transitions taking place among the excitonic levels of the dimer, split by $2R$ [see Fig.~\ref{fig:Fig1_Scheme}(b)]. 
We now show that, in this case, Eq.~\eqref{eq:EffectiveMasterEqsol} can be cast in a Lindblad form, providing valuable insights into the mechanism underlying the stabilization of entanglement.

\subsection{Antisymmetric transition $\Delta_a =R$}
By setting $\Delta_a = R$, the sum over $i$ and $j$ in Eq.~\eqref{eq:EffectiveMasterEqsol} will be dominated by terms in which $\omega_{ij} \approx R$. Following the notation that we used to define the eigenstates in Section~\ref{sec:dressed_system}, and the corresponding list of eigenvalues in Eq.~\eqref{eq:eigenvalues_dressed}, we observe that the set of pairs $(i,j)$ dominating the sum are 
\begin{equation}
\mu_A \equiv \{(1,2), (1,3), (2,4),(3,4) \},
\label{eq:mu_A}
\end{equation}
corresponding to the transitions $|+\rangle \rightarrow |A_2\rangle$, $|+\rangle \rightarrow |S_2\rangle$, $|A_2\rangle \rightarrow |-\rangle$ and $|S_2\rangle \rightarrow |-\rangle$, respectively. The rest of terms in the sum will be proportional to $\propto g^2/R$ and, based on the assumption $R\gg \kappa \gg g$, they can be neglected. Similarly, all the terms in the sum over $(m,n)$ that do not belong to the set $\mu_A$ will give rise to terms that, in the Heisenberg picture, will rotate with a frequency proportional to $R$. Based on the same assumptions, we can perform a rotating wave approximation and neglect all the terms in the sum over $(m,n)$ that do not belong to $\mu_A$. 

Since we will restrict the sum in Eq.~\eqref{eq:Redfield_greek} to $\alpha,\beta\in \mu_A$, the value of $(\Delta_a - \omega_\alpha)$ in the denominator of Eq.~\eqref{eq:a_alpha_beta} will, at most, proportional to $\Omega_\mathrm{2p}$. If $\kappa$ is such that the cavity cannot resolve the different transitions in $\mu_A$, i.e., if $\kappa\gg \Omega_\mathrm{2p}$, we may set 
\begin{equation}
a_{\alpha \nu} \approx \frac{4 g_\alpha g_\nu^*}{\kappa}.
\label{eq:a_matrix}
\end{equation}
Under this approximation, $a_{\alpha \nu}$ fulfills the condition in Eq.~\eqref{eq:condition_lindlad}, meaning we can write $\mathcal L_\mathrm{eff}$ in the first standard form.
The values of the coefficients $g_{ij}$ can be obtained from the expressions of the eigenstates of the laser-dressed dimer given in Eqs.~\eqref{eq:eigedressed} and \eqref{eq:s2a2eigenstates}.
To first order in $\beta$ (we remind the reader that this discussion applies to the case in which the emitters are strongly interacting, so $\beta \ll 1$)  these are:
\begin{align}
g_{1,2} &= g_{1,3} = g,\\
g_{2,4} &= -g_{3,4} \approx g\frac{\beta}{2},
\end{align}
which allows us to straightforwardly build the coefficient matrix $a_{\alpha \nu}$ using Eq.~\eqref{eq:a_matrix}. The resulting matrix has a single non-zero eigenvalue with value $\lambda =2\Gamma_{P}+ \mathcal O (\beta^2)$, where $\Gamma_P$ is the standard Purcell rate
\begin{equation}
\Gamma_P=  \frac{4g^2}{\kappa}.
\end{equation} 
The operator built from the corresponding eigenstate~\cite{BreuerTheoryOpen2007} is given by 
\begin{equation}
\hat \xi_A = \frac{1}{\sqrt 2}\left[\hat \sigma_{12} + \hat\sigma_{13} +\frac{\beta}{2}\left(\hat\sigma_{24} - \hat\sigma_{34} \right)  \right].
\end{equation}
Using again the expression of the eigenstates in Eqs.~\eqref{eq:eigedressed} and \eqref{eq:s2a2eigenstates}, we can see that $\hat\sigma_{12} + \hat\sigma_{13} = \sqrt{2}|gg\rangle\langle +|$, and $\hat\sigma_{24} - \hat\sigma_{34}  = -\sqrt{2} |-\rangle\langle ee|$, so 
\begin{equation}
\hat\xi_A = |gg\rangle\langle + | - \frac{\beta}{2}|-\rangle \langle ee|.
\label{eq:jump_xiA}
\end{equation}

As a result, we can write the effective dynamics induced by the cavity in a simple Lindblad form, which consists only of a single jump operator:
\begin{equation}
\mathcal L_\mathrm{eff}[\hat \rho] = \Gamma_P\mathcal D[\hat\xi_A ]\hat\rho.
\end{equation}

The combination of this jump operator and the two-photon driving eventually stabilizes the subradiant state $|-\rangle \approx |A\rangle$.
This can be intuitively understood in the following way: first, the two-photon excitation populates the state $|ee\rangle$; then, the term $\frac{\beta}{2}|-\rangle \langle ee|$ in the jump operator $\hat\xi_A$ takes the system from $|ee\rangle$ into $|-\rangle$. The latter process occurs with a rate 
\begin{equation}
\Gamma_{\mathrm I,A} = \beta^2\Gamma_P/2,
\end{equation}
which means that, since the system leaves state $|-\rangle$ with a rate $\gamma_- = \gamma-\gamma_{12}\cos \beta$, this state will be stabilized with high occupation provided that $\Gamma_{\mathrm I,A}  \gg \gamma_-$.

In the regime of parameters considered in this text we will typically find that  $\Omega_{\mathrm{2p}}\gg \Gamma_{\mathrm I,A}$. In this limit, the resulting rate equations can be reduced to that of a two-state system, and the evolution of the population $\rho_{-}\equiv \langle -|\hat\rho|-\rangle$ is well described by the equation
\begin{equation}
\rho_{-}(t)  = \rho_{A,\mathrm{ss}}\left[1-e^{-\frac{1}{2}(\gamma_- + \Gamma_{\mathrm I,A})t}\right],
\end{equation}
where the steady-state value is given by
\begin{equation}
\rho_{A,\mathrm{ss}} = \frac{\Gamma_{\mathrm I,A}}{\Gamma_{\mathrm I,A}+\gamma_-}.
\end{equation}
These equations provide the timescale of stabilization $\tau_{I,A}=2(\Gamma_{\mathrm{I},A}+\gamma_-)^{-1}$, which in the case $\Gamma_{\mathrm I,A}\gg \gamma_-$ the process is efficient, is given by
\begin{equation}
\tau_{\mathrm I,A}\approx 2/\Gamma_{\mathrm I,A} = \frac{4}{\beta^2 \Gamma_P}.
\end{equation}

Notice that the presence of the factor $\beta$ in the rate $\Gamma_{\mathrm I,A}$ indicates that a certain nonzero detuning $\delta$ is required to ensure that $\Gamma_{\mathrm I,A}\neq 0$. This finite detuning is necessary to prevent the state $|-\rangle$ from being entirely subradiant and uncoupled from the cavity.

\subsection{Symmetric transition $\Delta_a =-R$}
When the cavity is placed close to the symmetric resonance, the following set of transitions are enhanced.
\begin{equation}
\mu_S \equiv \{(2,1), (3,1), (4,2),(4,3) \}.
\label{eq:mu_S}
\end{equation}
Repeating the procedure described above for this set, we also obtain an effective dynamics described by a single Lindblad term with the following jump operator 
\begin{equation}
\hat\xi_S = \frac{1}{\sqrt 2}\left[\hat\sigma_{21} - \hat\sigma_{31}  +\frac{\beta}{2}\left(\hat\sigma_{42} + \hat\sigma_{43} \right)   \right].
\end{equation}
Using the expression of the eigenstates, one finds that  $\hat\sigma_{21} - \hat\sigma_{31} = -\sqrt{2}|+\rangle \langle ee|$, and $\hat\sigma_{42} + \hat\sigma_{43} = \sqrt{2}|gg\rangle \langle -|$, which allows us to write the jump operator as
\begin{equation}
\hat\xi_S = -|+\rangle \langle ee| + \frac{\beta}{2}|gg\rangle\langle -|.
\end{equation}
The effective dynamics is then given by
\begin{equation}
\mathcal L_\mathrm{eff}[\hat\rho]=\Gamma_P\mathcal D[\hat\xi_S]\hat\rho.
\end{equation}
The process of stabilization of the superradiant state $|+\rangle$ is analogous to the subradiant case, with the primary difference being the decay rate from the doubly excited state $|ee\rangle$, which is notably enhanced in the superradiant case and given by
\begin{equation}
\Gamma_{\mathrm I,S}=2\Gamma_P.
\end{equation}

Another difference  in this case is that, for the typical parameters considered, the induced decay rate from $|ee\rangle$ may be comparable to the two-photon driving strength  $\Gamma_{\mathrm I,S} \sim \Omega_{\mathrm{2p}}$ or even much larger $\Gamma_{\mathrm I,S} \gg \Omega_{\mathrm{2p}}$. In the second case, we may adiabatically eliminate $|ee\rangle$ from the dynamics and describe the combination of two-photon pumping and cavity-induced decay as an effective incoherent pumping of the state $|S\rangle$ from $|gg\rangle$, with a rate
\begin{equation}
P_S = \frac{4\Omega_\mathrm{2p}^2}{\Gamma_{\mathrm I,S}} = \frac{2\Omega_\mathrm{2p}^2}{\Gamma_P}.
\end{equation}
Since $\Gamma_{\mathrm I,S}\gg \gamma_-$, we can also neglect the subradiant state $|-\rangle$, and obtain a description of the dynamics corresponding to a incoherently-pumped two-level system---consisting of states  $|gg\rangle$ and $|S\rangle$---with pumping rate $P_S$ and decay rate $\gamma_+=\gamma+\gamma_{12}\cos \beta$. The resulting population of superradiant state $\rho_+ \equiv \langle +|\hat\rho|+\rangle$ as a function of time is then given by
\begin{equation}
\rho_+(t) = \rho_{S,\mathrm{ss}}\left[1-e^{- (\gamma_+ + P_{S})t} \right],
\end{equation}
where the steady-state value is 
\begin{equation}
\rho_{S,\mathrm{ss}} = \frac{P_S}{P_S + \gamma_+}.
\end{equation}
This steady state is stabilized in a timescale $\tau_{\mathrm I,S}$ given by 
\begin{equation}
\tau_{\mathrm I,S} = \frac{2}{\gamma_+ + P_S} \approx \frac{\Gamma_P}{\Omega_\mathrm{2p}^2},
\label{eq:tauIS-appendix}
\end{equation}
where the last inequality is taken in the limit $P_S \gg \gamma_+$ where the state is stabilized with high probability.

If we cannot use the approximation $\Gamma_{\mathrm I, S} \gg \Omega_\mathrm{2p}$, the dynamics cannot be straightforwardly reduced to that of a two-level system by adiabatically eliminating $|ee\rangle$. However, although more involved, a three-level system description of the problem (in which we only eliminated $|-\rangle$ based on the assumption $\Gamma_{\mathrm I,S}\gg\gamma_-$) can still give useful analytical insights. In particular, we find that a more general expression of the steady state is given by
\begin{equation}
\rho_{S,\mathrm{ss}} = \frac{1}{1 + \gamma_+\left(P_S^{-1}+\Gamma_P^{-1}\right)}.
\end{equation}
Full diagonalization of the	Liouvillian also gives a full expression of the timescale of the relaxation, given by:
\begin{equation}
\tau_{\mathrm I, S} =  
\frac{2}{\Gamma_P - \text{Re} \sqrt{\Gamma_P^2-4\Omega_\mathrm{2p}^2}}
, 
\end{equation}
which neatly recovers the result in Eq.~\eqref{eq:tauIS-appendix} in the limit $\Gamma_P \gg \Omega_\mathrm{2p}$.

\begin{figure}[b]
	\includegraphics[width=0.45\textwidth]{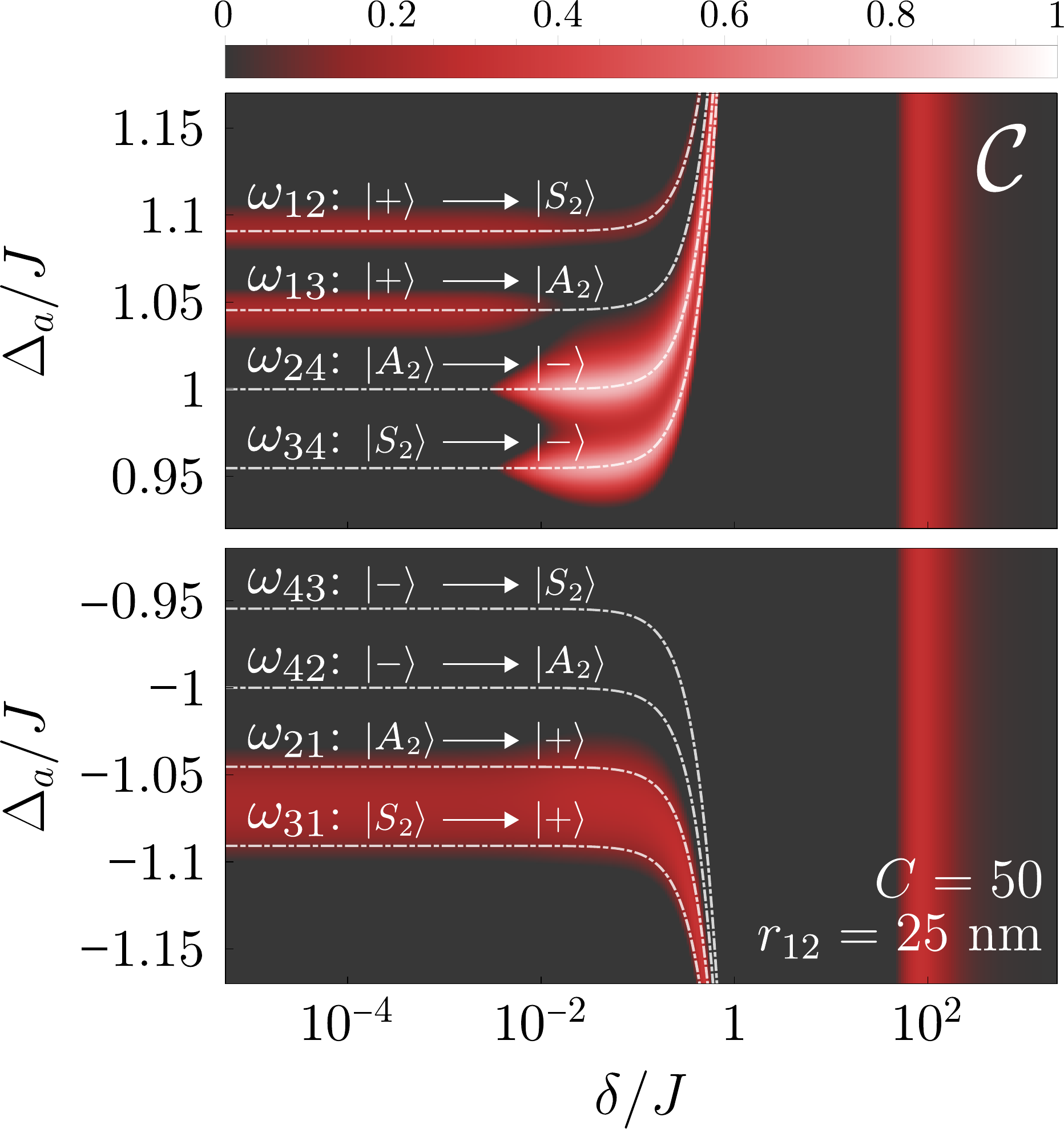}
	\caption{Formation of entanglement when $\kappa < \Omega_\mathrm{2p}$. Compared to the results shown in Fig.~\ref{fig:Fig4_ConcurrenceQubitCavityDetuning}(c), here the cavity is able to resolve the laser-dressed state structure of the driven dimer. Parameters: $r_{12}=2.5\ \text{nm}$, $k=2\pi/780\ \text{nm}^{-1}$, $J=9.18\times 10^4 \gamma$, $\gamma_{12}=0.999\gamma$, $\Delta=0$, $\Omega=10^4 \gamma$, $\kappa=1.25\times10^3 \gamma$, $g=10^{-1}\kappa$.}
	\label{fig:FigAppendix_ConcurrenceKappaLessOmega2p}
\end{figure}

\subsection{Resolved sideband regime $\kappa < \Omega_\mathrm{2p}$}
The dressing of the excitonic energy levels by the drive leads to the development of sidebands around the two main excitonic resonances at $\omega_0\pm R$, as seen e.g. in Fig.~\ref{fig:Fig2_EneryDiagramSpectrum}(b). In the perturbative regime $\Omega\ll R$, the separation between these sidebands is given by $\O mega_\mathrm{2p}$, since, from the expression of the eigenvalues in Eq.~\eqref{eq:eigenvalues_dressed}, we find:
\begin{subequations}
\begin{eqnarray}
\omega_{12} &=& -\omega_{21} = R + 2\Omega_\mathrm{2p},\\
\omega_{13} &=& -\omega_{31} = R + 4\Omega_\mathrm{2p},\\
\omega_{24} &=& -\omega_{42} = R, \\
\omega_{34} &=& -\omega_{43} = R - 2\Omega_\mathrm{2p}
\end{eqnarray}
\label{eq:transitions}
\end{subequations}
So far, we assumed that $\kappa \gg \Omega_\mathrm{2p}$, meaning that the cavity, once placed at one of the two resonances $\Delta_a = \pm R$, would not be able to resolve the different transitions in Eqs.~\eqref{eq:transitions} and enhanced them all equally. 

To evaluate to what extent this condition is necessary, we now compute numerically the degree of entanglement achieved in the regime in which $\kappa \not\gg \Omega_\mathrm{2p}$. Figure~\ref{fig:FigAppendix_ConcurrenceKappaLessOmega2p} shows the steady-state concurrence as a function of $\Delta_a$ and $\delta$, similarly to Fig.~\ref{fig:Fig4_ConcurrenceQubitCavityDetuning}, but for a value $\kappa = 1.25 \cdot 10^3\gamma$, corresponding to $C=50$ and $\kappa/\Omega_\mathrm{2p} \approx 0.57$. In this regime, one can see that the cavity can resolve the different sideband transitions within the sets $\mu_A$ and $\mu_S$, which manifest as resonant features in the entanglement as the cavity frequency is varied.

The entanglement resonances depend greatly on whether the cavity is in the region of antisymmetric transitions $\mu_A$ where $\Delta_a \approx R$, or symmetric transitions $\mu_S$ where $\Delta_a \approx -R$. Figure~\ref{fig:FigAppendix_ConcurrenceKappaLessOmega2p} reveals that, for the stabilization of the antisymmetric state $|A\rangle$, it is enough to selectively enhance only one of the transitions $\omega_{24}:|A_2\rangle\rightarrow |A\rangle$ or $\omega_{34}:|S_2\rangle\rightarrow |A\rangle$; each of them yielding a peak of maximum entanglement. In contrast, stabilizing $|S\rangle$ requires that the cavity enhances both the transitions $\omega_{21}:|A_2\rangle \rightarrow |S\rangle$ and $\omega_{31}:|S_2\rangle \rightarrow |S\rangle$. This conclusion is supported by the lack of well-resolved peaks in the entanglement corresponding to these two transitions, observing instead broader peak that encompasses both of them and that is much weaker than in the antisymmetric case. This indicates that both transitions must be enhanced to observe any effect, which necessarily requires $\kappa\gg\Omega_\mathrm{2p}$.

We interpret this asymmetric behavior as a result of the mechanism being activated when $J\gg\delta$, which is a limit where the target states are almost perfectly subradiant $|-\rangle \approx |A\rangle$ and super-radiant $|+\rangle \approx |S\rangle$ states, therefore having significantly different lifetimes. Due to the extremely long lifetime of the antisymmetric state $|A\rangle$, even a small enhancement of the dissipative transitions towards this state (which the cavity can provide when $C\gg 1$) contributes to a significant increase in the stationary population of this state.


\begin{figure}[b]
	\includegraphics[width=0.46\textwidth]{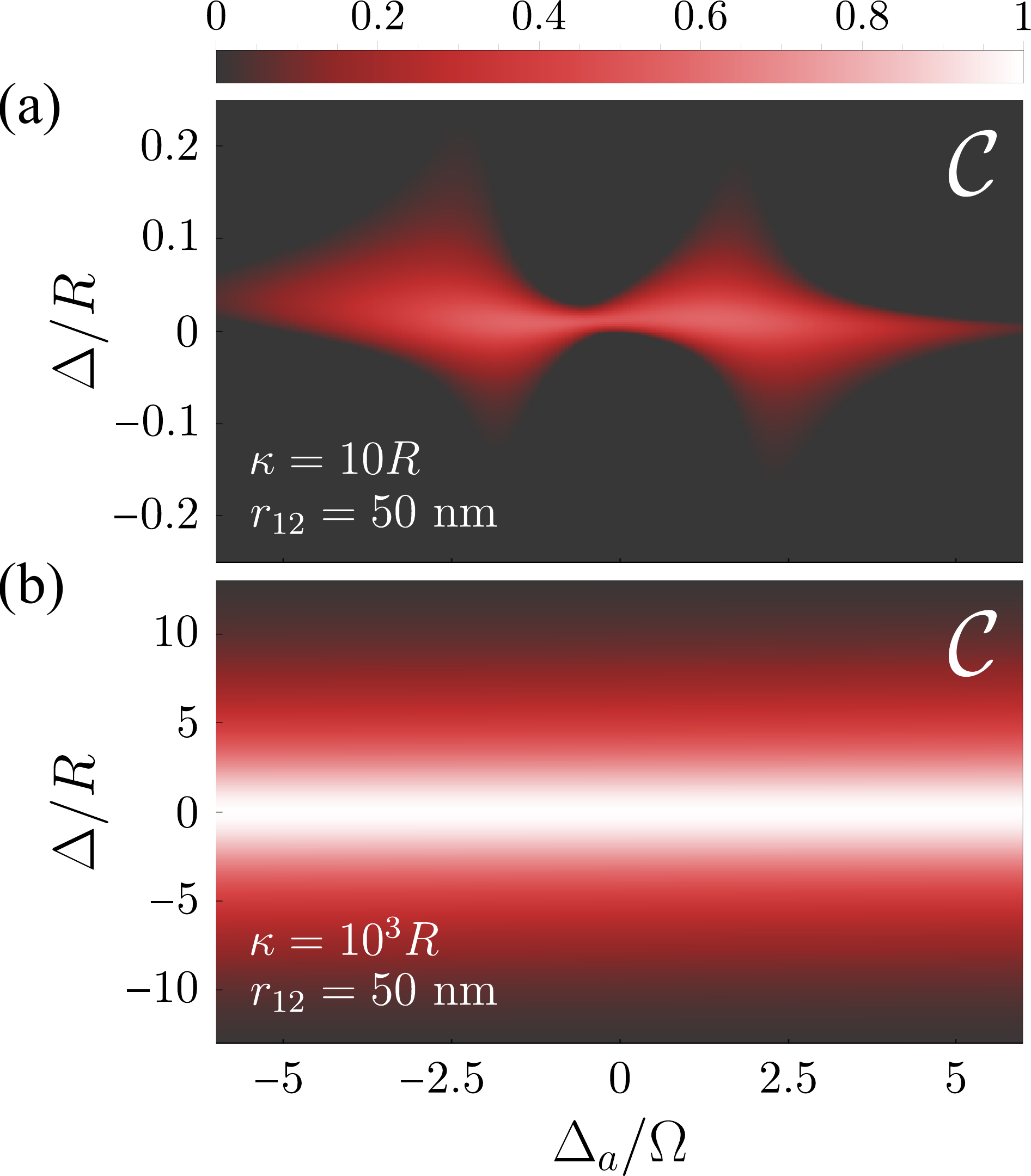}
	\caption{Stationary entanglement when $R/\kappa<1$.
		(a-b) Steady-state concurrence versus -laser cavity detuning $\Delta_a$ and laser-qubit detuning $\Delta$. Parameters:
		 (a-b)
		$r=50$ nm,
		$k=2\pi /780\ \text{nm}^{-1}$,
		$J=10.65\gamma$,
		$\gamma_{12}=0.967$, $\delta=10^2 J$, $R=10^3\gamma$,
		$\Omega=10^4 \gamma$, 
		 $g=10^{-1}\kappa$; (a) 
		$\kappa=10^4 \gamma$
		; (b) $\kappa=10^6 \gamma$	.}
	\label{fig:FigAppendixB_ConcurrenceQubitLaserDetuning}
\end{figure}

\section{STATIONARY ENTANGLEMENT  AT $\Delta=\pm R$ WHEN $R/\kappa < 1$}
\label{appendix:B}
In this appendix, we analyse the dependence of the qubit-laser detuning $\Delta$ in the generation of stationary entanglement when $R/\kappa < 1$, that is, when the Rabi frequency of the dipole-dipole coupling $R$ ceases to be the largest energy scale in the system. In our parametrization, this can be accomplished by placing the emitters further apart $r_{12}\rightarrow \infty$, resulting in a negligible dipole-dipole coupling $J$.

In the main text, we observed in Fig.~\ref{fig:Fig3_ConcurrenceQubitLaserDetuning} that the stationary concurrence features three main regions which depend on the qubit-laser detuning $\Delta$, corresponding to the two-photon resonance $\Delta=0$ and the symmetric/antisymmetric resonances $\Delta=\pm R$. 
However, when we increase the inter-emitter distance $r_{12}$, this pattern with three well-resolved resonances vanishes, and only the two-photon resonance survives, as depicted in Fig.~\ref{fig:FigAppendixB_ConcurrenceQubitLaserDetuning}(a). 
Since in the weak dipole-dipole coupling case the dimer eigenstates in Eq.~\eqref{eq:eigedressed}  tend to be those of independent, uncoupled emitters, $|-\rangle_{\beta\approx \pi/2} \approx |eg\rangle$ and $|+\rangle_{\beta\approx \pi/2} \approx |ge\rangle$, any one-photon resonance mechanism of entanglement generation gets disabled.  

Notably, if the cavity linewidth surpasses the maximum dressed energy transition $\kappa > R, \Omega$, i.e.,  when the system enters into regions characteristic of Mechanism II, Tab.~\ref{tab:EntanglementTab}, the cavity is not able to resolve any atomic state and thus, there is little distinction whether we place the laser frequency $\Delta=\{-R,0,R\}$. This is depicted in Fig. \ref{fig:FigAppendixB_ConcurrenceQubitLaserDetuning}(b), where we note nevertheless that the maximum is located around $\Delta=0$. We can thus conclude that the most relevant scenario for generating stationary entanglement occurs when we drive the system at the two-photon resonance $\Delta=0$.


\section{ANALYTICAL EXPRESSION FOR THE DENSITY MATRIX IN MECHANISM III$_\mathrm{sp}$}
\label{appendix:C}
In this appendix, we derive analytical expressions for the density matrix elements  of the dimer system when Mechanism III$_\mathrm{sp}$ is activated.
As we observed in Fig.~\ref{fig:Fig4_ConcurrenceQubitCavityDetuning}, a sizable degree of entanglement emerges at the onset of saturation  $\Omega_{\text{2p}}\sim\gamma$. This stationary concurrence is independent of the cavity frequency, suggesting that the cavity plays no role in its formation.

Based on this hypothesis, we describe the system by a reduced three level system composed of the ground and the doubly-excited states, and an intermediate single-photon state, as it is described in \cite{Vivas-VianaTwophotonResonance2021}. Under the assumptions: (i) $\Delta \approx 0$, (ii) $R\gg \Delta$, and (iii) $\Omega \ll R$, the state of the system $\xi^{\text{2p}}$ in the reduced space $\{ |gg\rangle, |ee\rangle, |1\rangle \}$ is described by the master equation 
\begin{equation}
	\partial_t \hat  \xi^{\text{2p}}=-i[\hat H_{\text{2p}},\hat  \xi^{\text{2p}}]
	+\frac{2\gamma}{2}\mathcal{D}[\sigma_\alpha]\hat \xi^{\text{2p}}+\frac{\gamma}{2}\mathcal{D}[\sigma_\beta,\sigma_\beta]\hat \xi^{\text{2p}},
	\label{eq:Model2p}
\end{equation}
where $\sigma_\alpha\equiv |1\rangle \langle ee|$ and $\sigma_\beta \equiv |gg\rangle \langle 1|$, and $\hat H_{\text{2p}}$ is the effective two-photon Hamiltonian 
\begin{multline}
	\hat H_{\text{2p}}=(2\Delta -\Omega_{\text{2p}})|ee\rangle \langle ee| -\Omega_{\text{2p}}|gg\rangle \langle gg| \\
	-\Omega_{\text{2p}}(|ee\rangle \langle gg|+ |gg\rangle \langle ee ),
\end{multline}
where $\Omega_{\text{2p}}\equiv 2\Omega^2/R \cos \beta$ can be understood as a two-photon Rabi frequency. Now, establishing the relations: $\rho_{ee,ee}^{(2)}\equiv \xi^\text{2p}_{ee,ee}$, $\rho_{gg,gg}^{(2)}\equiv \xi^\text{2p}_{gg,gg}$, $\rho_{gg,ee}^{(2)}\equiv \xi^\text{2p}_{gg,ee}$, and $\rho_{S,S}^{(2)}=\rho_{A,A}^{(2)}\equiv \xi^\text{2p}_{1,1}/2$, the steady-state solution of Eq.\eqref{eq:Model2p} yields,
\begin{align}
	\rho_{gg,gg}^{(2)}&=\frac{R^2(\gamma^2+4\Delta^2)+4\Omega^4 \cos^2 \beta}{R^2(\gamma^2+4\Delta^2)+16\Omega^4 \cos^2 \beta},  \\
	\rho_{ee,ee}^{(2)}&=\frac{4\Omega^4 \cos^2 \beta}{R^2(\gamma^2+4\Delta^2)+16\Omega^4 \cos^2 \beta}, \\
	\rho_{gg,ee}^{(2)}&=\frac{2R(-i\gamma+2\Delta)\Omega^2 \cos \beta }{R^2(\gamma^2+4\Delta^2)+16\Omega^4 \cos^2 \beta}, \\
	\rho_{S,S}^{(2)}&=\rho_{A,A}^{(2)}=\frac{4\Omega^4 \cos^2 \beta}{R^2(\gamma^2+4\Delta^2)+16\Omega^4 \cos^2 \beta} \, .
\end{align}
We note that $\rho_{i,j}^{(2)}$ denotes density matrix elements resulting from two-photon processes. 

In order to compute an analytical solution for the concurrence, we need to write this reduced density matrix in the full Hilbert space of the emitters. We do so by considering the following ansatz density matrix
\begin{equation}
\hat 	\rho \approx 
	\begin{pmatrix}
		\rho_{gg,gg}^{(2)} & 0 & 0 & \rho_{gg,ee}^{(2)} \\
		0								& 	\rho_{S,S}^{(2)} & 0 &0 \\
		0&0&	\rho_{A,A}^{(2)}&0 \\
		\rho_{ee,gg}^{(2)} &0 &0& \rho_{ee,ee}^{(2)}
	\end{pmatrix}.
\end{equation}

To quantify the degree of entanglement~\cite{WoottersEntanglementFormation1998,WoottersEntanglementFormation2001,PlenioIntroductionEntanglement2007,HorodeckiQuantumEntanglement2009}, one then needs to compute the eigenvalues of the matrix $\hat \rho  (\hat \sigma_y \otimes \hat \sigma_y)  \hat\rho^*( \hat \sigma_y \otimes \hat \sigma_y)$, where $\hat\rho^*$ is the element wise complex conjugate of $\hat\rho$ and $\hat \sigma_y\equiv i(\hat \sigma-\hat \sigma^\dagger) $ is the $y-$component of the Pauli matrices. 
The eigenvalues $ \lambda_i$, in decreasing order, are given by
\begin{align}
	\lambda_1&=\sqrt{	\rho_{gg,gg}^{(2)} 	\rho_{ee,ee}^{(2)}}+\text{Abs}[	\rho_{ee,gg}^{(2)}], \\
	\lambda_2&=\rho_{S,S}^{(2)} \rho_{S,S}^{(2)}, \\
	\lambda_3&= \rho_{A,A}^{(2)}  \rho_{A,A}^{(2)}, \\
	\lambda_4&=\sqrt{	\rho_{gg,gg}^{(2)} 	\rho_{ee,ee}^{(2)}}-\text{Abs}[	\rho_{ee,gg}^{(2)}], 
\end{align}
where $\text{Abs}[(*)]$ denotes the absolute value operation.
Since the region of applicability is located at $\delta > J$, we can approximate $\beta\approx \pi/2$ and obtain a simple expression for the concurrence, defined as $\mathcal{C}=\text{Max}[0,\sqrt{\lambda_1}-\sqrt{\lambda_2}-\sqrt{\lambda_3}-\sqrt{\lambda_4}]$,
\begin{equation}
	\mathcal{C}\approx \text{Max} \left[ 0,\frac{4J\Omega^2 (\gamma \delta^2 -2J\Omega^2)}{\gamma^2\delta^4+16J^2 \Omega^4} \right].
\end{equation}
This expression matches well our numerical results presented in the main text.  In addition, from this expression we can obtain the analytical value of the qubit-qubit detuning that maximizes the concurrence, 
\begin{equation}
\delta_{\text{max}}=\Omega\sqrt{2(1+\sqrt{5})J/\gamma}.
\end{equation}

\section{APPLICATION OF THE HIERARCHICAL ADIABATIC ELIMINATION}
\label{appendix:D}

Here, we show step-by-step the application of the hierarchical adiabatic elimination~\cite{Vivas-VianaUnconventionalMechanism2022} technique in order to obtain effective, analytical time-dependent formulae for the density matrix elements of the emitter-emitter system when Mechanism II is activated, c.f. Tab.~\ref{tab:EntanglementTab}. 
To proceed, we consider the following assumptions: (i) the dipole-dipole interaction rate is almost negligible in comparison with the other system terms $J\ll \delta, \Omega$, allowing us to set $J\approx0$; (ii) since $\Gamma_S\gg \Gamma_A$, the antisymmetric state is approximately decoupled from any dissipative channel, i.e., $\Gamma_A\approx 0$, and it is only coupled to the dynamics via the qubit-qubit detuning term $\delta$. We recall that $\Gamma_S=2\Gamma_P + \gamma+\gamma_{12}$, where $\Gamma_P$ si the Purcell rate.

Under these assumptions, the effective system is described by the following master equation
\begin{equation}
	\partial_t \hat \rho \approx -i[\hat H_{\text{eff}}, \hat \rho]	+ \frac{\Gamma_S}{2}\mathcal{D}[\tilde \sigma_{1}]\hat  \rho+ \frac{\Gamma_S}{2}\mathcal{D}[\tilde \sigma_{2} ] \hat \rho,
\end{equation}
where we defined the ladder operators $\tilde \sigma_{1} \equiv |gg\rangle \langle S|$ and $\tilde \sigma_{2} \equiv |S\rangle \langle ee|$, and the Hamiltonian
\begin{multline}
	\hat H_{\text{eff}}=\sqrt{2}\Omega ( |gg\rangle \langle S| + |S\rangle \langle ee| +  \text{H.c.}  )\\
	-\delta(|S\rangle\langle A|+\text{H.c.}).
\end{multline}

In the language of the hierarchical adiabatic elimination method \cite{Vivas-VianaUnconventionalMechanism2022}, we identify the ``real'' states $\{|gg\rangle, |S\rangle, |ee\rangle\}$, and the ``virtual'' state that mediates the interaction $|A\rangle = |V\rangle$. 

\subsection{First adiabatic elimination}
The first step of the hierarchical adiabatic elimination consists in adiabatically eliminating the ``virtual'' state, assuming the dynamics is completely contained within the ``real'' Hilbert space, $\mathcal{H}_R=\{|gg\rangle, |S\rangle, |ee\rangle\}$. In other words, in a timescale $\sim 1/\Gamma_S$ the antisymmetric state plays the role a virtual state: highly detuned from the subspace where the dynamics take place and weakly coupled to it. Therefore, the virtual state remains unpopulated and its only impact is to provide effective energy shifts to the states in the real subspace. Here, this definition translates into  $\delta \gg J$. Then, we can set $\dot \rho_{i,A}=0$ with $i=gg,S,ee$ and replace the virtual coherence terms by their steady-state values.
This results in the following set of differential equations
\begin{widetext}
	\begin{align}
		\dot \rho_{gg,S} & \approx -\frac{\Gamma_S (\delta^2+2\Omega^2)}{4\Omega^2}\rho_{gg,S}(t)+\frac{2\delta^2 }{\Gamma_S}(\rho_{S,ee}^*(t) -\rho_{gg,S}(t))  \notag \\
		&\qquad \qquad \qquad \qquad+\frac{i}{\sqrt{2}\Omega}[2\Omega^2 +\delta^2(\rho_{S,S}(t)-\rho_{A,A}(t))+2\Omega^2(\rho_{gg,ee}(t)-2\rho_{S,S}(t)-\rho_{A,A}(t)-\rho_{ee,ee}(t))], \\
		\dot \rho_{gg,ee} & \approx -\frac{1}{2}\Gamma_S \rho_{gg,ee}(t) + i\sqrt{2}\Omega (\rho_{gg,S}(t)-\rho_{S,ee}(t)) \\
		\dot \rho_{S,S} & \approx -\Gamma_S (\rho_{S,S}(t)-\rho_{ee,ee}(t))-\frac{\sqrt{2}}{\Omega} [(\delta^2-2\Omega^2)\text{Im}[\rho_{gg,S}(t)]+2\Omega^2\text{Im}[\rho_{S,ee}(t)]]  \\
		\dot \rho_{S,ee} & \approx \frac{1}{\Gamma_S}[2\delta^2 \rho_{gg,S}^*(t)-(\Gamma_S^2+2\delta^2)\rho_{S,ee}(t)-i\sqrt{2}\Gamma_S \Omega(\rho_{gg,ee}(t)-\rho_{S,S}(t)+\rho_{ee,ee}(t))] \\
		\dot \rho_{ee,ee} & \approx -\Gamma_S \rho_{ee,ee}(t) +2\sqrt{2}\Omega \text{Im}[\rho_{S,ee}(t)],
	\end{align}
	where $\text{Im}[(*)]$ denotes imaginary part.
	The differential equation that governs the dynamics of the antisymmetric state after the first adiabatic elimination takes the form
	\begin{equation}
		\dot \rho_{A,A}\approx \frac{\sqrt{2} \delta^2}{\Omega} \text{Im}[\rho_{gg,S}].
		\label{eq:DiffEqAntisymHAE1}
	\end{equation}

	In the limit $\Omega\gg \delta$, these equations can be simplified to
	\begin{align}
		\dot \rho_{gg,S} & \approx -\frac{\Gamma_S}{2}\rho_{gg,S}(t)+i\sqrt{2}\Omega[1+\rho_{gg,ee}(t)-2\rho_{S,S}(t)-\rho_{A,A}(t)-\rho_{ee,ee}(t)], \\
		\dot \rho_{gg,ee} & \approx -\frac{1}{2}\Gamma_S \rho_{gg,ee}(t) + i\sqrt{2}\Omega (\rho_{gg,S}(t)-\rho_{S,ee}(t)) \\
		\dot \rho_{S,S} & \approx -\Gamma_S (\rho_{S,S}(t)-\rho_{ee,ee}(t))+\sqrt{2}\Omega \text{Im}
		[\rho_{gg,S}(t)-\rho_{S,ee}(t)]  \\
		\dot \rho_{S,ee} & \approx -\Gamma_S \rho_{S,ee}(t)-i\sqrt{2}\Omega (\rho_{gg,ee}(t)-\rho_{S,S}(t)+\rho_{ee,ee}(t)) \\
		\dot \rho_{ee,ee} & \approx -\Gamma_S \rho_{ee,ee}(t) +2\sqrt{2}\Omega \text{Im}[\rho_{S,ee}(t)],
	\end{align}
	which describe a cascaded three level system driven with an Rabi frequency $\sqrt{2}\Omega$, and a decay rate $\Gamma_S$. The additional term $\rho_{A,A}$ evolves in a much slower timescale; so here it is considered as a time-independent parameter. For instance, we may set $\rho_{A,A}=0$ without loss of generality.

	\subsection{Second adiabatic elimination}
	In a timescale longer than $\sim 1/\Gamma_S$, the ``real'' variables $\rho_{i,j}$ with $i,j=gg,S,ee$ are considered as fast variables, since they relax to a steady-state in a very short time in comparison with the ``virtual state'' $\rho_{A,A}$. Therefore, we can perform a second adiabatic elimination.
	Assuming that the antisymmetric state is barely unchanged in a timescale $\sim 1/\Gamma_S$, we can compute a steady-state solution for the real variables that will depend on the fixed value of $\rho_{A,A}$. 
	In other words, we will obtain a steady-state solution that will follow any slow change of the antisymmetric state.
	The resulting $\rho_{A,A}$-dependent steady-state for the real variables is obtained by solving a linear system of the form $\hat M.\vec{\rho}+\vec{b}=0$ for the vector $\vec{\rho}=\{\rho_{gg,S}^{SS}, \rho_{gg,ee}^{SS}, \rho_{S,gg}^{SS}, \rho_{S,S}^{SS}, \rho_{S,ee}^{SS}, \rho_{ee,gg}^{SS}, \rho_{ee,S}^{SS}  ,\rho_{ee,ee}^{SS}\}$, where $\hat M$ and $\vec{b}$ are given by
	\begin{equation}
		M=	\left(
		\begin{array}{cccccccc}
			-\frac{\Gamma  \left(\delta ^2+2 \Omega ^2\right)}{4 \Omega ^2}-\frac{2
				\delta ^2}{\Gamma } & i \sqrt{2} \Omega  & 0 & \frac{i \left(\delta ^2-4
				\Omega ^2\right)}{\sqrt{2} \Omega } & 0 & 0 & \frac{2 \delta ^2}{\Gamma } &
			-i \sqrt{2} \Omega  \\
			i \sqrt{2} \Omega  & -\frac{\Gamma }{2} & 0 & 0 & -i \sqrt{2} \Omega  & 0 & 0
			& 0 \\
			0 & 0 & -\frac{\Gamma  \left(\delta ^2+2 \Omega ^2\right)}{4 \Omega
				^2}-\frac{2 \delta ^2}{\Gamma } & -\frac{i \left(\delta ^2-4 \Omega
				^2\right)}{\sqrt{2} \Omega } & \frac{2 \delta ^2}{\Gamma } & -i \sqrt{2}
			\Omega  & 0 & i \sqrt{2} \Omega  \\
			\frac{i \left(\delta ^2-2 \Omega ^2\right)}{\sqrt{2} \Omega } & 0 & -\frac{i
				\left(\delta ^2-2 \Omega ^2\right)}{\sqrt{2} \Omega } & -\Gamma  & i
			\sqrt{2} \Omega  & 0 & -i \sqrt{2} \Omega  & \Gamma  \\
			0 & -i \sqrt{2} \Omega  & \frac{2 \delta ^2}{\Gamma } & i \sqrt{2} \Omega  &
			-\frac{2 \delta ^2}{\Gamma }-\Gamma  & 0 & 0 & -i \sqrt{2} \Omega  \\
			0 & 0 & -i \sqrt{2} \Omega  & 0 & 0 & -\frac{\Gamma }{2} & i \sqrt{2} \Omega 
			& 0 \\
			\frac{2 \delta ^2}{\Gamma } & 0 & 0 & -i \sqrt{2} \Omega  & 0 & i \sqrt{2}
			\Omega  & -\frac{2 \delta ^2}{\Gamma }-\Gamma  & i \sqrt{2} \Omega  \\
			0 & 0 & 0 & 0 & -i \sqrt{2} \Omega  & 0 & i \sqrt{2} \Omega  & -\Gamma  \\
		\end{array}
		\right),
	\end{equation}
	\begin{equation}
		\vec{b}=\begin{pmatrix}
			-i\frac{\delta^2+2\Omega^2}{\sqrt{2}\Omega} \\
			0\\
			i\frac{\delta^2+2\Omega^2}{\sqrt{2}\Omega} \\
			0\\
			0\\
			0\\
			0\\
			0
		\end{pmatrix} \rho_{A,A}(t)+
		\begin{pmatrix}
			i\sqrt{2}\Omega \\
			0\\
			-i\sqrt{2}\Omega \\
			0\\
			0\\
			0\\
			0\\
			0
		\end{pmatrix}.
	\end{equation}
	Then, by solving this linear system, we obtain the analytical quasistationary $\rho_{A,A}$-dependent solution for the real variables,
	\begin{align}
		\rho_{gg,S}^{SS}[\rho_{A,A}(t)]&=\frac{4 i \sqrt{2} \Gamma  \Omega ^3 \left(\Gamma ^2+2 \delta ^2+8 \Omega
			^2\right)}{\chi}-\frac{2 i
			\sqrt{2} \Gamma  \Omega   \left(\delta ^2+2 \Omega ^2\right)
			\left(\Gamma ^2+2 \delta ^2+8 \Omega ^2\right)}{\chi}\rho _{A,A}(t) \\
		\rho_{gg,ee}^{SS}[\rho_{A,A}(t)]&=-\frac{16 \Omega ^4 \left(\Gamma ^2+6 \delta ^2\right)}{\chi}+\frac{8 \Omega ^2  \left(\Gamma ^2+6 \delta ^2\right) \left(\delta
			^2+2 \Omega ^2\right)}{\chi}\rho _{A,A}(t), \\
		\rho_{S,S}^{SS}[\rho_{A,A}(t)]&=-\frac{8 \Omega ^2 \left(\delta ^2-2 \Omega ^2\right)\left(\Gamma ^2+2 \delta ^2+8 \Omega ^2\right)}{\chi}+\frac{4  \left(\delta ^4-4 \Omega ^4\right) \left(\Gamma ^2+2
			\delta ^2+8 \Omega ^2\right)}{\chi} \rho _{A,A}(t), \\
		\rho_{S,ee}^{SS}[\rho_{A,A}(t)]&=	-\frac{16 i
			\sqrt{2} \Gamma  \Omega ^3 \left(\delta ^2-2 \Omega ^2\right)}{\chi}+\frac{8 i \sqrt{2} \Gamma  \Omega  \left(\delta ^4-4 \Omega
			^4\right)}{\chi}\rho _{A,A}(t) ,\\
		\rho_{ee,ee}^{SS}[\rho_{A,A}(t)]&=	-\frac{64 \Omega ^4 \left(\delta
			^2-2 \Omega ^2\right)}{\chi}+\frac{32 \Omega ^2  \left(\delta ^4-4 \Omega ^4\right)}{\chi}\rho _{A,A}(t) ,
	\end{align}
	with 
	\begin{equation}
		\chi\equiv \Gamma ^4 \left(\delta ^2+2 \Omega ^2\right)+\Gamma ^2 \left(6
		\delta ^4-4 \delta ^2 \Omega ^2+64 \Omega ^4\right)+8 \left(\delta ^6-4
		\delta ^4 \Omega ^2+4 \delta ^2 \Omega ^4+48 \Omega ^6\right).
	\end{equation}

\begin{figure}[t]
	\includegraphics[width=0.4\textwidth]{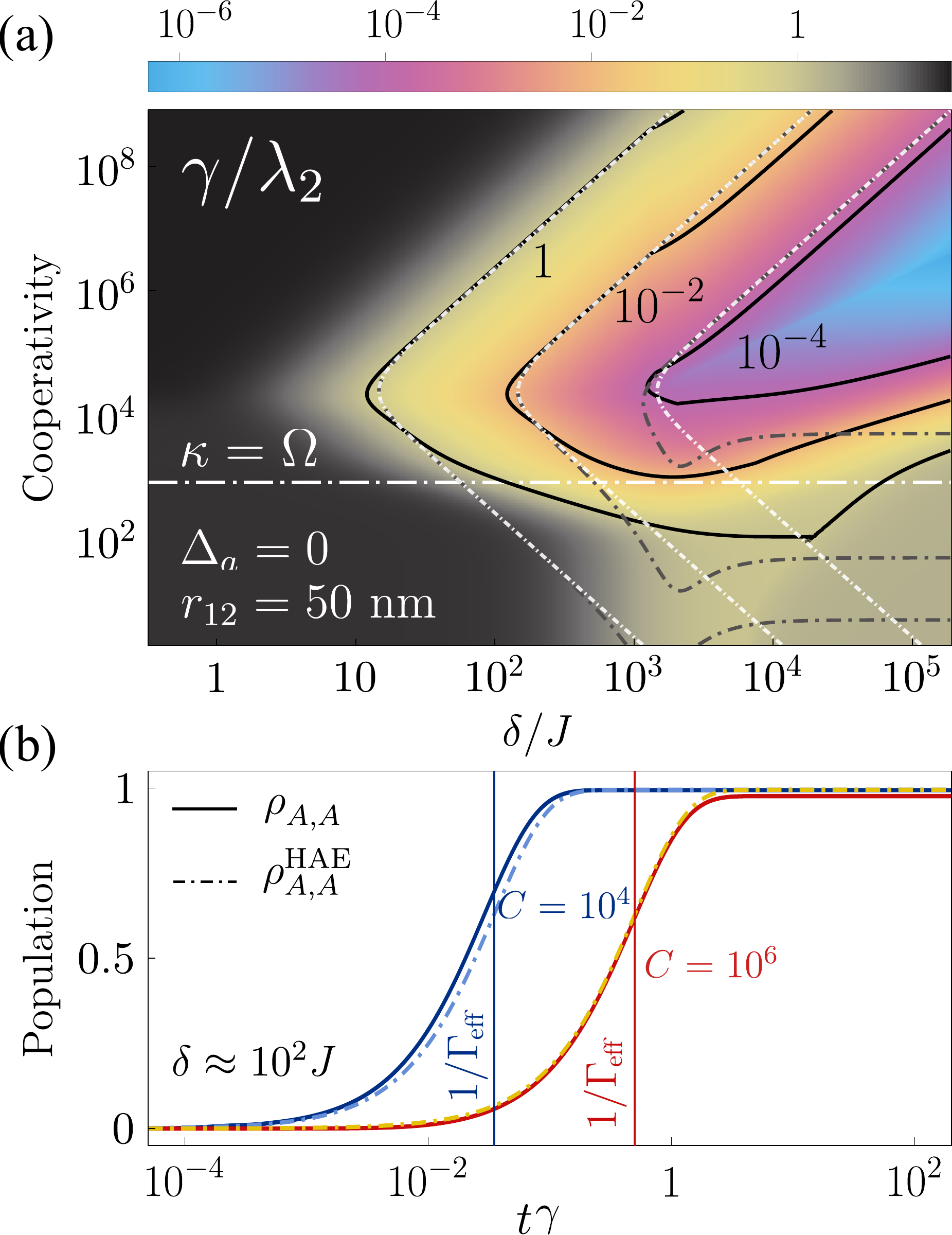}
	\caption{
		(a) Liouvillian gap in terms of the cooperativity $C$ and qubit-qubit detuning $\delta$. Solid black lines are exact computations, dashed black lines are analytical prediction from the relaxation rate, Eq.~\eqref{eq:RelaxRate}, and dashed white lines are analytical predictions from the approximated relaxation rate, Eq.~\eqref{eq:EffectiveRelaxRate}. (b) Dynamics of the antisymmetric state for two different values of cooperativity: $C=10^4$ (blue) and $C=10^6$ (red). Solid lines correspond to exact computations and blue dashed lines are analytical predictions from the hierarchical adiabatic elimination considering the full relaxation rate. 
		Parameters: 
		(a-b) 
		$r_{12}=50\ \text{nm}$,
		$k=2\pi /780\ \text{nm}^{-1}$,
		$J=10.65\gamma$,
		$\gamma_{12}=0.967$,
		$\Delta=0$,
		$\Omega=10^4 \gamma$,
		$g=10^{-1}\kappa$;
		(b)
		blue curve $C=10^4$, $\delta=10^3\gamma$;  red $C=10^6$, $\delta=10^3\gamma$.}
	\label{fig:FigAppendixD_HAE}
\end{figure}
	
	Now, we can substitute the pseudostationary value of $\rho_{gg,S}$ back into Eq.~\eqref{eq:DiffEqAntisymHAE1}, and obtain the effective differential equation for the antisymmetric state
	\begin{equation}
		\dot \rho_{A,A}(t)=
		\frac{8 \Gamma  \delta ^2 \Omega ^2 \left(\Gamma ^2+2 \delta ^2+8 \Omega
			^2\right)}{\chi} -\frac{4
			\Gamma  \delta ^2  \left(\delta ^2+2 \Omega ^2\right)
			\left(\Gamma ^2+2 \delta ^2+8 \Omega ^2\right)}{\chi}\rho _{A,A}(t).
	\end{equation}
	Solving this equation, we obtain the analytical expression for the evolution of $\rho_{A,A}(t)$
	\begin{equation}
		\rho_{A,A}(t)=\rho _{A,A}^{SS}(1-e^{-\Gamma_{\text{eff}}t}),
	\end{equation}
	where $\rho _{A,A}^{SS}$ is the steady-state value 
	\begin{equation}
		\rho _{A,A}^{SS}=\frac{2\Omega^2}{\delta^2+2\Omega^2},
	\end{equation}
	and $\Gamma_{\text{eff}}$ stands for the effective relaxation rate,
	\begin{equation}
		\Gamma_{\text{eff}}=\frac{4 \Gamma  \delta ^2 \left(\delta ^2+2 \Omega ^2\right) \left(\Gamma ^2+2
			\delta ^2+8 \Omega ^2\right)}{\Gamma ^4 \left(\delta ^2+2 \Omega
			^2\right)+\Gamma ^2 \left(6 \delta ^4-4 \delta ^2 \Omega ^2+64 \Omega
			^4\right)+8 \left(\delta ^6-4 \delta ^4 \Omega ^2+4 \delta ^2 \Omega ^4+48
			\Omega ^6\right)},
		\label{eq:RelaxRate}
	\end{equation}
	which serves as an analytical estimation of the Liouvillian gap. We can simplify this expression further under the assumption $\Omega\gg \delta$, which is a natural condition for this mechanism to take place.The relaxation rate then reduces to
	\begin{equation}
		\Gamma_{\text{eff}}=\frac{4\Gamma_S \delta^2}{\Gamma_S^2+24\Omega^2},
		\label{eq:EffectiveRelaxRate}
	\end{equation}
	which is the expression shown in the main text.
\end{widetext}
\subsection{Validity of the relaxation rate $\Gamma_{\text{eff}}$}
\begin{figure*}[t]
	\begin{center}
		\includegraphics[width=1\textwidth]{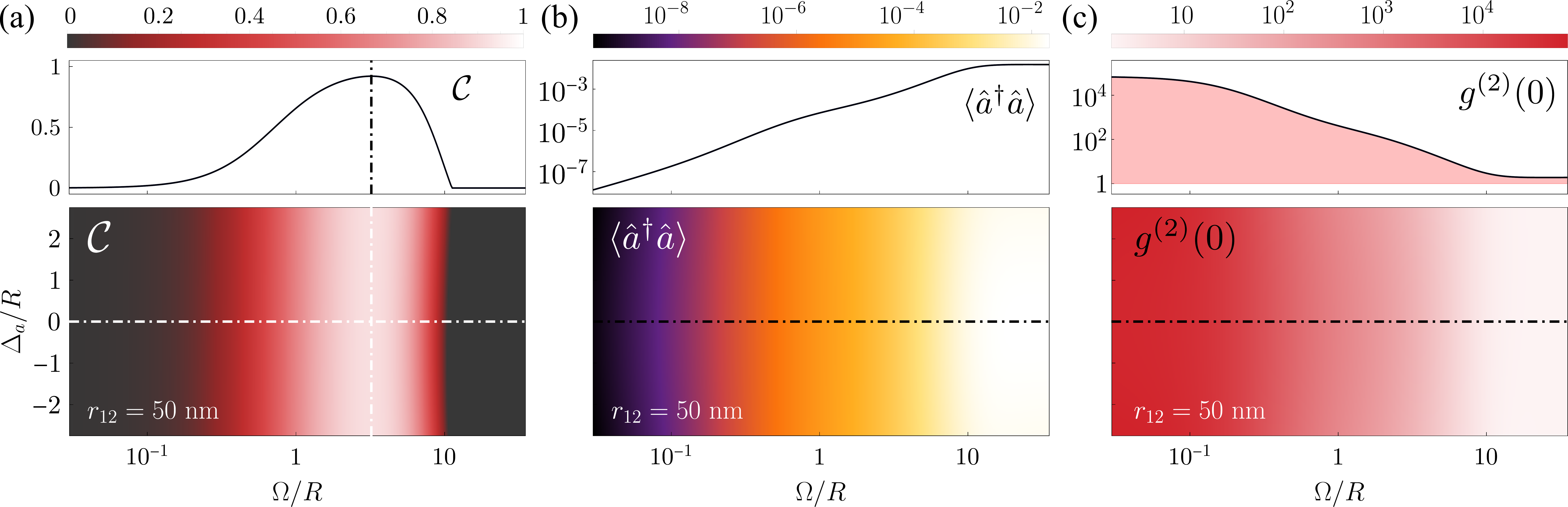}
	\end{center}
	\caption{
	Entanglement detection and control for Mechanism II: stationary concurrence and optical properties of the emitted light by the cavity.
	(a), (b), (c) Steady-state concurrence, transmission intensity and photon correlations versus the Rabi frequency of the laser $\Omega$ and the cavity-laser detuning $\Delta_a$. Top panels show a cut along $\Omega/\gamma$, marked by the horizontal white-dashed lines in the density plots. Parameters:  $r=50$ nm, $k=2\pi/780\ \text{nm}^{-1}$, $J=10.65\gamma$, $\gamma_{12}=0.967\gamma$,  $\Delta=0$, $\delta=10^2 J$, $R=10^3\gamma$, $\kappa=10^4 \gamma$, $g =10^{-1} \kappa$.
	}
	\label{fig:FigAppendixE_Observability_CollectivePurcell}
\end{figure*}
Finally, we analyze the validity of the analytical relaxation rates in Eq.~\eqref{eq:RelaxRate} and Eq.~\eqref{eq:EffectiveRelaxRate} obtained under the application of the HAE method to estimate the Liouvillian gap.

In Fig.~\ref{fig:FigAppendixD_HAE}(a) we show the exact computation of the Liouvillian gap in terms of the cooperativity and the qubit-qubit detuning $\delta$. Through the different contour lines included, we compare the result obtained with the full model Eq.~\eqref{eq:full_master_eq} (black straight lines) with the analytical estimations of the Liouvillian gap given by the full expression in Eq.~\eqref{eq:RelaxRate} (black dashed lines), and the simplified expression in Eq.~\eqref{eq:EffectiveRelaxRate} (white dashed lines).
Both analytical expression feature a good agreement with the numerical computation provided that $\kappa \geq \Omega$, which is precisely the condition required for Mechanism II to take place. Therefore, the deviations below this limit are expected.

In addition, we may observe the agreement between the exact (straight lines) and the analytical (dashed lines) computations in Fig.~\ref{fig:FigAppendixD_HAE}(b), where we depict the time evolution of the antisymmetric state for two values of the cooperativity: (i) $C=10^4$ (blue lines), and (ii) $C=10^6$ (red lines). We note that both the analytical expressions for the density matrix element for $\rho_{A,A}(t)$ and the relaxation rate $\Gamma_{\text{eff}}$ provide a good approximation to exact computation.  
 We note that $\gamma/\lambda_2\approx 1$ approximately coincides with the onset of entanglement generation.

\section{ADDITIONAL OBSERVATIONAL TESTS}
\label{appendix:E}
In this appendix we show the same optical magnitudes presented in Fig.~\ref{fig:Fig10_Observability}, measured in the regime where Mechanism II is active.
As we noted, in this particular regime the generation of stationary entanglement is independent of the cavity frequency for values $\Delta_a \ll \kappa$, resulting in optical observables that are nearly featureless. This is evident in Fig.~\ref{fig:FigAppendixE_Observability_CollectivePurcell}, where we depict the stationary concurrence (a), the mean intensity (b), and the second order correlations (c). 

We observe, as $\Omega$ is increased, the concurrence increases from zero, reaching a maximum value and eventually vanishing. Contrary to the case of Mechanism I, these features are not correlated with any of the observables shown here, and therefore we conclude that, by themselves, they do not serve as any type of indicator of the stabilization of entanglement. This motivates a proposal of an alternative approach based on the measurement of correlations between different spectral lines, see Fig.~\ref{fig:FreqResolved}.
	\let\oldaddcontentsline\addcontentsline
	\renewcommand{\addcontentsline}[3]{}
	\bibliography{Refs_AVV}

\begin{thebibliography}{104}%
\makeatletter
\providecommand \@ifxundefined [1]{%
 \@ifx{#1\undefined}
}%
\providecommand \@ifnum [1]{%
 \ifnum #1\expandafter \@firstoftwo
 \else \expandafter \@secondoftwo
 \fi
}%
\providecommand \@ifx [1]{%
 \ifx #1\expandafter \@firstoftwo
 \else \expandafter \@secondoftwo
 \fi
}%
\providecommand \natexlab [1]{#1}%
\providecommand \enquote  [1]{``#1''}%
\providecommand \bibnamefont  [1]{#1}%
\providecommand \bibfnamefont [1]{#1}%
\providecommand \citenamefont [1]{#1}%
\providecommand \href@noop [0]{\@secondoftwo}%
\providecommand \href [0]{\begingroup \@sanitize@url \@href}%
\providecommand \@href[1]{\@@startlink{#1}\@@href}%
\providecommand \@@href[1]{\endgroup#1\@@endlink}%
\providecommand \@sanitize@url [0]{\catcode `\\12\catcode `\$12\catcode
  `\&12\catcode `\#12\catcode `\^12\catcode `\_12\catcode `\%12\relax}%
\providecommand \@@startlink[1]{}%
\providecommand \@@endlink[0]{}%
\providecommand \url  [0]{\begingroup\@sanitize@url \@url }%
\providecommand \@url [1]{\endgroup\@href {#1}{\urlprefix }}%
\providecommand \urlprefix  [0]{URL }%
\providecommand \Eprint [0]{\href }%
\providecommand \doibase [0]{https://doi.org/}%
\providecommand \selectlanguage [0]{\@gobble}%
\providecommand \bibinfo  [0]{\@secondoftwo}%
\providecommand \bibfield  [0]{\@secondoftwo}%
\providecommand \translation [1]{[#1]}%
\providecommand \BibitemOpen [0]{}%
\providecommand \bibitemStop [0]{}%
\providecommand \bibitemNoStop [0]{.\EOS\space}%
\providecommand \EOS [0]{\spacefactor3000\relax}%
\providecommand \BibitemShut  [1]{\csname bibitem#1\endcsname}%
\let\auto@bib@innerbib\@empty
\bibitem [{\citenamefont {{Vivas-Via{\~n}a}}\ \emph {et~al.}(2023)\citenamefont
  {{Vivas-Via{\~n}a}}, \citenamefont {{Mart{\'i}n-Cano}},\ and\ \citenamefont
  {Mu{\~n}oz}}]{Vivas-VianaFrequencyresolvedPurcell2023}%
  \BibitemOpen
  \bibfield  {author} {\bibinfo {author} {\bibfnamefont {A.}~\bibnamefont
  {{Vivas-Via{\~n}a}}}, \bibinfo {author} {\bibfnamefont {D.}~\bibnamefont
  {{Mart{\'i}n-Cano}}},\ and\ \bibinfo {author} {\bibfnamefont {C.~S.}\
  \bibnamefont {Mu{\~n}oz}},\ }\href@noop {} {\bibinfo {title}
  {Frequency-resolved {{Purcell}} effect for the dissipative generation of
  steady-state entanglement}} (\bibinfo {year} {2023}),\ \Eprint
  {https://arxiv.org/abs/2312.12372} {arxiv:2312.12372} \BibitemShut {NoStop}%
\bibitem [{\citenamefont {Nielsen}\ and\ \citenamefont
  {Chuang}(2012)}]{NielsenQuantumComputation2012}%
  \BibitemOpen
  \bibfield  {author} {\bibinfo {author} {\bibfnamefont {M.~A.}\ \bibnamefont
  {Nielsen}}\ and\ \bibinfo {author} {\bibfnamefont {I.~L.}\ \bibnamefont
  {Chuang}},\ }\href {https://doi.org/10.1017/CBO9780511976667} {\emph
  {\bibinfo {title} {Quantum {{Computation}} and {{Quantum Information}}: 10th
  {{Anniversary Edition}}}}},\ \bibinfo {edition} {1st}\ ed.\ (\bibinfo
  {publisher} {{Cambridge University Press}},\ \bibinfo {year}
  {2012})\BibitemShut {NoStop}%
\bibitem [{\citenamefont {Kimble}(2008)}]{KimbleQuantumInternet2008}%
  \BibitemOpen
  \bibfield  {author} {\bibinfo {author} {\bibfnamefont {H.~J.}\ \bibnamefont
  {Kimble}},\ }\bibfield  {title} {\bibinfo {title} {The quantum internet},\
  }\href {https://doi.org/10.1038/nature07127} {\bibfield  {journal} {\bibinfo
  {journal} {Nature}\ }\textbf {\bibinfo {volume} {453}},\ \bibinfo {pages}
  {1023} (\bibinfo {year} {2008})}\BibitemShut {NoStop}%
\bibitem [{\citenamefont {Narla}\ \emph {et~al.}(2016)\citenamefont {Narla},
  \citenamefont {Shankar}, \citenamefont {Hatridge}, \citenamefont {Leghtas},
  \citenamefont {Sliwa}, \citenamefont {{Zalys-Geller}}, \citenamefont
  {Mundhada}, \citenamefont {Pfaff}, \citenamefont {Frunzio}, \citenamefont
  {Schoelkopf},\ and\ \citenamefont {Devoret}}]{NarlaRobustConcurrent2016}%
  \BibitemOpen
  \bibfield  {author} {\bibinfo {author} {\bibfnamefont {A.}~\bibnamefont
  {Narla}}, \bibinfo {author} {\bibfnamefont {S.}~\bibnamefont {Shankar}},
  \bibinfo {author} {\bibfnamefont {M.}~\bibnamefont {Hatridge}}, \bibinfo
  {author} {\bibfnamefont {Z.}~\bibnamefont {Leghtas}}, \bibinfo {author}
  {\bibfnamefont {K.~M.}\ \bibnamefont {Sliwa}}, \bibinfo {author}
  {\bibfnamefont {E.}~\bibnamefont {{Zalys-Geller}}}, \bibinfo {author}
  {\bibfnamefont {S.~O.}\ \bibnamefont {Mundhada}}, \bibinfo {author}
  {\bibfnamefont {W.}~\bibnamefont {Pfaff}}, \bibinfo {author} {\bibfnamefont
  {L.}~\bibnamefont {Frunzio}}, \bibinfo {author} {\bibfnamefont {R.~J.}\
  \bibnamefont {Schoelkopf}},\ and\ \bibinfo {author} {\bibfnamefont {M.~H.}\
  \bibnamefont {Devoret}},\ }\bibfield  {title} {\bibinfo {title} {Robust
  {{Concurrent Remote Entanglement Between Two Superconducting Qubits}}},\
  }\href {https://doi.org/10.1103/PhysRevX.6.031036} {\bibfield  {journal}
  {\bibinfo  {journal} {Physical Review X}\ }\textbf {\bibinfo {volume} {6}},\
  \bibinfo {pages} {031036} (\bibinfo {year} {2016})}\BibitemShut {NoStop}%
\bibitem [{\citenamefont {Pezz{\`e}}\ \emph {et~al.}(2018)\citenamefont
  {Pezz{\`e}}, \citenamefont {Smerzi}, \citenamefont {Oberthaler},
  \citenamefont {Schmied},\ and\ \citenamefont
  {Treutlein}}]{PezzeQuantumMetrology2018}%
  \BibitemOpen
  \bibfield  {author} {\bibinfo {author} {\bibfnamefont {L.}~\bibnamefont
  {Pezz{\`e}}}, \bibinfo {author} {\bibfnamefont {A.}~\bibnamefont {Smerzi}},
  \bibinfo {author} {\bibfnamefont {M.~K.}\ \bibnamefont {Oberthaler}},
  \bibinfo {author} {\bibfnamefont {R.}~\bibnamefont {Schmied}},\ and\ \bibinfo
  {author} {\bibfnamefont {P.}~\bibnamefont {Treutlein}},\ }\bibfield  {title}
  {\bibinfo {title} {Quantum metrology with nonclassical states of atomic
  ensembles},\ }\href {https://doi.org/10.1103/RevModPhys.90.035005} {\bibfield
   {journal} {\bibinfo  {journal} {Reviews of Modern Physics}\ }\textbf
  {\bibinfo {volume} {90}},\ \bibinfo {pages} {035005} (\bibinfo {year}
  {2018})}\BibitemShut {NoStop}%
\bibitem [{\citenamefont {Brask}\ \emph {et~al.}(2015)\citenamefont {Brask},
  \citenamefont {Chaves},\ and\ \citenamefont
  {Ko{\l}ody{\'n}ski}}]{BraskImprovedQuantum2015}%
  \BibitemOpen
  \bibfield  {author} {\bibinfo {author} {\bibfnamefont {J.~B.}\ \bibnamefont
  {Brask}}, \bibinfo {author} {\bibfnamefont {R.}~\bibnamefont {Chaves}},\ and\
  \bibinfo {author} {\bibfnamefont {J.}~\bibnamefont {Ko{\l}ody{\'n}ski}},\
  }\bibfield  {title} {\bibinfo {title} {Improved {{Quantum Magnetometry}}
  beyond the {{Standard Quantum Limit}}},\ }\href
  {https://doi.org/10.1103/PhysRevX.5.031010} {\bibfield  {journal} {\bibinfo
  {journal} {Physical Review X}\ }\textbf {\bibinfo {volume} {5}},\ \bibinfo
  {pages} {031010} (\bibinfo {year} {2015})}\BibitemShut {NoStop}%
\bibitem [{\citenamefont {Poyatos}\ \emph {et~al.}(1996)\citenamefont
  {Poyatos}, \citenamefont {Cirac},\ and\ \citenamefont
  {Zoller}}]{PoyatosQuantumReservoir1996}%
  \BibitemOpen
  \bibfield  {author} {\bibinfo {author} {\bibfnamefont {J.~F.}\ \bibnamefont
  {Poyatos}}, \bibinfo {author} {\bibfnamefont {J.~I.}\ \bibnamefont {Cirac}},\
  and\ \bibinfo {author} {\bibfnamefont {P.}~\bibnamefont {Zoller}},\
  }\bibfield  {title} {\bibinfo {title} {Quantum {{Reservoir Engineering}} with
  {{Laser Cooled Trapped Ions}}},\ }\href
  {https://doi.org/10.1103/PhysRevLett.77.4728} {\bibfield  {journal} {\bibinfo
   {journal} {Physical Review Letters}\ }\textbf {\bibinfo {volume} {77}},\
  \bibinfo {pages} {4728} (\bibinfo {year} {1996})}\BibitemShut {NoStop}%
\bibitem [{\citenamefont {Verstraete}\ \emph {et~al.}(2009)\citenamefont
  {Verstraete}, \citenamefont {Wolf},\ and\ \citenamefont
  {Ignacio~Cirac}}]{VerstraeteQuantumComputation2009}%
  \BibitemOpen
  \bibfield  {author} {\bibinfo {author} {\bibfnamefont {F.}~\bibnamefont
  {Verstraete}}, \bibinfo {author} {\bibfnamefont {M.~M.}\ \bibnamefont
  {Wolf}},\ and\ \bibinfo {author} {\bibfnamefont {J.}~\bibnamefont
  {Ignacio~Cirac}},\ }\bibfield  {title} {\bibinfo {title} {Quantum computation
  and quantum-state engineering driven by dissipation},\ }\href
  {https://doi.org/10.1038/nphys1342} {\bibfield  {journal} {\bibinfo
  {journal} {Nature Physics}\ }\textbf {\bibinfo {volume} {5}},\ \bibinfo
  {pages} {633} (\bibinfo {year} {2009})}\BibitemShut {NoStop}%
\bibitem [{\citenamefont {Liu}\ \emph {et~al.}(2016)\citenamefont {Liu},
  \citenamefont {Shankar}, \citenamefont {Ofek}, \citenamefont {Hatridge},
  \citenamefont {Narla}, \citenamefont {Sliwa}, \citenamefont {Frunzio},
  \citenamefont {Schoelkopf},\ and\ \citenamefont
  {Devoret}}]{LiuComparingCombining2016}%
  \BibitemOpen
  \bibfield  {author} {\bibinfo {author} {\bibfnamefont {Y.}~\bibnamefont
  {Liu}}, \bibinfo {author} {\bibfnamefont {S.}~\bibnamefont {Shankar}},
  \bibinfo {author} {\bibfnamefont {N.}~\bibnamefont {Ofek}}, \bibinfo {author}
  {\bibfnamefont {M.}~\bibnamefont {Hatridge}}, \bibinfo {author}
  {\bibfnamefont {A.}~\bibnamefont {Narla}}, \bibinfo {author} {\bibfnamefont
  {K.~M.}\ \bibnamefont {Sliwa}}, \bibinfo {author} {\bibfnamefont
  {L.}~\bibnamefont {Frunzio}}, \bibinfo {author} {\bibfnamefont {R.~J.}\
  \bibnamefont {Schoelkopf}},\ and\ \bibinfo {author} {\bibfnamefont {M.~H.}\
  \bibnamefont {Devoret}},\ }\bibfield  {title} {\bibinfo {title} {Comparing
  and {{Combining Measurement-Based}} and {{Driven-Dissipative Entanglement
  Stabilization}}},\ }\href {https://doi.org/10.1103/PhysRevX.6.011022}
  {\bibfield  {journal} {\bibinfo  {journal} {Physical Review X}\ }\textbf
  {\bibinfo {volume} {6}},\ \bibinfo {pages} {011022} (\bibinfo {year}
  {2016})}\BibitemShut {NoStop}%
\bibitem [{\citenamefont {Leghtas}\ \emph {et~al.}(2013)\citenamefont
  {Leghtas}, \citenamefont {Vool}, \citenamefont {Shankar}, \citenamefont
  {Hatridge}, \citenamefont {Girvin}, \citenamefont {Devoret},\ and\
  \citenamefont {Mirrahimi}}]{LeghtasStabilizingBell2013}%
  \BibitemOpen
  \bibfield  {author} {\bibinfo {author} {\bibfnamefont {Z.}~\bibnamefont
  {Leghtas}}, \bibinfo {author} {\bibfnamefont {U.}~\bibnamefont {Vool}},
  \bibinfo {author} {\bibfnamefont {S.}~\bibnamefont {Shankar}}, \bibinfo
  {author} {\bibfnamefont {M.}~\bibnamefont {Hatridge}}, \bibinfo {author}
  {\bibfnamefont {S.~M.}\ \bibnamefont {Girvin}}, \bibinfo {author}
  {\bibfnamefont {M.~H.}\ \bibnamefont {Devoret}},\ and\ \bibinfo {author}
  {\bibfnamefont {M.}~\bibnamefont {Mirrahimi}},\ }\bibfield  {title} {\bibinfo
  {title} {Stabilizing a {{Bell}} state of two superconducting qubits by
  dissipation engineering},\ }\href
  {https://doi.org/10.1103/PhysRevA.88.023849} {\bibfield  {journal} {\bibinfo
  {journal} {Physical Review A}\ }\textbf {\bibinfo {volume} {88}},\ \bibinfo
  {pages} {023849} (\bibinfo {year} {2013})}\BibitemShut {NoStop}%
\bibitem [{\citenamefont {Shankar}\ \emph {et~al.}(2013)\citenamefont
  {Shankar}, \citenamefont {Hatridge}, \citenamefont {Leghtas}, \citenamefont
  {Sliwa}, \citenamefont {Narla}, \citenamefont {Vool}, \citenamefont {Girvin},
  \citenamefont {Frunzio}, \citenamefont {Mirrahimi},\ and\ \citenamefont
  {Devoret}}]{ShankarAutonomouslyStabilized2013}%
  \BibitemOpen
  \bibfield  {author} {\bibinfo {author} {\bibfnamefont {S.}~\bibnamefont
  {Shankar}}, \bibinfo {author} {\bibfnamefont {M.}~\bibnamefont {Hatridge}},
  \bibinfo {author} {\bibfnamefont {Z.}~\bibnamefont {Leghtas}}, \bibinfo
  {author} {\bibfnamefont {K.~M.}\ \bibnamefont {Sliwa}}, \bibinfo {author}
  {\bibfnamefont {A.}~\bibnamefont {Narla}}, \bibinfo {author} {\bibfnamefont
  {U.}~\bibnamefont {Vool}}, \bibinfo {author} {\bibfnamefont {S.~M.}\
  \bibnamefont {Girvin}}, \bibinfo {author} {\bibfnamefont {L.}~\bibnamefont
  {Frunzio}}, \bibinfo {author} {\bibfnamefont {M.}~\bibnamefont {Mirrahimi}},\
  and\ \bibinfo {author} {\bibfnamefont {M.~H.}\ \bibnamefont {Devoret}},\
  }\bibfield  {title} {\bibinfo {title} {Autonomously stabilized entanglement
  between two superconducting quantum bits},\ }\href
  {https://doi.org/10.1038/nature12802} {\bibfield  {journal} {\bibinfo
  {journal} {Nature}\ }\textbf {\bibinfo {volume} {504}},\ \bibinfo {pages}
  {419} (\bibinfo {year} {2013})}\BibitemShut {NoStop}%
\bibitem [{\citenamefont {Krauter}\ \emph {et~al.}(2011)\citenamefont
  {Krauter}, \citenamefont {Muschik}, \citenamefont {Jensen}, \citenamefont
  {Wasilewski}, \citenamefont {Petersen}, \citenamefont {Cirac},\ and\
  \citenamefont {Polzik}}]{KrauterEntanglementGenerated2011}%
  \BibitemOpen
  \bibfield  {author} {\bibinfo {author} {\bibfnamefont {H.}~\bibnamefont
  {Krauter}}, \bibinfo {author} {\bibfnamefont {C.~A.}\ \bibnamefont
  {Muschik}}, \bibinfo {author} {\bibfnamefont {K.}~\bibnamefont {Jensen}},
  \bibinfo {author} {\bibfnamefont {W.}~\bibnamefont {Wasilewski}}, \bibinfo
  {author} {\bibfnamefont {J.~M.}\ \bibnamefont {Petersen}}, \bibinfo {author}
  {\bibfnamefont {J.~I.}\ \bibnamefont {Cirac}},\ and\ \bibinfo {author}
  {\bibfnamefont {E.~S.}\ \bibnamefont {Polzik}},\ }\bibfield  {title}
  {\bibinfo {title} {Entanglement {{Generated}} by {{Dissipation}} and {{Steady
  State Entanglement}} of {{Two Macroscopic Objects}}},\ }\href
  {https://doi.org/10.1103/PhysRevLett.107.080503} {\bibfield  {journal}
  {\bibinfo  {journal} {Physical Review Letters}\ }\textbf {\bibinfo {volume}
  {107}},\ \bibinfo {pages} {080503} (\bibinfo {year} {2011})}\BibitemShut
  {NoStop}%
\bibitem [{\citenamefont {Lin}\ \emph {et~al.}(2013)\citenamefont {Lin},
  \citenamefont {Gaebler}, \citenamefont {Reiter}, \citenamefont {Tan},
  \citenamefont {Bowler}, \citenamefont {S{\o}rensen}, \citenamefont
  {Leibfried},\ and\ \citenamefont {Wineland}}]{LinDissipativeProduction2013}%
  \BibitemOpen
  \bibfield  {author} {\bibinfo {author} {\bibfnamefont {Y.}~\bibnamefont
  {Lin}}, \bibinfo {author} {\bibfnamefont {J.~P.}\ \bibnamefont {Gaebler}},
  \bibinfo {author} {\bibfnamefont {F.}~\bibnamefont {Reiter}}, \bibinfo
  {author} {\bibfnamefont {T.~R.}\ \bibnamefont {Tan}}, \bibinfo {author}
  {\bibfnamefont {R.}~\bibnamefont {Bowler}}, \bibinfo {author} {\bibfnamefont
  {A.~S.}\ \bibnamefont {S{\o}rensen}}, \bibinfo {author} {\bibfnamefont
  {D.}~\bibnamefont {Leibfried}},\ and\ \bibinfo {author} {\bibfnamefont
  {D.~J.}\ \bibnamefont {Wineland}},\ }\bibfield  {title} {\bibinfo {title}
  {Dissipative production of a maximally entangled steady state of two quantum
  bits},\ }\href {https://doi.org/10.1038/nature12801} {\bibfield  {journal}
  {\bibinfo  {journal} {Nature}\ }\textbf {\bibinfo {volume} {504}},\ \bibinfo
  {pages} {415} (\bibinfo {year} {2013})}\BibitemShut {NoStop}%
\bibitem [{\citenamefont {Lodahl}\ \emph {et~al.}(2015)\citenamefont {Lodahl},
  \citenamefont {Mahmoodian},\ and\ \citenamefont
  {Stobbe}}]{LodahlInterfacingSingle2015}%
  \BibitemOpen
  \bibfield  {author} {\bibinfo {author} {\bibfnamefont {P.}~\bibnamefont
  {Lodahl}}, \bibinfo {author} {\bibfnamefont {S.}~\bibnamefont {Mahmoodian}},\
  and\ \bibinfo {author} {\bibfnamefont {S.}~\bibnamefont {Stobbe}},\
  }\bibfield  {title} {\bibinfo {title} {Interfacing single photons and single
  quantum dots with photonic nanostructures},\ }\href
  {https://doi.org/10.1103/RevModPhys.87.347} {\bibfield  {journal} {\bibinfo
  {journal} {Reviews of Modern Physics}\ }\textbf {\bibinfo {volume} {87}},\
  \bibinfo {pages} {347} (\bibinfo {year} {2015})}\BibitemShut {NoStop}%
\bibitem [{\citenamefont {Toninelli}\ \emph {et~al.}(2021)\citenamefont
  {Toninelli}, \citenamefont {Gerhardt}, \citenamefont {Clark}, \citenamefont
  {{Reserbat-Plantey}}, \citenamefont {G{\"o}tzinger}, \citenamefont
  {Ristanovi{\'c}}, \citenamefont {Colautti}, \citenamefont {Lombardi},
  \citenamefont {Major}, \citenamefont {Deperasi{\'n}ska}, \citenamefont
  {Pernice}, \citenamefont {Koppens}, \citenamefont {Kozankiewicz},
  \citenamefont {Gourdon}, \citenamefont {Sandoghdar},\ and\ \citenamefont
  {Orrit}}]{ToninelliSingleOrganic2021}%
  \BibitemOpen
  \bibfield  {author} {\bibinfo {author} {\bibfnamefont {C.}~\bibnamefont
  {Toninelli}}, \bibinfo {author} {\bibfnamefont {I.}~\bibnamefont {Gerhardt}},
  \bibinfo {author} {\bibfnamefont {A.~S.}\ \bibnamefont {Clark}}, \bibinfo
  {author} {\bibfnamefont {A.}~\bibnamefont {{Reserbat-Plantey}}}, \bibinfo
  {author} {\bibfnamefont {S.}~\bibnamefont {G{\"o}tzinger}}, \bibinfo {author}
  {\bibfnamefont {Z.}~\bibnamefont {Ristanovi{\'c}}}, \bibinfo {author}
  {\bibfnamefont {M.}~\bibnamefont {Colautti}}, \bibinfo {author}
  {\bibfnamefont {P.}~\bibnamefont {Lombardi}}, \bibinfo {author}
  {\bibfnamefont {K.~D.}\ \bibnamefont {Major}}, \bibinfo {author}
  {\bibfnamefont {I.}~\bibnamefont {Deperasi{\'n}ska}}, \bibinfo {author}
  {\bibfnamefont {W.~H.}\ \bibnamefont {Pernice}}, \bibinfo {author}
  {\bibfnamefont {F.~H.~L.}\ \bibnamefont {Koppens}}, \bibinfo {author}
  {\bibfnamefont {B.}~\bibnamefont {Kozankiewicz}}, \bibinfo {author}
  {\bibfnamefont {A.}~\bibnamefont {Gourdon}}, \bibinfo {author} {\bibfnamefont
  {V.}~\bibnamefont {Sandoghdar}},\ and\ \bibinfo {author} {\bibfnamefont
  {M.}~\bibnamefont {Orrit}},\ }\bibfield  {title} {\bibinfo {title} {Single
  organic molecules for photonic quantum technologies},\ }\href
  {https://doi.org/10.1038/s41563-021-00987-4} {\bibfield  {journal} {\bibinfo
  {journal} {Nature Materials}\ }\textbf {\bibinfo {volume} {20}},\ \bibinfo
  {pages} {1615} (\bibinfo {year} {2021})}\BibitemShut {NoStop}%
\bibitem [{\citenamefont {Sipahigil}\ \emph {et~al.}(2016)\citenamefont
  {Sipahigil}, \citenamefont {Evans}, \citenamefont {Sukachev}, \citenamefont
  {Burek}, \citenamefont {Borregaard}, \citenamefont {Bhaskar}, \citenamefont
  {Nguyen}, \citenamefont {Pacheco}, \citenamefont {Atikian}, \citenamefont
  {Meuwly}, \citenamefont {Camacho}, \citenamefont {Jelezko}, \citenamefont
  {Bielejec}, \citenamefont {Park}, \citenamefont {Lon{\v c}ar},\ and\
  \citenamefont {Lukin}}]{SipahigilIntegratedDiamond2016}%
  \BibitemOpen
  \bibfield  {author} {\bibinfo {author} {\bibfnamefont {A.}~\bibnamefont
  {Sipahigil}}, \bibinfo {author} {\bibfnamefont {R.~E.}\ \bibnamefont
  {Evans}}, \bibinfo {author} {\bibfnamefont {D.~D.}\ \bibnamefont {Sukachev}},
  \bibinfo {author} {\bibfnamefont {M.~J.}\ \bibnamefont {Burek}}, \bibinfo
  {author} {\bibfnamefont {J.}~\bibnamefont {Borregaard}}, \bibinfo {author}
  {\bibfnamefont {M.~K.}\ \bibnamefont {Bhaskar}}, \bibinfo {author}
  {\bibfnamefont {C.~T.}\ \bibnamefont {Nguyen}}, \bibinfo {author}
  {\bibfnamefont {J.~L.}\ \bibnamefont {Pacheco}}, \bibinfo {author}
  {\bibfnamefont {H.~A.}\ \bibnamefont {Atikian}}, \bibinfo {author}
  {\bibfnamefont {C.}~\bibnamefont {Meuwly}}, \bibinfo {author} {\bibfnamefont
  {R.~M.}\ \bibnamefont {Camacho}}, \bibinfo {author} {\bibfnamefont
  {F.}~\bibnamefont {Jelezko}}, \bibinfo {author} {\bibfnamefont
  {E.}~\bibnamefont {Bielejec}}, \bibinfo {author} {\bibfnamefont
  {H.}~\bibnamefont {Park}}, \bibinfo {author} {\bibfnamefont {M.}~\bibnamefont
  {Lon{\v c}ar}},\ and\ \bibinfo {author} {\bibfnamefont {M.~D.}\ \bibnamefont
  {Lukin}},\ }\bibfield  {title} {\bibinfo {title} {An integrated diamond
  nanophotonics platform for quantum-optical networks},\ }\href
  {https://doi.org/10.1126/science.aah6875} {\bibfield  {journal} {\bibinfo
  {journal} {Science}\ }\textbf {\bibinfo {volume} {354}},\ \bibinfo {pages}
  {847} (\bibinfo {year} {2016})}\BibitemShut {NoStop}%
\bibitem [{\citenamefont {Awschalom}\ \emph {et~al.}(2018)\citenamefont
  {Awschalom}, \citenamefont {Hanson}, \citenamefont {Wrachtrup},\ and\
  \citenamefont {Zhou}}]{AwschalomQuantumTechnologies2018}%
  \BibitemOpen
  \bibfield  {author} {\bibinfo {author} {\bibfnamefont {D.~D.}\ \bibnamefont
  {Awschalom}}, \bibinfo {author} {\bibfnamefont {R.}~\bibnamefont {Hanson}},
  \bibinfo {author} {\bibfnamefont {J.}~\bibnamefont {Wrachtrup}},\ and\
  \bibinfo {author} {\bibfnamefont {B.~B.}\ \bibnamefont {Zhou}},\ }\bibfield
  {title} {\bibinfo {title} {Quantum technologies with optically interfaced
  solid-state spins},\ }\href {https://doi.org/10.1038/s41566-018-0232-2}
  {\bibfield  {journal} {\bibinfo  {journal} {Nature Photonics}\ }\textbf
  {\bibinfo {volume} {12}},\ \bibinfo {pages} {516} (\bibinfo {year}
  {2018})}\BibitemShut {NoStop}%
\bibitem [{\citenamefont {O'Brien}\ \emph {et~al.}(2009)\citenamefont
  {O'Brien}, \citenamefont {Furusawa},\ and\ \citenamefont {Vu{\v
  c}kovi{\'c}}}]{OBrienPhotonicQuantum2009}%
  \BibitemOpen
  \bibfield  {author} {\bibinfo {author} {\bibfnamefont {J.~L.}\ \bibnamefont
  {O'Brien}}, \bibinfo {author} {\bibfnamefont {A.}~\bibnamefont {Furusawa}},\
  and\ \bibinfo {author} {\bibfnamefont {J.}~\bibnamefont {Vu{\v
  c}kovi{\'c}}},\ }\bibfield  {title} {\bibinfo {title} {Photonic quantum
  technologies},\ }\href {https://doi.org/10.1038/nphoton.2009.229} {\bibfield
  {journal} {\bibinfo  {journal} {Nature Photonics}\ }\textbf {\bibinfo
  {volume} {3}},\ \bibinfo {pages} {687} (\bibinfo {year} {2009})}\BibitemShut
  {NoStop}%
\bibitem [{\citenamefont {Wang}\ \emph {et~al.}(2020)\citenamefont {Wang},
  \citenamefont {Sciarrino}, \citenamefont {Laing},\ and\ \citenamefont
  {Thompson}}]{WangIntegratedPhotonic2020}%
  \BibitemOpen
  \bibfield  {author} {\bibinfo {author} {\bibfnamefont {J.}~\bibnamefont
  {Wang}}, \bibinfo {author} {\bibfnamefont {F.}~\bibnamefont {Sciarrino}},
  \bibinfo {author} {\bibfnamefont {A.}~\bibnamefont {Laing}},\ and\ \bibinfo
  {author} {\bibfnamefont {M.~G.}\ \bibnamefont {Thompson}},\ }\bibfield
  {title} {\bibinfo {title} {Integrated photonic quantum technologies},\ }\href
  {https://doi.org/10.1038/s41566-019-0532-1} {\bibfield  {journal} {\bibinfo
  {journal} {Nature Photonics}\ }\textbf {\bibinfo {volume} {14}},\ \bibinfo
  {pages} {273} (\bibinfo {year} {2020})}\BibitemShut {NoStop}%
\bibitem [{\citenamefont {{Gonzalez-Tudela}}\ \emph {et~al.}(2011)\citenamefont
  {{Gonzalez-Tudela}}, \citenamefont {{Martin-Cano}}, \citenamefont {Moreno},
  \citenamefont {{Martin-Moreno}}, \citenamefont {Tejedor},\ and\ \citenamefont
  {{Garcia-Vidal}}}]{Gonzalez-TudelaEntanglementTwo2011}%
  \BibitemOpen
  \bibfield  {author} {\bibinfo {author} {\bibfnamefont {A.}~\bibnamefont
  {{Gonzalez-Tudela}}}, \bibinfo {author} {\bibfnamefont {D.}~\bibnamefont
  {{Martin-Cano}}}, \bibinfo {author} {\bibfnamefont {E.}~\bibnamefont
  {Moreno}}, \bibinfo {author} {\bibfnamefont {L.}~\bibnamefont
  {{Martin-Moreno}}}, \bibinfo {author} {\bibfnamefont {C.}~\bibnamefont
  {Tejedor}},\ and\ \bibinfo {author} {\bibfnamefont {F.~J.}\ \bibnamefont
  {{Garcia-Vidal}}},\ }\bibfield  {title} {\bibinfo {title} {Entanglement of
  {{Two Qubits Mediated}} by {{One-Dimensional Plasmonic Waveguides}}},\ }\href
  {https://doi.org/10.1103/PhysRevLett.106.020501} {\bibfield  {journal}
  {\bibinfo  {journal} {Physical Review Letters}\ }\textbf {\bibinfo {volume}
  {106}},\ \bibinfo {pages} {020501} (\bibinfo {year} {2011})}\BibitemShut
  {NoStop}%
\bibitem [{\citenamefont {{Mart{\'i}n-Cano}}\ \emph {et~al.}(2011)\citenamefont
  {{Mart{\'i}n-Cano}}, \citenamefont {{Gonz{\'a}lez-Tudela}}, \citenamefont
  {{Mart{\'i}n-Moreno}}, \citenamefont {{Garc{\'i}a-Vidal}}, \citenamefont
  {Tejedor},\ and\ \citenamefont
  {Moreno}}]{Martin-CanoDissipationdrivenGeneration2011}%
  \BibitemOpen
  \bibfield  {author} {\bibinfo {author} {\bibfnamefont {D.}~\bibnamefont
  {{Mart{\'i}n-Cano}}}, \bibinfo {author} {\bibfnamefont {A.}~\bibnamefont
  {{Gonz{\'a}lez-Tudela}}}, \bibinfo {author} {\bibfnamefont {L.}~\bibnamefont
  {{Mart{\'i}n-Moreno}}}, \bibinfo {author} {\bibfnamefont {F.~J.}\
  \bibnamefont {{Garc{\'i}a-Vidal}}}, \bibinfo {author} {\bibfnamefont
  {C.}~\bibnamefont {Tejedor}},\ and\ \bibinfo {author} {\bibfnamefont
  {E.}~\bibnamefont {Moreno}},\ }\bibfield  {title} {\bibinfo {title}
  {Dissipation-driven generation of two-qubit entanglement mediated by
  plasmonic waveguides},\ }\href {https://doi.org/10.1103/PhysRevB.84.235306}
  {\bibfield  {journal} {\bibinfo  {journal} {Physical Review B}\ }\textbf
  {\bibinfo {volume} {84}},\ \bibinfo {pages} {235306} (\bibinfo {year}
  {2011})}\BibitemShut {NoStop}%
\bibitem [{\citenamefont {{Gonz{\'a}lez-Tudela}}\ and\ \citenamefont
  {Porras}(2013)}]{Gonzalez-TudelaMesoscopicEntanglement2013}%
  \BibitemOpen
  \bibfield  {author} {\bibinfo {author} {\bibfnamefont {A.}~\bibnamefont
  {{Gonz{\'a}lez-Tudela}}}\ and\ \bibinfo {author} {\bibfnamefont
  {D.}~\bibnamefont {Porras}},\ }\bibfield  {title} {\bibinfo {title}
  {Mesoscopic {{Entanglement Induced}} by {{Spontaneous Emission}} in
  {{Solid-State Quantum Optics}}},\ }\href
  {https://doi.org/10.1103/PhysRevLett.110.080502} {\bibfield  {journal}
  {\bibinfo  {journal} {Physical Review Letters}\ }\textbf {\bibinfo {volume}
  {110}},\ \bibinfo {pages} {080502} (\bibinfo {year} {2013})}\BibitemShut
  {NoStop}%
\bibitem [{\citenamefont {Ramos}\ \emph {et~al.}(2014)\citenamefont {Ramos},
  \citenamefont {Pichler}, \citenamefont {Daley},\ and\ \citenamefont
  {Zoller}}]{RamosQuantumSpin2014}%
  \BibitemOpen
  \bibfield  {author} {\bibinfo {author} {\bibfnamefont {T.}~\bibnamefont
  {Ramos}}, \bibinfo {author} {\bibfnamefont {H.}~\bibnamefont {Pichler}},
  \bibinfo {author} {\bibfnamefont {A.~J.}\ \bibnamefont {Daley}},\ and\
  \bibinfo {author} {\bibfnamefont {P.}~\bibnamefont {Zoller}},\ }\bibfield
  {title} {\bibinfo {title} {Quantum {{Spin Dimers}} from {{Chiral
  Dissipation}} in {{Cold-Atom Chains}}},\ }\href
  {https://doi.org/10.1103/PhysRevLett.113.237203} {\bibfield  {journal}
  {\bibinfo  {journal} {Physical Review Letters}\ }\textbf {\bibinfo {volume}
  {113}},\ \bibinfo {pages} {237203} (\bibinfo {year} {2014})}\BibitemShut
  {NoStop}%
\bibitem [{\citenamefont {Pichler}\ \emph {et~al.}(2015)\citenamefont
  {Pichler}, \citenamefont {Ramos}, \citenamefont {Daley},\ and\ \citenamefont
  {Zoller}}]{PichlerQuantumOptics2015}%
  \BibitemOpen
  \bibfield  {author} {\bibinfo {author} {\bibfnamefont {H.}~\bibnamefont
  {Pichler}}, \bibinfo {author} {\bibfnamefont {T.}~\bibnamefont {Ramos}},
  \bibinfo {author} {\bibfnamefont {A.~J.}\ \bibnamefont {Daley}},\ and\
  \bibinfo {author} {\bibfnamefont {P.}~\bibnamefont {Zoller}},\ }\bibfield
  {title} {\bibinfo {title} {Quantum optics of chiral spin networks},\ }\href
  {https://doi.org/10.1103/PhysRevA.91.042116} {\bibfield  {journal} {\bibinfo
  {journal} {Physical Review A}\ }\textbf {\bibinfo {volume} {91}},\ \bibinfo
  {pages} {042116} (\bibinfo {year} {2015})}\BibitemShut {NoStop}%
\bibitem [{\citenamefont {Haakh}\ and\ \citenamefont
  {{Mart{\'i}n-Cano}}(2015)}]{HaakhSqueezedLight2015}%
  \BibitemOpen
  \bibfield  {author} {\bibinfo {author} {\bibfnamefont {H.~R.}\ \bibnamefont
  {Haakh}}\ and\ \bibinfo {author} {\bibfnamefont {D.}~\bibnamefont
  {{Mart{\'i}n-Cano}}},\ }\bibfield  {title} {\bibinfo {title} {Squeezed
  {{Light}} from {{Entangled Nonidentical Emitters}} via {{Nanophotonic
  Environments}}},\ }\href {https://doi.org/10.1021/acsphotonics.5b00585}
  {\bibfield  {journal} {\bibinfo  {journal} {ACS Photonics}\ }\textbf
  {\bibinfo {volume} {2}},\ \bibinfo {pages} {1686} (\bibinfo {year}
  {2015})}\BibitemShut {NoStop}%
\bibitem [{\citenamefont {Chang}\ \emph {et~al.}(2018)\citenamefont {Chang},
  \citenamefont {Douglas}, \citenamefont {{Gonz{\'a}lez-Tudela}}, \citenamefont
  {Hung},\ and\ \citenamefont {Kimble}}]{ChangColloquiumQuantum2018}%
  \BibitemOpen
  \bibfield  {author} {\bibinfo {author} {\bibfnamefont {D.~E.}\ \bibnamefont
  {Chang}}, \bibinfo {author} {\bibfnamefont {J.~S.}\ \bibnamefont {Douglas}},
  \bibinfo {author} {\bibfnamefont {A.}~\bibnamefont {{Gonz{\'a}lez-Tudela}}},
  \bibinfo {author} {\bibfnamefont {C.-L.}\ \bibnamefont {Hung}},\ and\
  \bibinfo {author} {\bibfnamefont {H.~J.}\ \bibnamefont {Kimble}},\ }\bibfield
   {title} {\bibinfo {title} {{\emph{Colloquium}} : {{Quantum}} matter built
  from nanoscopic lattices of atoms and photons},\ }\href
  {https://doi.org/10.1103/RevModPhys.90.031002} {\bibfield  {journal}
  {\bibinfo  {journal} {Reviews of Modern Physics}\ }\textbf {\bibinfo {volume}
  {90}},\ \bibinfo {pages} {031002} (\bibinfo {year} {2018})}\BibitemShut
  {NoStop}%
\bibitem [{\citenamefont {Reitz}\ \emph {et~al.}(2022)\citenamefont {Reitz},
  \citenamefont {Sommer},\ and\ \citenamefont
  {Genes}}]{ReitzCooperativeQuantum2022}%
  \BibitemOpen
  \bibfield  {author} {\bibinfo {author} {\bibfnamefont {M.}~\bibnamefont
  {Reitz}}, \bibinfo {author} {\bibfnamefont {C.}~\bibnamefont {Sommer}},\ and\
  \bibinfo {author} {\bibfnamefont {C.}~\bibnamefont {Genes}},\ }\bibfield
  {title} {\bibinfo {title} {Cooperative {{Quantum Phenomena}} in
  {{Light-Matter Platforms}}},\ }\href
  {https://doi.org/10.1103/PRXQuantum.3.010201} {\bibfield  {journal} {\bibinfo
   {journal} {PRX Quantum}\ }\textbf {\bibinfo {volume} {3}},\ \bibinfo {pages}
  {010201} (\bibinfo {year} {2022})}\BibitemShut {NoStop}%
\bibitem [{\citenamefont {Hettich}\ \emph {et~al.}(2002)\citenamefont
  {Hettich}, \citenamefont {Schmitt}, \citenamefont {Zitzmann}, \citenamefont
  {K{\"u}hn}, \citenamefont {Gerhardt},\ and\ \citenamefont
  {Sandoghdar}}]{HettichNanometerResolution2002}%
  \BibitemOpen
  \bibfield  {author} {\bibinfo {author} {\bibfnamefont {C.}~\bibnamefont
  {Hettich}}, \bibinfo {author} {\bibfnamefont {C.}~\bibnamefont {Schmitt}},
  \bibinfo {author} {\bibfnamefont {J.}~\bibnamefont {Zitzmann}}, \bibinfo
  {author} {\bibfnamefont {S.}~\bibnamefont {K{\"u}hn}}, \bibinfo {author}
  {\bibfnamefont {I.}~\bibnamefont {Gerhardt}},\ and\ \bibinfo {author}
  {\bibfnamefont {V.}~\bibnamefont {Sandoghdar}},\ }\bibfield  {title}
  {\bibinfo {title} {Nanometer {{Resolution}} and {{Coherent Optical Dipole
  Coupling}} of {{Two Individual Molecules}}},\ }\href
  {https://doi.org/10.1126/science.1075606} {\bibfield  {journal} {\bibinfo
  {journal} {Science}\ }\textbf {\bibinfo {volume} {298}},\ \bibinfo {pages}
  {385} (\bibinfo {year} {2002})}\BibitemShut {NoStop}%
\bibitem [{\citenamefont {Trebbia}\ \emph {et~al.}(2022)\citenamefont
  {Trebbia}, \citenamefont {Deplano}, \citenamefont {Tamarat},\ and\
  \citenamefont {Lounis}}]{TrebbiaTailoringSuperradiant2022}%
  \BibitemOpen
  \bibfield  {author} {\bibinfo {author} {\bibfnamefont {J.-B.}\ \bibnamefont
  {Trebbia}}, \bibinfo {author} {\bibfnamefont {Q.}~\bibnamefont {Deplano}},
  \bibinfo {author} {\bibfnamefont {P.}~\bibnamefont {Tamarat}},\ and\ \bibinfo
  {author} {\bibfnamefont {B.}~\bibnamefont {Lounis}},\ }\bibfield  {title}
  {\bibinfo {title} {Tailoring the superradiant and subradiant nature of two
  coherently coupled quantum emitters},\ }\href
  {https://doi.org/10.1038/s41467-022-30672-2} {\bibfield  {journal} {\bibinfo
  {journal} {Nature Communications}\ }\textbf {\bibinfo {volume} {13}},\
  \bibinfo {pages} {2962} (\bibinfo {year} {2022})}\BibitemShut {NoStop}%
\bibitem [{\citenamefont {Reitzenstein}\ \emph {et~al.}(2006)\citenamefont
  {Reitzenstein}, \citenamefont {L{\"o}ffler}, \citenamefont {Hofmann},
  \citenamefont {Kubanek}, \citenamefont {Kamp}, \citenamefont {Reithmaier},
  \citenamefont {Forchel}, \citenamefont {Kulakovskii}, \citenamefont
  {Keldysh}, \citenamefont {Ponomarev},\ and\ \citenamefont
  {Reinecke}}]{ReitzensteinCoherentPhotonic2006}%
  \BibitemOpen
  \bibfield  {author} {\bibinfo {author} {\bibfnamefont {S.}~\bibnamefont
  {Reitzenstein}}, \bibinfo {author} {\bibfnamefont {A.}~\bibnamefont
  {L{\"o}ffler}}, \bibinfo {author} {\bibfnamefont {C.}~\bibnamefont
  {Hofmann}}, \bibinfo {author} {\bibfnamefont {A.}~\bibnamefont {Kubanek}},
  \bibinfo {author} {\bibfnamefont {M.}~\bibnamefont {Kamp}}, \bibinfo {author}
  {\bibfnamefont {J.~P.}\ \bibnamefont {Reithmaier}}, \bibinfo {author}
  {\bibfnamefont {A.}~\bibnamefont {Forchel}}, \bibinfo {author} {\bibfnamefont
  {V.~D.}\ \bibnamefont {Kulakovskii}}, \bibinfo {author} {\bibfnamefont
  {L.~V.}\ \bibnamefont {Keldysh}}, \bibinfo {author} {\bibfnamefont {I.~V.}\
  \bibnamefont {Ponomarev}},\ and\ \bibinfo {author} {\bibfnamefont {T.~L.}\
  \bibnamefont {Reinecke}},\ }\bibfield  {title} {\bibinfo {title} {Coherent
  photonic coupling of semiconductor quantum dots},\ }\href
  {https://doi.org/10.1364/OL.31.001738} {\bibfield  {journal} {\bibinfo
  {journal} {Optics Letters}\ }\textbf {\bibinfo {volume} {31}},\ \bibinfo
  {pages} {1738} (\bibinfo {year} {2006})}\BibitemShut {NoStop}%
\bibitem [{\citenamefont {Laucht}\ \emph {et~al.}(2010)\citenamefont {Laucht},
  \citenamefont {{Villas-B{\^o}as}}, \citenamefont {Stobbe}, \citenamefont
  {Hauke}, \citenamefont {Hofbauer}, \citenamefont {B{\"o}hm}, \citenamefont
  {Lodahl}, \citenamefont {Amann}, \citenamefont {Kaniber},\ and\ \citenamefont
  {Finley}}]{LauchtMutualCoupling2010}%
  \BibitemOpen
  \bibfield  {author} {\bibinfo {author} {\bibfnamefont {A.}~\bibnamefont
  {Laucht}}, \bibinfo {author} {\bibfnamefont {J.~M.}\ \bibnamefont
  {{Villas-B{\^o}as}}}, \bibinfo {author} {\bibfnamefont {S.}~\bibnamefont
  {Stobbe}}, \bibinfo {author} {\bibfnamefont {N.}~\bibnamefont {Hauke}},
  \bibinfo {author} {\bibfnamefont {F.}~\bibnamefont {Hofbauer}}, \bibinfo
  {author} {\bibfnamefont {G.}~\bibnamefont {B{\"o}hm}}, \bibinfo {author}
  {\bibfnamefont {P.}~\bibnamefont {Lodahl}}, \bibinfo {author} {\bibfnamefont
  {M.-C.}\ \bibnamefont {Amann}}, \bibinfo {author} {\bibfnamefont
  {M.}~\bibnamefont {Kaniber}},\ and\ \bibinfo {author} {\bibfnamefont {J.~J.}\
  \bibnamefont {Finley}},\ }\bibfield  {title} {\bibinfo {title} {Mutual
  coupling of two semiconductor quantum dots via an optical nanocavity},\
  }\href {https://doi.org/10.1103/PhysRevB.82.075305} {\bibfield  {journal}
  {\bibinfo  {journal} {Physical Review B}\ }\textbf {\bibinfo {volume} {82}},\
  \bibinfo {pages} {075305} (\bibinfo {year} {2010})}\BibitemShut {NoStop}%
\bibitem [{\citenamefont {Kim}\ \emph {et~al.}(2018)\citenamefont {Kim},
  \citenamefont {Aghaeimeibodi}, \citenamefont {Richardson}, \citenamefont
  {Leavitt},\ and\ \citenamefont {Waks}}]{KimSuperRadiantEmission2018}%
  \BibitemOpen
  \bibfield  {author} {\bibinfo {author} {\bibfnamefont {J.-H.}\ \bibnamefont
  {Kim}}, \bibinfo {author} {\bibfnamefont {S.}~\bibnamefont {Aghaeimeibodi}},
  \bibinfo {author} {\bibfnamefont {C.~J.~K.}\ \bibnamefont {Richardson}},
  \bibinfo {author} {\bibfnamefont {R.~P.}\ \bibnamefont {Leavitt}},\ and\
  \bibinfo {author} {\bibfnamefont {E.}~\bibnamefont {Waks}},\ }\bibfield
  {title} {\bibinfo {title} {Super-{{Radiant Emission}} from {{Quantum Dots}}
  in a {{Nanophotonic Waveguide}}},\ }\href
  {https://doi.org/10.1021/acs.nanolett.8b01133} {\bibfield  {journal}
  {\bibinfo  {journal} {Nano Letters}\ }\textbf {\bibinfo {volume} {18}},\
  \bibinfo {pages} {4734} (\bibinfo {year} {2018})}\BibitemShut {NoStop}%
\bibitem [{\citenamefont {Grim}\ \emph {et~al.}(2019)\citenamefont {Grim},
  \citenamefont {Bracker}, \citenamefont {Zalalutdinov}, \citenamefont
  {Carter}, \citenamefont {Kozen}, \citenamefont {Kim}, \citenamefont {Kim},
  \citenamefont {Mlack}, \citenamefont {Yakes}, \citenamefont {Lee},\ and\
  \citenamefont {Gammon}}]{GrimScalableOperando2019}%
  \BibitemOpen
  \bibfield  {author} {\bibinfo {author} {\bibfnamefont {J.~Q.}\ \bibnamefont
  {Grim}}, \bibinfo {author} {\bibfnamefont {A.~S.}\ \bibnamefont {Bracker}},
  \bibinfo {author} {\bibfnamefont {M.}~\bibnamefont {Zalalutdinov}}, \bibinfo
  {author} {\bibfnamefont {S.~G.}\ \bibnamefont {Carter}}, \bibinfo {author}
  {\bibfnamefont {A.~C.}\ \bibnamefont {Kozen}}, \bibinfo {author}
  {\bibfnamefont {M.}~\bibnamefont {Kim}}, \bibinfo {author} {\bibfnamefont
  {C.~S.}\ \bibnamefont {Kim}}, \bibinfo {author} {\bibfnamefont {J.~T.}\
  \bibnamefont {Mlack}}, \bibinfo {author} {\bibfnamefont {M.}~\bibnamefont
  {Yakes}}, \bibinfo {author} {\bibfnamefont {B.}~\bibnamefont {Lee}},\ and\
  \bibinfo {author} {\bibfnamefont {D.}~\bibnamefont {Gammon}},\ }\bibfield
  {title} {\bibinfo {title} {Scalable in operando strain tuning in nanophotonic
  waveguides enabling three-quantum-dot superradiance},\ }\href
  {https://doi.org/10.1038/s41563-019-0418-0} {\bibfield  {journal} {\bibinfo
  {journal} {Nature Materials}\ }\textbf {\bibinfo {volume} {18}},\ \bibinfo
  {pages} {963} (\bibinfo {year} {2019})}\BibitemShut {NoStop}%
\bibitem [{\citenamefont {Tiranov}\ \emph {et~al.}(2023)\citenamefont
  {Tiranov}, \citenamefont {Angelopoulou}, \citenamefont {{van Diepen}},
  \citenamefont {Schrinski}, \citenamefont {Sandberg}, \citenamefont {Wang},
  \citenamefont {Midolo}, \citenamefont {Scholz}, \citenamefont {Wieck},
  \citenamefont {Ludwig}, \citenamefont {S{\o}rensen},\ and\ \citenamefont
  {Lodahl}}]{TiranovCollectiveSuper2023}%
  \BibitemOpen
  \bibfield  {author} {\bibinfo {author} {\bibfnamefont {A.}~\bibnamefont
  {Tiranov}}, \bibinfo {author} {\bibfnamefont {V.}~\bibnamefont
  {Angelopoulou}}, \bibinfo {author} {\bibfnamefont {C.~J.}\ \bibnamefont {{van
  Diepen}}}, \bibinfo {author} {\bibfnamefont {B.}~\bibnamefont {Schrinski}},
  \bibinfo {author} {\bibfnamefont {O.~A.~D.}\ \bibnamefont {Sandberg}},
  \bibinfo {author} {\bibfnamefont {Y.}~\bibnamefont {Wang}}, \bibinfo {author}
  {\bibfnamefont {L.}~\bibnamefont {Midolo}}, \bibinfo {author} {\bibfnamefont
  {S.}~\bibnamefont {Scholz}}, \bibinfo {author} {\bibfnamefont {A.~D.}\
  \bibnamefont {Wieck}}, \bibinfo {author} {\bibfnamefont {A.}~\bibnamefont
  {Ludwig}}, \bibinfo {author} {\bibfnamefont {A.~S.}\ \bibnamefont
  {S{\o}rensen}},\ and\ \bibinfo {author} {\bibfnamefont {P.}~\bibnamefont
  {Lodahl}},\ }\bibfield  {title} {\bibinfo {title} {Collective super- and
  subradiant dynamics between distant optical quantum emitters},\ }\href
  {https://doi.org/10.1126/science.ade9324} {\bibfield  {journal} {\bibinfo
  {journal} {Science}\ }\textbf {\bibinfo {volume} {379}},\ \bibinfo {pages}
  {389} (\bibinfo {year} {2023})}\BibitemShut {NoStop}%
\bibitem [{\citenamefont {Varada}\ and\ \citenamefont
  {Agarwal}(1992)}]{VaradaTwophotonResonance1992}%
  \BibitemOpen
  \bibfield  {author} {\bibinfo {author} {\bibfnamefont {G.~V.}\ \bibnamefont
  {Varada}}\ and\ \bibinfo {author} {\bibfnamefont {G.~S.}\ \bibnamefont
  {Agarwal}},\ }\bibfield  {title} {\bibinfo {title} {Two-photon resonance
  induced by the dipole-dipole interaction},\ }\href
  {https://doi.org/10.1103/PhysRevA.45.6721} {\bibfield  {journal} {\bibinfo
  {journal} {Physical Review A}\ }\textbf {\bibinfo {volume} {45}},\ \bibinfo
  {pages} {6721} (\bibinfo {year} {1992})}\BibitemShut {NoStop}%
\bibitem [{\citenamefont {{Vivas-Via{\~n}a}}\ \emph {et~al.}(2022)\citenamefont
  {{Vivas-Via{\~n}a}}, \citenamefont {{Gonz{\'a}lez-Tudela}},\ and\
  \citenamefont {Mu{\~n}oz}}]{Vivas-VianaUnconventionalMechanism2022}%
  \BibitemOpen
  \bibfield  {author} {\bibinfo {author} {\bibfnamefont {A.}~\bibnamefont
  {{Vivas-Via{\~n}a}}}, \bibinfo {author} {\bibfnamefont {A.}~\bibnamefont
  {{Gonz{\'a}lez-Tudela}}},\ and\ \bibinfo {author} {\bibfnamefont {C.~S.}\
  \bibnamefont {Mu{\~n}oz}},\ }\bibfield  {title} {\bibinfo {title}
  {Unconventional mechanism of virtual-state population through dissipation},\
  }\href {https://doi.org/10.1103/PhysRevA.106.012217} {\bibfield  {journal}
  {\bibinfo  {journal} {Physical Review A}\ }\textbf {\bibinfo {volume}
  {106}},\ \bibinfo {pages} {012217} (\bibinfo {year} {2022})}\BibitemShut
  {NoStop}%
\bibitem [{\citenamefont {Patel}\ \emph {et~al.}(2010)\citenamefont {Patel},
  \citenamefont {Bennett}, \citenamefont {Farrer}, \citenamefont {Nicoll},
  \citenamefont {Ritchie},\ and\ \citenamefont
  {Shields}}]{PatelTwophotonInterference2010}%
  \BibitemOpen
  \bibfield  {author} {\bibinfo {author} {\bibfnamefont {R.~B.}\ \bibnamefont
  {Patel}}, \bibinfo {author} {\bibfnamefont {A.~J.}\ \bibnamefont {Bennett}},
  \bibinfo {author} {\bibfnamefont {I.}~\bibnamefont {Farrer}}, \bibinfo
  {author} {\bibfnamefont {C.~A.}\ \bibnamefont {Nicoll}}, \bibinfo {author}
  {\bibfnamefont {D.~A.}\ \bibnamefont {Ritchie}},\ and\ \bibinfo {author}
  {\bibfnamefont {A.~J.}\ \bibnamefont {Shields}},\ }\bibfield  {title}
  {\bibinfo {title} {Two-photon interference of the emission from electrically
  tunable remote quantum dots},\ }\href
  {https://doi.org/10.1038/nphoton.2010.161} {\bibfield  {journal} {\bibinfo
  {journal} {Nature Photonics}\ }\textbf {\bibinfo {volume} {4}},\ \bibinfo
  {pages} {632} (\bibinfo {year} {2010})}\BibitemShut {NoStop}%
\bibitem [{\citenamefont {Giesz}\ \emph {et~al.}(2016)\citenamefont {Giesz},
  \citenamefont {Somaschi}, \citenamefont {Hornecker}, \citenamefont {Grange},
  \citenamefont {Reznychenko}, \citenamefont {De~Santis}, \citenamefont
  {Demory}, \citenamefont {Gomez}, \citenamefont {Sagnes}, \citenamefont
  {Lema{\^i}tre}, \citenamefont {Krebs}, \citenamefont {{Lanzillotti-Kimura}},
  \citenamefont {Lanco}, \citenamefont {Auffeves},\ and\ \citenamefont
  {Senellart}}]{GieszCoherentManipulation2016}%
  \BibitemOpen
  \bibfield  {author} {\bibinfo {author} {\bibfnamefont {V.}~\bibnamefont
  {Giesz}}, \bibinfo {author} {\bibfnamefont {N.}~\bibnamefont {Somaschi}},
  \bibinfo {author} {\bibfnamefont {G.}~\bibnamefont {Hornecker}}, \bibinfo
  {author} {\bibfnamefont {T.}~\bibnamefont {Grange}}, \bibinfo {author}
  {\bibfnamefont {B.}~\bibnamefont {Reznychenko}}, \bibinfo {author}
  {\bibfnamefont {L.}~\bibnamefont {De~Santis}}, \bibinfo {author}
  {\bibfnamefont {J.}~\bibnamefont {Demory}}, \bibinfo {author} {\bibfnamefont
  {C.}~\bibnamefont {Gomez}}, \bibinfo {author} {\bibfnamefont
  {I.}~\bibnamefont {Sagnes}}, \bibinfo {author} {\bibfnamefont
  {A.}~\bibnamefont {Lema{\^i}tre}}, \bibinfo {author} {\bibfnamefont
  {O.}~\bibnamefont {Krebs}}, \bibinfo {author} {\bibfnamefont {N.~D.}\
  \bibnamefont {{Lanzillotti-Kimura}}}, \bibinfo {author} {\bibfnamefont
  {L.}~\bibnamefont {Lanco}}, \bibinfo {author} {\bibfnamefont
  {A.}~\bibnamefont {Auffeves}},\ and\ \bibinfo {author} {\bibfnamefont
  {P.}~\bibnamefont {Senellart}},\ }\bibfield  {title} {\bibinfo {title}
  {Coherent manipulation of a solid-state artificial atom with few photons},\
  }\href {https://doi.org/10.1038/ncomms11986} {\bibfield  {journal} {\bibinfo
  {journal} {Nature Communications}\ }\textbf {\bibinfo {volume} {7}},\
  \bibinfo {pages} {11986} (\bibinfo {year} {2016})}\BibitemShut {NoStop}%
\bibitem [{\citenamefont {Evans}\ \emph {et~al.}(2018)\citenamefont {Evans},
  \citenamefont {Bhaskar}, \citenamefont {Sukachev}, \citenamefont {Nguyen},
  \citenamefont {Sipahigil}, \citenamefont {Burek}, \citenamefont {Machielse},
  \citenamefont {Zhang}, \citenamefont {Zibrov}, \citenamefont {Bielejec},
  \citenamefont {Park}, \citenamefont {Lon{\v c}ar},\ and\ \citenamefont
  {Lukin}}]{EvansPhotonmediatedInteractions2018}%
  \BibitemOpen
  \bibfield  {author} {\bibinfo {author} {\bibfnamefont {R.~E.}\ \bibnamefont
  {Evans}}, \bibinfo {author} {\bibfnamefont {M.~K.}\ \bibnamefont {Bhaskar}},
  \bibinfo {author} {\bibfnamefont {D.~D.}\ \bibnamefont {Sukachev}}, \bibinfo
  {author} {\bibfnamefont {C.~T.}\ \bibnamefont {Nguyen}}, \bibinfo {author}
  {\bibfnamefont {A.}~\bibnamefont {Sipahigil}}, \bibinfo {author}
  {\bibfnamefont {M.~J.}\ \bibnamefont {Burek}}, \bibinfo {author}
  {\bibfnamefont {B.}~\bibnamefont {Machielse}}, \bibinfo {author}
  {\bibfnamefont {G.~H.}\ \bibnamefont {Zhang}}, \bibinfo {author}
  {\bibfnamefont {A.~S.}\ \bibnamefont {Zibrov}}, \bibinfo {author}
  {\bibfnamefont {E.}~\bibnamefont {Bielejec}}, \bibinfo {author}
  {\bibfnamefont {H.}~\bibnamefont {Park}}, \bibinfo {author} {\bibfnamefont
  {M.}~\bibnamefont {Lon{\v c}ar}},\ and\ \bibinfo {author} {\bibfnamefont
  {M.~D.}\ \bibnamefont {Lukin}},\ }\bibfield  {title} {\bibinfo {title}
  {Photon-mediated interactions between quantum emitters in a diamond
  nanocavity},\ }\href {https://doi.org/10.1126/science.aau4691} {\bibfield
  {journal} {\bibinfo  {journal} {Science}\ }\textbf {\bibinfo {volume}
  {362}},\ \bibinfo {pages} {662} (\bibinfo {year} {2018})}\BibitemShut
  {NoStop}%
\bibitem [{\citenamefont {Lukin}\ \emph {et~al.}(2020)\citenamefont {Lukin},
  \citenamefont {Guidry},\ and\ \citenamefont {Vu{\v
  c}kovi{\'c}}}]{LukinIntegratedQuantum2020}%
  \BibitemOpen
  \bibfield  {author} {\bibinfo {author} {\bibfnamefont {D.~M.}\ \bibnamefont
  {Lukin}}, \bibinfo {author} {\bibfnamefont {M.~A.}\ \bibnamefont {Guidry}},\
  and\ \bibinfo {author} {\bibfnamefont {J.}~\bibnamefont {Vu{\v
  c}kovi{\'c}}},\ }\bibfield  {title} {\bibinfo {title} {Integrated {{Quantum
  Photonics}} with {{Silicon Carbide}}: {{Challenges}} and {{Prospects}}},\
  }\href {https://doi.org/10.1103/PRXQuantum.1.020102} {\bibfield  {journal}
  {\bibinfo  {journal} {PRX Quantum}\ }\textbf {\bibinfo {volume} {1}},\
  \bibinfo {pages} {020102} (\bibinfo {year} {2020})}\BibitemShut {NoStop}%
\bibitem [{\citenamefont {Lukin}\ \emph {et~al.}(2023)\citenamefont {Lukin},
  \citenamefont {Guidry}, \citenamefont {Yang}, \citenamefont {Ghezellou},
  \citenamefont {Deb~Mishra}, \citenamefont {Abe}, \citenamefont {Ohshima},
  \citenamefont {{Ul-Hassan}},\ and\ \citenamefont {Vu{\v
  c}kovi{\'c}}}]{LukinTwoEmitterMultimode2023}%
  \BibitemOpen
  \bibfield  {author} {\bibinfo {author} {\bibfnamefont {D.~M.}\ \bibnamefont
  {Lukin}}, \bibinfo {author} {\bibfnamefont {M.~A.}\ \bibnamefont {Guidry}},
  \bibinfo {author} {\bibfnamefont {J.}~\bibnamefont {Yang}}, \bibinfo {author}
  {\bibfnamefont {M.}~\bibnamefont {Ghezellou}}, \bibinfo {author}
  {\bibfnamefont {S.}~\bibnamefont {Deb~Mishra}}, \bibinfo {author}
  {\bibfnamefont {H.}~\bibnamefont {Abe}}, \bibinfo {author} {\bibfnamefont
  {T.}~\bibnamefont {Ohshima}}, \bibinfo {author} {\bibfnamefont
  {J.}~\bibnamefont {{Ul-Hassan}}},\ and\ \bibinfo {author} {\bibfnamefont
  {J.}~\bibnamefont {Vu{\v c}kovi{\'c}}},\ }\bibfield  {title} {\bibinfo
  {title} {Two-{{Emitter Multimode Cavity Quantum Electrodynamics}} in
  {{Thin-Film Silicon Carbide Photonics}}},\ }\href
  {https://doi.org/10.1103/PhysRevX.13.011005} {\bibfield  {journal} {\bibinfo
  {journal} {Physical Review X}\ }\textbf {\bibinfo {volume} {13}},\ \bibinfo
  {pages} {011005} (\bibinfo {year} {2023})}\BibitemShut {NoStop}%
\bibitem [{\citenamefont {Gardiner}\ and\ \citenamefont
  {Zoller}(2004)}]{GardinerQuantumNoise2004}%
  \BibitemOpen
  \bibfield  {author} {\bibinfo {author} {\bibfnamefont {C.~W.}\ \bibnamefont
  {Gardiner}}\ and\ \bibinfo {author} {\bibfnamefont {P.}~\bibnamefont
  {Zoller}},\ }\href@noop {} {\emph {\bibinfo {title} {Quantum Noise: A
  Handbook of {{Markovian}} and Non-{{Markovian}} Quantum Stochastic Methods
  with Applications to Quantum Optics}}},\ \bibinfo {edition} {3rd}\ ed.,\
  Springer Series in Synergetics\ (\bibinfo  {publisher} {{Springer}},\
  \bibinfo {address} {{Berlin ; New York}},\ \bibinfo {year}
  {2004})\BibitemShut {NoStop}%
\bibitem [{\citenamefont {Ficek}\ and\ \citenamefont
  {Swain}(2005)}]{FicekQuantumInterference2005}%
  \BibitemOpen
  \bibfield  {author} {\bibinfo {author} {\bibfnamefont {Z.}~\bibnamefont
  {Ficek}}\ and\ \bibinfo {author} {\bibfnamefont {S.}~\bibnamefont {Swain}},\
  }\href {https://doi.org/10.1007/b100106} {\emph {\bibinfo {title} {Quantum
  {{Interference}} and {{Coherence}}}}},\ \bibinfo {series} {Springer
  {{Series}} in {{Optical Sciences}}}, Vol.\ \bibinfo {volume} {100}\ (\bibinfo
   {publisher} {{Springer-Verlag}},\ \bibinfo {address} {{New York}},\ \bibinfo
  {year} {2005})\BibitemShut {NoStop}%
\bibitem [{\citenamefont {Breuer}\ and\ \citenamefont
  {Petruccione}(2007)}]{BreuerTheoryOpen2007}%
  \BibitemOpen
  \bibfield  {author} {\bibinfo {author} {\bibfnamefont {H.~P.}\ \bibnamefont
  {Breuer}}\ and\ \bibinfo {author} {\bibfnamefont {F.}~\bibnamefont
  {Petruccione}},\ }\href
  {https://doi.org/10.1093/acprof:oso/9780199213900.001.0001} {\emph {\bibinfo
  {title} {The {{Theory}} of {{Open Quantum Systems}}}}},\ Vol.\ \bibinfo
  {volume} {9780199213}\ (\bibinfo  {publisher} {{Oxford University Press}},\
  \bibinfo {year} {2007})\BibitemShut {NoStop}%
\bibitem [{\citenamefont
  {Garc{\'i}a~Ripoll}(2022)}]{GarciaRipollQuantumInformation2022}%
  \BibitemOpen
  \bibfield  {author} {\bibinfo {author} {\bibfnamefont {J.~J.}\ \bibnamefont
  {Garc{\'i}a~Ripoll}},\ }\href {https://doi.org/10.1017/9781316779460} {\emph
  {\bibinfo {title} {Quantum {{Information}} and {{Quantum Optics}} with
  {{Superconducting Circuits}}}}},\ \bibinfo {edition} {1st}\ ed.\ (\bibinfo
  {publisher} {{Cambridge University Press}},\ \bibinfo {year}
  {2022})\BibitemShut {NoStop}%
\bibitem [{\citenamefont
  {Carmichael}(1999)}]{CarmichaelStatisticalMethods1999}%
  \BibitemOpen
  \bibfield  {author} {\bibinfo {author} {\bibfnamefont {H.~J.}\ \bibnamefont
  {Carmichael}},\ }\href {https://doi.org/10.1007/978-3-662-03875-8} {\emph
  {\bibinfo {title} {Statistical {{Methods}} in {{Quantum Optics}} 1}}}\
  (\bibinfo  {publisher} {{Springer Berlin Heidelberg}},\ \bibinfo {address}
  {{Berlin, Heidelberg}},\ \bibinfo {year} {1999})\BibitemShut {NoStop}%
\bibitem [{\citenamefont {{Miguel-Torcal}}\ \emph {et~al.}(2022)\citenamefont
  {{Miguel-Torcal}}, \citenamefont {{Abad-Arredondo}}, \citenamefont
  {{Garc{\'i}a-Vidal}},\ and\ \citenamefont
  {{Fern{\'a}ndez-Dom{\'i}nguez}}}]{Miguel-TorcalInversedesignedDielectric2022}%
  \BibitemOpen
  \bibfield  {author} {\bibinfo {author} {\bibfnamefont {A.}~\bibnamefont
  {{Miguel-Torcal}}}, \bibinfo {author} {\bibfnamefont {J.}~\bibnamefont
  {{Abad-Arredondo}}}, \bibinfo {author} {\bibfnamefont {F.~J.}\ \bibnamefont
  {{Garc{\'i}a-Vidal}}},\ and\ \bibinfo {author} {\bibfnamefont {A.~I.}\
  \bibnamefont {{Fern{\'a}ndez-Dom{\'i}nguez}}},\ }\bibfield  {title} {\bibinfo
  {title} {Inverse-designed dielectric cloaks for entanglement generation},\
  }\href {https://doi.org/10.1515/nanoph-2022-0231} {\bibfield  {journal}
  {\bibinfo  {journal} {Nanophotonics}\ }\textbf {\bibinfo {volume} {11}},\
  \bibinfo {pages} {4387} (\bibinfo {year} {2022})}\BibitemShut {NoStop}%
\bibitem [{\citenamefont {{Vivas-Via{\~n}a}}\ and\ \citenamefont
  {S{\'a}nchez~Mu{\~n}oz}(2021)}]{Vivas-VianaTwophotonResonance2021}%
  \BibitemOpen
  \bibfield  {author} {\bibinfo {author} {\bibfnamefont {A.}~\bibnamefont
  {{Vivas-Via{\~n}a}}}\ and\ \bibinfo {author} {\bibfnamefont {C.}~\bibnamefont
  {S{\'a}nchez~Mu{\~n}oz}},\ }\bibfield  {title} {\bibinfo {title} {Two-photon
  resonance fluorescence of two interacting nonidentical quantum emitters},\
  }\href {https://doi.org/10.1103/PhysRevResearch.3.033136} {\bibfield
  {journal} {\bibinfo  {journal} {Physical Review Research}\ }\textbf {\bibinfo
  {volume} {3}},\ \bibinfo {pages} {33136} (\bibinfo {year}
  {2021})}\BibitemShut {NoStop}%
\bibitem [{\citenamefont {S{\'a}nchez~Mu{\~n}oz}\ and\ \citenamefont
  {Schlawin}(2020)}]{SanchezMunozPhotonCorrelation2020}%
  \BibitemOpen
  \bibfield  {author} {\bibinfo {author} {\bibfnamefont {C.}~\bibnamefont
  {S{\'a}nchez~Mu{\~n}oz}}\ and\ \bibinfo {author} {\bibfnamefont
  {F.}~\bibnamefont {Schlawin}},\ }\bibfield  {title} {\bibinfo {title} {Photon
  correlation spectroscopy as a witness for quantum coherence},\ }\href
  {https://doi.org/10.1103/PhysRevLett.124.203601} {\bibfield  {journal}
  {\bibinfo  {journal} {Physical Review Letters}\ }\textbf {\bibinfo {volume}
  {124}},\ \bibinfo {pages} {203601} (\bibinfo {year} {2020})}\BibitemShut
  {NoStop}%
\bibitem [{\citenamefont {Darsheshdar}\ \emph {et~al.}(2021)\citenamefont
  {Darsheshdar}, \citenamefont {Hugbart}, \citenamefont {Bachelard},\ and\
  \citenamefont {{Villas-Boas}}}]{DarsheshdarPhotonphotonCorrelations2021}%
  \BibitemOpen
  \bibfield  {author} {\bibinfo {author} {\bibfnamefont {E.}~\bibnamefont
  {Darsheshdar}}, \bibinfo {author} {\bibfnamefont {M.}~\bibnamefont
  {Hugbart}}, \bibinfo {author} {\bibfnamefont {R.}~\bibnamefont {Bachelard}},\
  and\ \bibinfo {author} {\bibfnamefont {C.~J.}\ \bibnamefont
  {{Villas-Boas}}},\ }\bibfield  {title} {\bibinfo {title} {Photon-photon
  correlations from a pair of strongly coupled two-level emitters},\ }\href
  {https://doi.org/10.1103/PhysRevA.103.053702} {\bibfield  {journal} {\bibinfo
   {journal} {Physical Review A}\ }\textbf {\bibinfo {volume} {103}},\ \bibinfo
  {pages} {053702} (\bibinfo {year} {2021})}\BibitemShut {NoStop}%
\bibitem [{\citenamefont {Mollow}(1969)}]{MollowPowerSpectrum1969}%
  \BibitemOpen
  \bibfield  {author} {\bibinfo {author} {\bibfnamefont {B.~R.}\ \bibnamefont
  {Mollow}},\ }\bibfield  {title} {\bibinfo {title} {Power {{Spectrum}} of
  {{Light Scattered}} by {{Two-Level Systems}}},\ }\href
  {https://doi.org/10.1103/PhysRev.188.1969} {\bibfield  {journal} {\bibinfo
  {journal} {Physical Review}\ }\textbf {\bibinfo {volume} {188}},\ \bibinfo
  {pages} {1969} (\bibinfo {year} {1969})}\BibitemShut {NoStop}%
\bibitem [{\citenamefont {Cohen-Tannoudji}\ \emph {et~al.}(1998)\citenamefont
  {Cohen-Tannoudji}, \citenamefont {Dupont-Roc},\ and\ \citenamefont
  {Grynberg}}]{Cohen-TannoudjiAtomPhotonInteractions1998}%
  \BibitemOpen
  \bibfield  {author} {\bibinfo {author} {\bibfnamefont {C.}~\bibnamefont
  {Cohen-Tannoudji}}, \bibinfo {author} {\bibfnamefont {J.}~\bibnamefont
  {Dupont-Roc}},\ and\ \bibinfo {author} {\bibfnamefont {G.}~\bibnamefont
  {Grynberg}},\ }\href {https://doi.org/10.1002/9783527617197} {\emph {\bibinfo
  {title} {Atom-{{Photon Interactions}}}}}\ (\bibinfo  {publisher} {{Wiley}},\
  \bibinfo {year} {1998})\BibitemShut {NoStop}%
\bibitem [{\citenamefont {Ardelt}\ \emph {et~al.}(2016)\citenamefont {Ardelt},
  \citenamefont {Koller}, \citenamefont {Simmet}, \citenamefont {Hanschke},
  \citenamefont {Bechtold}, \citenamefont {Regler}, \citenamefont
  {Wierzbowski}, \citenamefont {Riedl}, \citenamefont {Finley},\ and\
  \citenamefont {M{\"u}ller}}]{ArdeltOpticalControl2016}%
  \BibitemOpen
  \bibfield  {author} {\bibinfo {author} {\bibfnamefont {P.-L.}\ \bibnamefont
  {Ardelt}}, \bibinfo {author} {\bibfnamefont {M.}~\bibnamefont {Koller}},
  \bibinfo {author} {\bibfnamefont {T.}~\bibnamefont {Simmet}}, \bibinfo
  {author} {\bibfnamefont {L.}~\bibnamefont {Hanschke}}, \bibinfo {author}
  {\bibfnamefont {A.}~\bibnamefont {Bechtold}}, \bibinfo {author}
  {\bibfnamefont {A.}~\bibnamefont {Regler}}, \bibinfo {author} {\bibfnamefont
  {J.}~\bibnamefont {Wierzbowski}}, \bibinfo {author} {\bibfnamefont
  {H.}~\bibnamefont {Riedl}}, \bibinfo {author} {\bibfnamefont {J.~J.}\
  \bibnamefont {Finley}},\ and\ \bibinfo {author} {\bibfnamefont
  {K.}~\bibnamefont {M{\"u}ller}},\ }\bibfield  {title} {\bibinfo {title}
  {Optical control of nonlinearly dressed states in an individual quantum
  dot},\ }\href {https://doi.org/10.1103/PhysRevB.93.165305} {\bibfield
  {journal} {\bibinfo  {journal} {Physical Review B}\ }\textbf {\bibinfo
  {volume} {93}},\ \bibinfo {pages} {165305} (\bibinfo {year}
  {2016})}\BibitemShut {NoStop}%
\bibitem [{\citenamefont {Hargart}\ \emph {et~al.}(2016)\citenamefont
  {Hargart}, \citenamefont {M{\"u}ller}, \citenamefont {{Roy-Choudhury}},
  \citenamefont {Portalupi}, \citenamefont {Schneider}, \citenamefont
  {H{\"o}fling}, \citenamefont {Kamp}, \citenamefont {Hughes},\ and\
  \citenamefont {Michler}}]{HargartCavityenhancedSimultaneous2016}%
  \BibitemOpen
  \bibfield  {author} {\bibinfo {author} {\bibfnamefont {F.}~\bibnamefont
  {Hargart}}, \bibinfo {author} {\bibfnamefont {M.}~\bibnamefont {M{\"u}ller}},
  \bibinfo {author} {\bibfnamefont {K.}~\bibnamefont {{Roy-Choudhury}}},
  \bibinfo {author} {\bibfnamefont {S.~L.}\ \bibnamefont {Portalupi}}, \bibinfo
  {author} {\bibfnamefont {C.}~\bibnamefont {Schneider}}, \bibinfo {author}
  {\bibfnamefont {S.}~\bibnamefont {H{\"o}fling}}, \bibinfo {author}
  {\bibfnamefont {M.}~\bibnamefont {Kamp}}, \bibinfo {author} {\bibfnamefont
  {S.}~\bibnamefont {Hughes}},\ and\ \bibinfo {author} {\bibfnamefont
  {P.}~\bibnamefont {Michler}},\ }\bibfield  {title} {\bibinfo {title}
  {Cavity-enhanced simultaneous dressing of quantum dot exciton and biexciton
  states},\ }\href {https://doi.org/10.1103/PhysRevB.93.115308} {\bibfield
  {journal} {\bibinfo  {journal} {Physical Review B}\ }\textbf {\bibinfo
  {volume} {93}},\ \bibinfo {pages} {115308} (\bibinfo {year}
  {2016})}\BibitemShut {NoStop}%
\bibitem [{\citenamefont {Scully}\ and\ \citenamefont
  {Zubairy}(1997)}]{ScullyQuantumOptics1997}%
  \BibitemOpen
  \bibfield  {author} {\bibinfo {author} {\bibfnamefont {M.~O.}\ \bibnamefont
  {Scully}}\ and\ \bibinfo {author} {\bibfnamefont {M.~S.}\ \bibnamefont
  {Zubairy}},\ }\href {https://doi.org/10.1017/CBO9780511813993} {\emph
  {\bibinfo {title} {Quantum {{Optics}}}}},\ \bibinfo {edition} {1st}\ ed.\
  (\bibinfo  {publisher} {{Cambridge University Press}},\ \bibinfo {year}
  {1997})\BibitemShut {NoStop}%
\bibitem [{\citenamefont {Novotny}\ and\ \citenamefont
  {Hecht}(2012)}]{NovotnyPrinciplesNanoOptics2012}%
  \BibitemOpen
  \bibfield  {author} {\bibinfo {author} {\bibfnamefont {L.}~\bibnamefont
  {Novotny}}\ and\ \bibinfo {author} {\bibfnamefont {B.}~\bibnamefont
  {Hecht}},\ }\href {https://doi.org/10.1017/CBO9780511794193} {\emph {\bibinfo
  {title} {Principles of {{Nano-Optics}}}}},\ \bibinfo {edition} {2nd}\ ed.\
  (\bibinfo  {publisher} {{Cambridge University Press}},\ \bibinfo {year}
  {2012})\BibitemShut {NoStop}%
\bibitem [{\citenamefont {Savage}(1988)}]{SavageStationaryTwolevel1988}%
  \BibitemOpen
  \bibfield  {author} {\bibinfo {author} {\bibfnamefont {C.~M.}\ \bibnamefont
  {Savage}},\ }\bibfield  {title} {\bibinfo {title} {Stationary two-level
  atomic inversion in a quantized cavity field},\ }\href
  {https://doi.org/10.1103/PhysRevLett.60.1828} {\bibfield  {journal} {\bibinfo
   {journal} {Physical Review Letters}\ }\textbf {\bibinfo {volume} {60}},\
  \bibinfo {pages} {1828} (\bibinfo {year} {1988})}\BibitemShut {NoStop}%
\bibitem [{\citenamefont {Cirac}(1992)}]{CiracInteractionTwolevel1992}%
  \BibitemOpen
  \bibfield  {author} {\bibinfo {author} {\bibfnamefont {J.~I.}\ \bibnamefont
  {Cirac}},\ }\bibfield  {title} {\bibinfo {title} {Interaction of a two-level
  atom with a cavity mode in the bad-cavity limit},\ }\href
  {https://doi.org/10.1103/PhysRevA.46.4354} {\bibfield  {journal} {\bibinfo
  {journal} {Physical Review A}\ }\textbf {\bibinfo {volume} {46}},\ \bibinfo
  {pages} {4354} (\bibinfo {year} {1992})}\BibitemShut {NoStop}%
\bibitem [{\citenamefont {Zhou}\ and\ \citenamefont
  {Swain}(1998)}]{ZhouDynamicsDriven1998}%
  \BibitemOpen
  \bibfield  {author} {\bibinfo {author} {\bibfnamefont {P.}~\bibnamefont
  {Zhou}}\ and\ \bibinfo {author} {\bibfnamefont {S.}~\bibnamefont {Swain}},\
  }\bibfield  {title} {\bibinfo {title} {Dynamics of a driven two-level atom
  coupled to a frequency-tunable cavity},\ }\href
  {https://doi.org/10.1103/PhysRevA.58.1515} {\bibfield  {journal} {\bibinfo
  {journal} {Physical Review A}\ }\textbf {\bibinfo {volume} {58}},\ \bibinfo
  {pages} {1515} (\bibinfo {year} {1998})}\BibitemShut {NoStop}%
\bibitem [{\citenamefont {Rivas}\ and\ \citenamefont
  {Huelga}(2012)}]{RivasOpenQuantum2012}%
  \BibitemOpen
  \bibfield  {author} {\bibinfo {author} {\bibfnamefont {{\'A}.}~\bibnamefont
  {Rivas}}\ and\ \bibinfo {author} {\bibfnamefont {S.~F.}\ \bibnamefont
  {Huelga}},\ }\href {https://doi.org/10.1007/978-3-642-23354-8} {\emph
  {\bibinfo {title} {Open {{Quantum Systems}}}}},\ {{SpringerBriefs}} in
  {{Physics}}\ (\bibinfo  {publisher} {{Springer Berlin Heidelberg}},\ \bibinfo
  {address} {{Berlin, Heidelberg}},\ \bibinfo {year} {2012})\BibitemShut
  {NoStop}%
\bibitem [{\citenamefont {Whitney}(2008)}]{WhitneyStayingPositive2008}%
  \BibitemOpen
  \bibfield  {author} {\bibinfo {author} {\bibfnamefont {R.~S.}\ \bibnamefont
  {Whitney}},\ }\bibfield  {title} {\bibinfo {title} {Staying positive:
  {{Going}} beyond {{Lindblad}} with perturbative master equations},\ }\href
  {https://doi.org/10.1088/1751-8113/41/17/175304} {\bibfield  {journal}
  {\bibinfo  {journal} {Journal of Physics A: Mathematical and Theoretical}\
  }\textbf {\bibinfo {volume} {41}},\ \bibinfo {pages} {175304} (\bibinfo
  {year} {2008})}\BibitemShut {NoStop}%
\bibitem [{\citenamefont {Jeske}\ \emph {et~al.}(2015)\citenamefont {Jeske},
  \citenamefont {Ing}, \citenamefont {Plenio}, \citenamefont {Huelga},\ and\
  \citenamefont {Cole}}]{JeskeBlochRedfieldEquations2015}%
  \BibitemOpen
  \bibfield  {author} {\bibinfo {author} {\bibfnamefont {J.}~\bibnamefont
  {Jeske}}, \bibinfo {author} {\bibfnamefont {D.~J.}\ \bibnamefont {Ing}},
  \bibinfo {author} {\bibfnamefont {M.~B.}\ \bibnamefont {Plenio}}, \bibinfo
  {author} {\bibfnamefont {S.~F.}\ \bibnamefont {Huelga}},\ and\ \bibinfo
  {author} {\bibfnamefont {J.~H.}\ \bibnamefont {Cole}},\ }\bibfield  {title}
  {\bibinfo {title} {Bloch-{{Redfield}} equations for modeling light-harvesting
  complexes},\ }\href {https://doi.org/10.1063/1.4907370} {\bibfield  {journal}
  {\bibinfo  {journal} {The Journal of Chemical Physics}\ }\textbf {\bibinfo
  {volume} {142}},\ \bibinfo {pages} {064104} (\bibinfo {year}
  {2015})}\BibitemShut {NoStop}%
\bibitem [{\citenamefont {Purcell}\ \emph {et~al.}(1946)\citenamefont
  {Purcell}, \citenamefont {Torrey},\ and\ \citenamefont
  {Pound}}]{PurcellResonanceAbsorption1946}%
  \BibitemOpen
  \bibfield  {author} {\bibinfo {author} {\bibfnamefont {E.~M.}\ \bibnamefont
  {Purcell}}, \bibinfo {author} {\bibfnamefont {H.~C.}\ \bibnamefont
  {Torrey}},\ and\ \bibinfo {author} {\bibfnamefont {R.~V.}\ \bibnamefont
  {Pound}},\ }\bibfield  {title} {\bibinfo {title} {Resonance {{Absorption}} by
  {{Nuclear Magnetic Moments}} in a {{Solid}}},\ }\href
  {https://doi.org/10.1103/PhysRev.69.37} {\bibfield  {journal} {\bibinfo
  {journal} {Physical Review}\ }\textbf {\bibinfo {volume} {69}},\ \bibinfo
  {pages} {37} (\bibinfo {year} {1946})}\BibitemShut {NoStop}%
\bibitem [{\citenamefont {Kavokin}\ \emph {et~al.}(2017)\citenamefont
  {Kavokin}, \citenamefont {Baumberg}, \citenamefont {Malpuech},\ and\
  \citenamefont {Laussy}}]{KavokinMicrocavities2017}%
  \BibitemOpen
  \bibfield  {author} {\bibinfo {author} {\bibfnamefont {A.~V.}\ \bibnamefont
  {Kavokin}}, \bibinfo {author} {\bibfnamefont {J.~J.}\ \bibnamefont
  {Baumberg}}, \bibinfo {author} {\bibfnamefont {G.}~\bibnamefont {Malpuech}},\
  and\ \bibinfo {author} {\bibfnamefont {F.~P.}\ \bibnamefont {Laussy}},\
  }\href {https://doi.org/10.1093/oso/9780198782995.001.0001} {\emph {\bibinfo
  {title} {Microcavities}}},\ \bibinfo {edition} {2nd}\ ed.\ (\bibinfo
  {publisher} {{Oxford University PressOxford}},\ \bibinfo {year}
  {2017})\BibitemShut {NoStop}%
\bibitem [{\citenamefont {Wootters}(1998)}]{WoottersEntanglementFormation1998}%
  \BibitemOpen
  \bibfield  {author} {\bibinfo {author} {\bibfnamefont {W.~K.}\ \bibnamefont
  {Wootters}},\ }\bibfield  {title} {\bibinfo {title} {Entanglement of
  formation of an arbitrary state of two qubits},\ }\href
  {https://doi.org/10.1103/PhysRevLett.80.2245} {\bibfield  {journal} {\bibinfo
   {journal} {Physical Review Letters}\ }\textbf {\bibinfo {volume} {80}},\
  \bibinfo {pages} {2245} (\bibinfo {year} {1998})}\BibitemShut {NoStop}%
\bibitem [{\citenamefont {Wootters}(2001)}]{WoottersEntanglementFormation2001}%
  \BibitemOpen
  \bibfield  {author} {\bibinfo {author} {\bibfnamefont {W.~K.}\ \bibnamefont
  {Wootters}},\ }\bibfield  {title} {\bibinfo {title} {Entanglement of
  formation and concurrence},\ }\href {https://doi.org/10.26421/QIC1.1-3}
  {\bibfield  {journal} {\bibinfo  {journal} {Quantum Information and
  Computation}\ }\textbf {\bibinfo {volume} {1}},\ \bibinfo {pages} {27}
  (\bibinfo {year} {2001})}\BibitemShut {NoStop}%
\bibitem [{\citenamefont {Plenio}\ and\ \citenamefont
  {Virmani}(2007)}]{PlenioIntroductionEntanglement2007}%
  \BibitemOpen
  \bibfield  {author} {\bibinfo {author} {\bibfnamefont {M.}~\bibnamefont
  {Plenio}}\ and\ \bibinfo {author} {\bibfnamefont {S.}~\bibnamefont
  {Virmani}},\ }\bibfield  {title} {\bibinfo {title} {An introduction to
  entanglement measures},\ }\href {https://doi.org/10.26421/QIC7.1-2-1}
  {\bibfield  {journal} {\bibinfo  {journal} {Quantum Information and
  Computation}\ }\textbf {\bibinfo {volume} {7}},\ \bibinfo {pages} {1}
  (\bibinfo {year} {2007})}\BibitemShut {NoStop}%
\bibitem [{\citenamefont {Horodecki}\ \emph {et~al.}(2009)\citenamefont
  {Horodecki}, \citenamefont {Horodecki}, \citenamefont {Horodecki},\ and\
  \citenamefont {Horodecki}}]{HorodeckiQuantumEntanglement2009}%
  \BibitemOpen
  \bibfield  {author} {\bibinfo {author} {\bibfnamefont {R.}~\bibnamefont
  {Horodecki}}, \bibinfo {author} {\bibfnamefont {P.}~\bibnamefont
  {Horodecki}}, \bibinfo {author} {\bibfnamefont {M.}~\bibnamefont
  {Horodecki}},\ and\ \bibinfo {author} {\bibfnamefont {K.}~\bibnamefont
  {Horodecki}},\ }\bibfield  {title} {\bibinfo {title} {Quantum entanglement},\
  }\href {https://doi.org/10.1103/RevModPhys.81.865} {\bibfield  {journal}
  {\bibinfo  {journal} {Reviews of Modern Physics}\ }\textbf {\bibinfo {volume}
  {81}},\ \bibinfo {pages} {865} (\bibinfo {year} {2009})}\BibitemShut
  {NoStop}%
\bibitem [{\citenamefont {Seidelmann}\ \emph
  {et~al.}(2021{\natexlab{a}})\citenamefont {Seidelmann}, \citenamefont
  {Cosacchi}, \citenamefont {Cygorek}, \citenamefont {Reiter}, \citenamefont
  {Vagov},\ and\ \citenamefont {Axt}}]{SeidelmannDifferentTypes2021}%
  \BibitemOpen
  \bibfield  {author} {\bibinfo {author} {\bibfnamefont {T.}~\bibnamefont
  {Seidelmann}}, \bibinfo {author} {\bibfnamefont {M.}~\bibnamefont
  {Cosacchi}}, \bibinfo {author} {\bibfnamefont {M.}~\bibnamefont {Cygorek}},
  \bibinfo {author} {\bibfnamefont {D.~E.}\ \bibnamefont {Reiter}}, \bibinfo
  {author} {\bibfnamefont {A.}~\bibnamefont {Vagov}},\ and\ \bibinfo {author}
  {\bibfnamefont {V.~M.}\ \bibnamefont {Axt}},\ }\bibfield  {title} {\bibinfo
  {title} {Different {{Types}} of {{Photon Entanglement}} from a {{Constantly
  Driven Quantum Emitter Inside}} a {{Cavity}}},\ }\href
  {https://doi.org/10.1002/qute.202000108} {\bibfield  {journal} {\bibinfo
  {journal} {Advanced Quantum Technologies}\ }\textbf {\bibinfo {volume} {4}},\
  \bibinfo {pages} {2000108} (\bibinfo {year}
  {2021}{\natexlab{a}})}\BibitemShut {NoStop}%
\bibitem [{\citenamefont {Seidelmann}\ \emph
  {et~al.}(2021{\natexlab{b}})\citenamefont {Seidelmann}, \citenamefont
  {Reiter}, \citenamefont {Cosacchi}, \citenamefont {Cygorek}, \citenamefont
  {Vagov},\ and\ \citenamefont {Axt}}]{SeidelmannTimedependentSwitching2021}%
  \BibitemOpen
  \bibfield  {author} {\bibinfo {author} {\bibfnamefont {T.}~\bibnamefont
  {Seidelmann}}, \bibinfo {author} {\bibfnamefont {D.~E.}\ \bibnamefont
  {Reiter}}, \bibinfo {author} {\bibfnamefont {M.}~\bibnamefont {Cosacchi}},
  \bibinfo {author} {\bibfnamefont {M.}~\bibnamefont {Cygorek}}, \bibinfo
  {author} {\bibfnamefont {A.}~\bibnamefont {Vagov}},\ and\ \bibinfo {author}
  {\bibfnamefont {V.~M.}\ \bibnamefont {Axt}},\ }\bibfield  {title} {\bibinfo
  {title} {Time-dependent switching of the photon entanglement type using a
  driven quantum {{Emitter}}{\textendash}{{Cavity}} system},\ }\href
  {https://doi.org/10.1063/5.0045377} {\bibfield  {journal} {\bibinfo
  {journal} {Applied Physics Letters}\ }\textbf {\bibinfo {volume} {118}},\
  \bibinfo {pages} {164001} (\bibinfo {year} {2021}{\natexlab{b}})}\BibitemShut
  {NoStop}%
\bibitem [{\citenamefont {Seidelmann}\ \emph {et~al.}(2023)\citenamefont
  {Seidelmann}, \citenamefont {Cosacchi}, \citenamefont {Cygorek},
  \citenamefont {Reiter}, \citenamefont {Vagov},\ and\ \citenamefont
  {Axt}}]{SeidelmannPhononinducedTransition2023}%
  \BibitemOpen
  \bibfield  {author} {\bibinfo {author} {\bibfnamefont {T.}~\bibnamefont
  {Seidelmann}}, \bibinfo {author} {\bibfnamefont {M.}~\bibnamefont
  {Cosacchi}}, \bibinfo {author} {\bibfnamefont {M.}~\bibnamefont {Cygorek}},
  \bibinfo {author} {\bibfnamefont {D.~E.}\ \bibnamefont {Reiter}}, \bibinfo
  {author} {\bibfnamefont {A.}~\bibnamefont {Vagov}},\ and\ \bibinfo {author}
  {\bibfnamefont {V.~M.}\ \bibnamefont {Axt}},\ }\bibfield  {title} {\bibinfo
  {title} {Phonon-induced transition between entangled and nonentangled photon
  emission in constantly driven {{Quantum-Dot}}{\textendash}{{Cavity}}
  systems},\ }\href {https://doi.org/10.1103/PhysRevB.107.075301} {\bibfield
  {journal} {\bibinfo  {journal} {Physical Review B}\ }\textbf {\bibinfo
  {volume} {107}},\ \bibinfo {pages} {075301} (\bibinfo {year}
  {2023})}\BibitemShut {NoStop}%
\bibitem [{\citenamefont {Diehl}\ \emph {et~al.}(2014)\citenamefont {Diehl},
  \citenamefont {Roos}, \citenamefont {Duymaz}, \citenamefont {Lunkenheimer},
  \citenamefont {K{\"o}hn},\ and\ \citenamefont
  {Basch{\'e}}}]{DiehlEmergenceCoherence2014}%
  \BibitemOpen
  \bibfield  {author} {\bibinfo {author} {\bibfnamefont {F.~P.}\ \bibnamefont
  {Diehl}}, \bibinfo {author} {\bibfnamefont {C.}~\bibnamefont {Roos}},
  \bibinfo {author} {\bibfnamefont {A.}~\bibnamefont {Duymaz}}, \bibinfo
  {author} {\bibfnamefont {B.}~\bibnamefont {Lunkenheimer}}, \bibinfo {author}
  {\bibfnamefont {A.}~\bibnamefont {K{\"o}hn}},\ and\ \bibinfo {author}
  {\bibfnamefont {T.}~\bibnamefont {Basch{\'e}}},\ }\bibfield  {title}
  {\bibinfo {title} {Emergence of {{Coherence}} through {{Variation}} of
  {{Intermolecular Distances}} in a {{Series}} of {{Molecular Dimers}}},\
  }\href {https://doi.org/10.1021/jz402512g} {\bibfield  {journal} {\bibinfo
  {journal} {The Journal of Physical Chemistry Letters}\ }\textbf {\bibinfo
  {volume} {5}},\ \bibinfo {pages} {262} (\bibinfo {year} {2014})}\BibitemShut
  {NoStop}%
\bibitem [{\citenamefont {Krenner}\ \emph {et~al.}(2005)\citenamefont
  {Krenner}, \citenamefont {Sabathil}, \citenamefont {Clark}, \citenamefont
  {Kress}, \citenamefont {Schuh}, \citenamefont {Bichler}, \citenamefont
  {Abstreiter},\ and\ \citenamefont {Finley}}]{KrennerDirectObservation2005}%
  \BibitemOpen
  \bibfield  {author} {\bibinfo {author} {\bibfnamefont {H.~J.}\ \bibnamefont
  {Krenner}}, \bibinfo {author} {\bibfnamefont {M.}~\bibnamefont {Sabathil}},
  \bibinfo {author} {\bibfnamefont {E.~C.}\ \bibnamefont {Clark}}, \bibinfo
  {author} {\bibfnamefont {A.}~\bibnamefont {Kress}}, \bibinfo {author}
  {\bibfnamefont {D.}~\bibnamefont {Schuh}}, \bibinfo {author} {\bibfnamefont
  {M.}~\bibnamefont {Bichler}}, \bibinfo {author} {\bibfnamefont
  {G.}~\bibnamefont {Abstreiter}},\ and\ \bibinfo {author} {\bibfnamefont
  {J.~J.}\ \bibnamefont {Finley}},\ }\bibfield  {title} {\bibinfo {title}
  {Direct {{Observation}} of {{Controlled Coupling}} in an {{Individual Quantum
  Dot Molecule}}},\ }\href {https://doi.org/10.1103/PhysRevLett.94.057402}
  {\bibfield  {journal} {\bibinfo  {journal} {Physical Review Letters}\
  }\textbf {\bibinfo {volume} {94}},\ \bibinfo {pages} {057402} (\bibinfo
  {year} {2005})}\BibitemShut {NoStop}%
\bibitem [{\citenamefont {Oliveira}\ \emph {et~al.}(2023)\citenamefont
  {Oliveira}, \citenamefont {Higgins}, \citenamefont {Zhang}, \citenamefont
  {Predojevi{\'c}}, \citenamefont {Hennrich}, \citenamefont {Bachelard},\ and\
  \citenamefont {{Villas-Boas}}}]{OliveiraSteadystateEntanglement2023}%
  \BibitemOpen
  \bibfield  {author} {\bibinfo {author} {\bibfnamefont {M.~H.}\ \bibnamefont
  {Oliveira}}, \bibinfo {author} {\bibfnamefont {G.}~\bibnamefont {Higgins}},
  \bibinfo {author} {\bibfnamefont {C.}~\bibnamefont {Zhang}}, \bibinfo
  {author} {\bibfnamefont {A.}~\bibnamefont {Predojevi{\'c}}}, \bibinfo
  {author} {\bibfnamefont {M.}~\bibnamefont {Hennrich}}, \bibinfo {author}
  {\bibfnamefont {R.}~\bibnamefont {Bachelard}},\ and\ \bibinfo {author}
  {\bibfnamefont {C.~J.}\ \bibnamefont {{Villas-Boas}}},\ }\bibfield  {title}
  {\bibinfo {title} {Steady-state entanglement generation for nondegenerate
  qubits},\ }\href {https://doi.org/10.1103/PhysRevA.107.023706} {\bibfield
  {journal} {\bibinfo  {journal} {Physical Review A}\ }\textbf {\bibinfo
  {volume} {107}},\ \bibinfo {pages} {023706} (\bibinfo {year}
  {2023})}\BibitemShut {NoStop}%
\bibitem [{\citenamefont {Sheremet}\ \emph {et~al.}(2023)\citenamefont
  {Sheremet}, \citenamefont {Petrov}, \citenamefont {Iorsh}, \citenamefont
  {Poshakinskiy},\ and\ \citenamefont
  {Poddubny}}]{SheremetWaveguideQuantum2023}%
  \BibitemOpen
  \bibfield  {author} {\bibinfo {author} {\bibfnamefont {A.~S.}\ \bibnamefont
  {Sheremet}}, \bibinfo {author} {\bibfnamefont {M.~I.}\ \bibnamefont
  {Petrov}}, \bibinfo {author} {\bibfnamefont {I.~V.}\ \bibnamefont {Iorsh}},
  \bibinfo {author} {\bibfnamefont {A.~V.}\ \bibnamefont {Poshakinskiy}},\ and\
  \bibinfo {author} {\bibfnamefont {A.~N.}\ \bibnamefont {Poddubny}},\
  }\bibfield  {title} {\bibinfo {title} {Waveguide quantum electrodynamics:
  {{Collective}} radiance and photon-photon correlations},\ }\href
  {https://doi.org/10.1103/RevModPhys.95.015002} {\bibfield  {journal}
  {\bibinfo  {journal} {Reviews of Modern Physics}\ }\textbf {\bibinfo {volume}
  {95}},\ \bibinfo {pages} {015002} (\bibinfo {year} {2023})}\BibitemShut
  {NoStop}%
\bibitem [{\citenamefont {Arcari}\ \emph {et~al.}(2014)\citenamefont {Arcari},
  \citenamefont {S{\"o}llner}, \citenamefont {Javadi}, \citenamefont
  {Lindskov~Hansen}, \citenamefont {Mahmoodian}, \citenamefont {Liu},
  \citenamefont {Thyrrestrup}, \citenamefont {Lee}, \citenamefont {Song},
  \citenamefont {Stobbe},\ and\ \citenamefont
  {Lodahl}}]{ArcariNearUnityCoupling2014}%
  \BibitemOpen
  \bibfield  {author} {\bibinfo {author} {\bibfnamefont {M.}~\bibnamefont
  {Arcari}}, \bibinfo {author} {\bibfnamefont {I.}~\bibnamefont {S{\"o}llner}},
  \bibinfo {author} {\bibfnamefont {A.}~\bibnamefont {Javadi}}, \bibinfo
  {author} {\bibfnamefont {S.}~\bibnamefont {Lindskov~Hansen}}, \bibinfo
  {author} {\bibfnamefont {S.}~\bibnamefont {Mahmoodian}}, \bibinfo {author}
  {\bibfnamefont {J.}~\bibnamefont {Liu}}, \bibinfo {author} {\bibfnamefont
  {H.}~\bibnamefont {Thyrrestrup}}, \bibinfo {author} {\bibfnamefont {E.~H.}\
  \bibnamefont {Lee}}, \bibinfo {author} {\bibfnamefont {J.~D.}\ \bibnamefont
  {Song}}, \bibinfo {author} {\bibfnamefont {S.}~\bibnamefont {Stobbe}},\ and\
  \bibinfo {author} {\bibfnamefont {P.}~\bibnamefont {Lodahl}},\ }\bibfield
  {title} {\bibinfo {title} {Near-{{Unity Coupling Efficiency}} of a {{Quantum
  Emitter}} to a {{Photonic Crystal Waveguide}}},\ }\href
  {https://doi.org/10.1103/PhysRevLett.113.093603} {\bibfield  {journal}
  {\bibinfo  {journal} {Physical Review Letters}\ }\textbf {\bibinfo {volume}
  {113}},\ \bibinfo {pages} {093603} (\bibinfo {year} {2014})}\BibitemShut
  {NoStop}%
\bibitem [{\citenamefont {Koong}\ \emph {et~al.}(2022)\citenamefont {Koong},
  \citenamefont {Cygorek}, \citenamefont {Scerri}, \citenamefont {Santana},
  \citenamefont {Park}, \citenamefont {Song}, \citenamefont {Gauger},\ and\
  \citenamefont {Gerardot}}]{KoongCoherenceCooperative2022}%
  \BibitemOpen
  \bibfield  {author} {\bibinfo {author} {\bibfnamefont {Z.~X.}\ \bibnamefont
  {Koong}}, \bibinfo {author} {\bibfnamefont {M.}~\bibnamefont {Cygorek}},
  \bibinfo {author} {\bibfnamefont {E.}~\bibnamefont {Scerri}}, \bibinfo
  {author} {\bibfnamefont {T.~S.}\ \bibnamefont {Santana}}, \bibinfo {author}
  {\bibfnamefont {S.~I.}\ \bibnamefont {Park}}, \bibinfo {author}
  {\bibfnamefont {J.~D.}\ \bibnamefont {Song}}, \bibinfo {author}
  {\bibfnamefont {E.~M.}\ \bibnamefont {Gauger}},\ and\ \bibinfo {author}
  {\bibfnamefont {B.~D.}\ \bibnamefont {Gerardot}},\ }\bibfield  {title}
  {\bibinfo {title} {Coherence in cooperative photon emission from
  indistinguishable quantum emitters},\ }\href
  {https://doi.org/10.1126/sciadv.abm8171} {\bibfield  {journal} {\bibinfo
  {journal} {Science Advances}\ }\textbf {\bibinfo {volume} {8}},\ \bibinfo
  {pages} {eabm8171} (\bibinfo {year} {2022})}\BibitemShut {NoStop}%
\bibitem [{\citenamefont {S{\'a}nchez~Mu{\~n}oz}\ \emph
  {et~al.}(2019)\citenamefont {S{\'a}nchez~Mu{\~n}oz}, \citenamefont {Bu{\v
  c}a}, \citenamefont {Tindall}, \citenamefont {{Gonz{\'a}lez-Tudela}},
  \citenamefont {Jaksch},\ and\ \citenamefont
  {Porras}}]{SanchezMunozSymmetriesConservation2019}%
  \BibitemOpen
  \bibfield  {author} {\bibinfo {author} {\bibfnamefont {C.}~\bibnamefont
  {S{\'a}nchez~Mu{\~n}oz}}, \bibinfo {author} {\bibfnamefont {B.}~\bibnamefont
  {Bu{\v c}a}}, \bibinfo {author} {\bibfnamefont {J.}~\bibnamefont {Tindall}},
  \bibinfo {author} {\bibfnamefont {A.}~\bibnamefont {{Gonz{\'a}lez-Tudela}}},
  \bibinfo {author} {\bibfnamefont {D.}~\bibnamefont {Jaksch}},\ and\ \bibinfo
  {author} {\bibfnamefont {D.}~\bibnamefont {Porras}},\ }\bibfield  {title}
  {\bibinfo {title} {Symmetries and conservation laws in quantum trajectories:
  {{Dissipative}} freezing},\ }\href
  {https://doi.org/10.1103/PhysRevA.100.042113} {\bibfield  {journal} {\bibinfo
   {journal} {Physical Review A}\ }\textbf {\bibinfo {volume} {100}},\ \bibinfo
  {pages} {042113} (\bibinfo {year} {2019})}\BibitemShut {NoStop}%
\bibitem [{\citenamefont {Flagg}\ \emph {et~al.}(2009)\citenamefont {Flagg},
  \citenamefont {Muller}, \citenamefont {Robertson}, \citenamefont {Founta},
  \citenamefont {Deppe}, \citenamefont {Xiao}, \citenamefont {Ma},
  \citenamefont {Salamo},\ and\ \citenamefont
  {Shih}}]{FlaggResonantlyDriven2009}%
  \BibitemOpen
  \bibfield  {author} {\bibinfo {author} {\bibfnamefont {E.~B.}\ \bibnamefont
  {Flagg}}, \bibinfo {author} {\bibfnamefont {A.}~\bibnamefont {Muller}},
  \bibinfo {author} {\bibfnamefont {J.~W.}\ \bibnamefont {Robertson}}, \bibinfo
  {author} {\bibfnamefont {S.}~\bibnamefont {Founta}}, \bibinfo {author}
  {\bibfnamefont {D.~G.}\ \bibnamefont {Deppe}}, \bibinfo {author}
  {\bibfnamefont {M.}~\bibnamefont {Xiao}}, \bibinfo {author} {\bibfnamefont
  {W.}~\bibnamefont {Ma}}, \bibinfo {author} {\bibfnamefont {G.~J.}\
  \bibnamefont {Salamo}},\ and\ \bibinfo {author} {\bibfnamefont {C.~K.}\
  \bibnamefont {Shih}},\ }\bibfield  {title} {\bibinfo {title} {Resonantly
  driven coherent oscillations in a solid-state quantum emitter},\ }\href
  {https://doi.org/10.1038/nphys1184} {\bibfield  {journal} {\bibinfo
  {journal} {Nature Physics}\ }\textbf {\bibinfo {volume} {5}},\ \bibinfo
  {pages} {203} (\bibinfo {year} {2009})}\BibitemShut {NoStop}%
\bibitem [{\citenamefont {Somaschi}\ \emph {et~al.}(2016)\citenamefont
  {Somaschi}, \citenamefont {Giesz}, \citenamefont {De~Santis}, \citenamefont
  {Loredo}, \citenamefont {Almeida}, \citenamefont {Hornecker}, \citenamefont
  {Portalupi}, \citenamefont {Grange}, \citenamefont {Ant{\'o}n}, \citenamefont
  {Demory}, \citenamefont {G{\'o}mez}, \citenamefont {Sagnes}, \citenamefont
  {{Lanzillotti-Kimura}}, \citenamefont {Lema{\'i}tre}, \citenamefont
  {Auffeves}, \citenamefont {White}, \citenamefont {Lanco},\ and\ \citenamefont
  {Senellart}}]{SomaschiNearoptimalSinglephoton2016}%
  \BibitemOpen
  \bibfield  {author} {\bibinfo {author} {\bibfnamefont {N.}~\bibnamefont
  {Somaschi}}, \bibinfo {author} {\bibfnamefont {V.}~\bibnamefont {Giesz}},
  \bibinfo {author} {\bibfnamefont {L.}~\bibnamefont {De~Santis}}, \bibinfo
  {author} {\bibfnamefont {J.~C.}\ \bibnamefont {Loredo}}, \bibinfo {author}
  {\bibfnamefont {M.~P.}\ \bibnamefont {Almeida}}, \bibinfo {author}
  {\bibfnamefont {G.}~\bibnamefont {Hornecker}}, \bibinfo {author}
  {\bibfnamefont {S.~L.}\ \bibnamefont {Portalupi}}, \bibinfo {author}
  {\bibfnamefont {T.}~\bibnamefont {Grange}}, \bibinfo {author} {\bibfnamefont
  {C.}~\bibnamefont {Ant{\'o}n}}, \bibinfo {author} {\bibfnamefont
  {J.}~\bibnamefont {Demory}}, \bibinfo {author} {\bibfnamefont
  {C.}~\bibnamefont {G{\'o}mez}}, \bibinfo {author} {\bibfnamefont
  {I.}~\bibnamefont {Sagnes}}, \bibinfo {author} {\bibfnamefont {N.~D.}\
  \bibnamefont {{Lanzillotti-Kimura}}}, \bibinfo {author} {\bibfnamefont
  {A.}~\bibnamefont {Lema{\'i}tre}}, \bibinfo {author} {\bibfnamefont
  {A.}~\bibnamefont {Auffeves}}, \bibinfo {author} {\bibfnamefont {A.~G.}\
  \bibnamefont {White}}, \bibinfo {author} {\bibfnamefont {L.}~\bibnamefont
  {Lanco}},\ and\ \bibinfo {author} {\bibfnamefont {P.}~\bibnamefont
  {Senellart}},\ }\bibfield  {title} {\bibinfo {title} {Near-optimal
  single-photon sources in the solid state},\ }\href
  {https://doi.org/10.1038/nphoton.2016.23} {\bibfield  {journal} {\bibinfo
  {journal} {Nature Photonics}\ }\textbf {\bibinfo {volume} {10}},\ \bibinfo
  {pages} {340} (\bibinfo {year} {2016})}\BibitemShut {NoStop}%
\bibitem [{\citenamefont {Fischer}\ \emph {et~al.}(2016)\citenamefont
  {Fischer}, \citenamefont {M{\"u}ller}, \citenamefont {Rundquist},
  \citenamefont {Sarmiento}, \citenamefont {Piggott}, \citenamefont {Kelaita},
  \citenamefont {Dory}, \citenamefont {Lagoudakis},\ and\ \citenamefont {Vu{\v
  c}kovi{\'c}}}]{FischerSelfhomodyneMeasurement2016}%
  \BibitemOpen
  \bibfield  {author} {\bibinfo {author} {\bibfnamefont {K.~A.}\ \bibnamefont
  {Fischer}}, \bibinfo {author} {\bibfnamefont {K.}~\bibnamefont {M{\"u}ller}},
  \bibinfo {author} {\bibfnamefont {A.}~\bibnamefont {Rundquist}}, \bibinfo
  {author} {\bibfnamefont {T.}~\bibnamefont {Sarmiento}}, \bibinfo {author}
  {\bibfnamefont {A.~Y.}\ \bibnamefont {Piggott}}, \bibinfo {author}
  {\bibfnamefont {Y.}~\bibnamefont {Kelaita}}, \bibinfo {author} {\bibfnamefont
  {C.}~\bibnamefont {Dory}}, \bibinfo {author} {\bibfnamefont {K.~G.}\
  \bibnamefont {Lagoudakis}},\ and\ \bibinfo {author} {\bibfnamefont
  {J.}~\bibnamefont {Vu{\v c}kovi{\'c}}},\ }\bibfield  {title} {\bibinfo
  {title} {Self-homodyne measurement of a dynamic {{Mollow}} triplet in the
  solid state},\ }\href {https://doi.org/10.1038/nphoton.2015.276} {\bibfield
  {journal} {\bibinfo  {journal} {Nature Photonics}\ }\textbf {\bibinfo
  {volume} {10}},\ \bibinfo {pages} {163} (\bibinfo {year} {2016})}\BibitemShut
  {NoStop}%
\bibitem [{\citenamefont {Fischer}\ \emph {et~al.}(2017)\citenamefont
  {Fischer}, \citenamefont {Kelaita}, \citenamefont {Sapra}, \citenamefont
  {Dory}, \citenamefont {Lagoudakis}, \citenamefont {M{\"u}ller},\ and\
  \citenamefont {Vu{\v c}kovi{\'c}}}]{FischerOnChipArchitecture2017}%
  \BibitemOpen
  \bibfield  {author} {\bibinfo {author} {\bibfnamefont {K.~A.}\ \bibnamefont
  {Fischer}}, \bibinfo {author} {\bibfnamefont {Y.~A.}\ \bibnamefont
  {Kelaita}}, \bibinfo {author} {\bibfnamefont {N.~V.}\ \bibnamefont {Sapra}},
  \bibinfo {author} {\bibfnamefont {C.}~\bibnamefont {Dory}}, \bibinfo {author}
  {\bibfnamefont {K.~G.}\ \bibnamefont {Lagoudakis}}, \bibinfo {author}
  {\bibfnamefont {K.}~\bibnamefont {M{\"u}ller}},\ and\ \bibinfo {author}
  {\bibfnamefont {J.}~\bibnamefont {Vu{\v c}kovi{\'c}}},\ }\bibfield  {title}
  {\bibinfo {title} {On-{{Chip Architecture}} for {{Self-Homodyned Nonclassical
  Light}}},\ }\href {https://doi.org/10.1103/PhysRevApplied.7.044002}
  {\bibfield  {journal} {\bibinfo  {journal} {Physical Review Applied}\
  }\textbf {\bibinfo {volume} {7}},\ \bibinfo {pages} {044002} (\bibinfo {year}
  {2017})}\BibitemShut {NoStop}%
\bibitem [{\citenamefont {Hanschke}\ \emph {et~al.}(2020)\citenamefont
  {Hanschke}, \citenamefont {Schweickert}, \citenamefont {Carre{\~n}o},
  \citenamefont {Sch{\"o}ll}, \citenamefont {Zeuner}, \citenamefont {Lettner},
  \citenamefont {Casalengua}, \citenamefont {Reindl}, \citenamefont {{da
  Silva}}, \citenamefont {Trotta}, \citenamefont {Finley}, \citenamefont
  {Rastelli}, \citenamefont {{del Valle}}, \citenamefont {Laussy},
  \citenamefont {Zwiller}, \citenamefont {M{\"u}ller},\ and\ \citenamefont
  {J{\"o}ns}}]{HanschkeOriginAntibunching2020}%
  \BibitemOpen
  \bibfield  {author} {\bibinfo {author} {\bibfnamefont {L.}~\bibnamefont
  {Hanschke}}, \bibinfo {author} {\bibfnamefont {L.}~\bibnamefont
  {Schweickert}}, \bibinfo {author} {\bibfnamefont {J.~C.~L.}\ \bibnamefont
  {Carre{\~n}o}}, \bibinfo {author} {\bibfnamefont {E.}~\bibnamefont
  {Sch{\"o}ll}}, \bibinfo {author} {\bibfnamefont {K.~D.}\ \bibnamefont
  {Zeuner}}, \bibinfo {author} {\bibfnamefont {T.}~\bibnamefont {Lettner}},
  \bibinfo {author} {\bibfnamefont {E.~Z.}\ \bibnamefont {Casalengua}},
  \bibinfo {author} {\bibfnamefont {M.}~\bibnamefont {Reindl}}, \bibinfo
  {author} {\bibfnamefont {S.~F.~C.}\ \bibnamefont {{da Silva}}}, \bibinfo
  {author} {\bibfnamefont {R.}~\bibnamefont {Trotta}}, \bibinfo {author}
  {\bibfnamefont {J.~J.}\ \bibnamefont {Finley}}, \bibinfo {author}
  {\bibfnamefont {A.}~\bibnamefont {Rastelli}}, \bibinfo {author}
  {\bibfnamefont {E.}~\bibnamefont {{del Valle}}}, \bibinfo {author}
  {\bibfnamefont {F.~P.}\ \bibnamefont {Laussy}}, \bibinfo {author}
  {\bibfnamefont {V.}~\bibnamefont {Zwiller}}, \bibinfo {author} {\bibfnamefont
  {K.}~\bibnamefont {M{\"u}ller}},\ and\ \bibinfo {author} {\bibfnamefont
  {K.~D.}\ \bibnamefont {J{\"o}ns}},\ }\bibfield  {title} {\bibinfo {title}
  {Origin of {{Antibunching}} in {{Resonance Fluorescence}}},\ }\href
  {https://doi.org/10.1103/PhysRevLett.125.170402} {\bibfield  {journal}
  {\bibinfo  {journal} {Physical Review Letters}\ }\textbf {\bibinfo {volume}
  {125}},\ \bibinfo {pages} {170402} (\bibinfo {year} {2020})}\BibitemShut
  {NoStop}%
\bibitem [{\citenamefont {Cygorek}\ \emph {et~al.}(2023)\citenamefont
  {Cygorek}, \citenamefont {Scerri}, \citenamefont {Santana}, \citenamefont
  {Koong}, \citenamefont {Gerardot},\ and\ \citenamefont
  {Gauger}}]{CygorekSignaturesCooperative2023}%
  \BibitemOpen
  \bibfield  {author} {\bibinfo {author} {\bibfnamefont {M.}~\bibnamefont
  {Cygorek}}, \bibinfo {author} {\bibfnamefont {E.~D.}\ \bibnamefont {Scerri}},
  \bibinfo {author} {\bibfnamefont {T.~S.}\ \bibnamefont {Santana}}, \bibinfo
  {author} {\bibfnamefont {Z.~X.}\ \bibnamefont {Koong}}, \bibinfo {author}
  {\bibfnamefont {B.~D.}\ \bibnamefont {Gerardot}},\ and\ \bibinfo {author}
  {\bibfnamefont {E.~M.}\ \bibnamefont {Gauger}},\ }\bibfield  {title}
  {\bibinfo {title} {Signatures of cooperative emission in photon coincidence:
  {{Superradiance}} versus measurement-induced cooperativity},\ }\href
  {https://doi.org/10.1103/PhysRevA.107.023718} {\bibfield  {journal} {\bibinfo
   {journal} {Physical Review A}\ }\textbf {\bibinfo {volume} {107}},\ \bibinfo
  {pages} {023718} (\bibinfo {year} {2023})}\BibitemShut {NoStop}%
\bibitem [{\citenamefont {Del~Valle}\ \emph {et~al.}(2012)\citenamefont
  {Del~Valle}, \citenamefont {{Gonzalez-Tudela}}, \citenamefont {Laussy},
  \citenamefont {Tejedor},\ and\ \citenamefont
  {Hartmann}}]{DelValleTheoryFrequencyFiltered2012}%
  \BibitemOpen
  \bibfield  {author} {\bibinfo {author} {\bibfnamefont {E.}~\bibnamefont
  {Del~Valle}}, \bibinfo {author} {\bibfnamefont {A.}~\bibnamefont
  {{Gonzalez-Tudela}}}, \bibinfo {author} {\bibfnamefont {F.~P.}\ \bibnamefont
  {Laussy}}, \bibinfo {author} {\bibfnamefont {C.}~\bibnamefont {Tejedor}},\
  and\ \bibinfo {author} {\bibfnamefont {M.~J.}\ \bibnamefont {Hartmann}},\
  }\bibfield  {title} {\bibinfo {title} {Theory of {{Frequency-Filtered}} and
  {{Time-Resolved N}} -{{Photon Correlations}}},\ }\href
  {https://doi.org/10.1103/PhysRevLett.109.183601} {\bibfield  {journal}
  {\bibinfo  {journal} {Physical Review Letters}\ }\textbf {\bibinfo {volume}
  {109}},\ \bibinfo {pages} {183601} (\bibinfo {year} {2012})}\BibitemShut
  {NoStop}%
\bibitem [{\citenamefont {{Gonzalez-Tudela}}\ \emph {et~al.}(2013)\citenamefont
  {{Gonzalez-Tudela}}, \citenamefont {Laussy}, \citenamefont {Tejedor},
  \citenamefont {Hartmann},\ and\ \citenamefont {{del
  Valle}}}]{Gonzalez-TudelaTwophotonSpectra2013}%
  \BibitemOpen
  \bibfield  {author} {\bibinfo {author} {\bibfnamefont {A.}~\bibnamefont
  {{Gonzalez-Tudela}}}, \bibinfo {author} {\bibfnamefont {F.~P.}\ \bibnamefont
  {Laussy}}, \bibinfo {author} {\bibfnamefont {C.}~\bibnamefont {Tejedor}},
  \bibinfo {author} {\bibfnamefont {M.~J.}\ \bibnamefont {Hartmann}},\ and\
  \bibinfo {author} {\bibfnamefont {E.}~\bibnamefont {{del Valle}}},\
  }\bibfield  {title} {\bibinfo {title} {Two-photon spectra of quantum
  emitters},\ }\href {https://doi.org/10.1088/1367-2630/15/3/033036} {\bibfield
   {journal} {\bibinfo  {journal} {New Journal of Physics}\ }\textbf {\bibinfo
  {volume} {15}},\ \bibinfo {pages} {033036} (\bibinfo {year}
  {2013})}\BibitemShut {NoStop}%
\bibitem [{\citenamefont {Peiris}\ \emph {et~al.}(2015)\citenamefont {Peiris},
  \citenamefont {Petrak}, \citenamefont {Konthasinghe}, \citenamefont {Yu},
  \citenamefont {Niu},\ and\ \citenamefont
  {Muller}}]{PeirisTwocolorPhoton2015}%
  \BibitemOpen
  \bibfield  {author} {\bibinfo {author} {\bibfnamefont {M.}~\bibnamefont
  {Peiris}}, \bibinfo {author} {\bibfnamefont {B.}~\bibnamefont {Petrak}},
  \bibinfo {author} {\bibfnamefont {K.}~\bibnamefont {Konthasinghe}}, \bibinfo
  {author} {\bibfnamefont {Y.}~\bibnamefont {Yu}}, \bibinfo {author}
  {\bibfnamefont {Z.~C.}\ \bibnamefont {Niu}},\ and\ \bibinfo {author}
  {\bibfnamefont {A.}~\bibnamefont {Muller}},\ }\bibfield  {title} {\bibinfo
  {title} {Two-color photon correlations of the light scattered by a quantum
  dot},\ }\href {https://doi.org/10.1103/PhysRevB.91.195125} {\bibfield
  {journal} {\bibinfo  {journal} {Physical Review B}\ }\textbf {\bibinfo
  {volume} {91}},\ \bibinfo {pages} {195125} (\bibinfo {year}
  {2015})}\BibitemShut {NoStop}%
\bibitem [{\citenamefont {Silva}\ \emph {et~al.}(2016)\citenamefont {Silva},
  \citenamefont {S{\'a}nchez~Mu{\~n}oz}, \citenamefont {Ballarini},
  \citenamefont {{Gonz{\'a}lez-Tudela}}, \citenamefont {De~Giorgi},
  \citenamefont {Gigli}, \citenamefont {West}, \citenamefont {Pfeiffer},
  \citenamefont {Del~Valle}, \citenamefont {Sanvitto},\ and\ \citenamefont
  {Laussy}}]{SilvaColoredHanbury2016}%
  \BibitemOpen
  \bibfield  {author} {\bibinfo {author} {\bibfnamefont {B.}~\bibnamefont
  {Silva}}, \bibinfo {author} {\bibfnamefont {C.}~\bibnamefont
  {S{\'a}nchez~Mu{\~n}oz}}, \bibinfo {author} {\bibfnamefont {D.}~\bibnamefont
  {Ballarini}}, \bibinfo {author} {\bibfnamefont {A.}~\bibnamefont
  {{Gonz{\'a}lez-Tudela}}}, \bibinfo {author} {\bibfnamefont {M.}~\bibnamefont
  {De~Giorgi}}, \bibinfo {author} {\bibfnamefont {G.}~\bibnamefont {Gigli}},
  \bibinfo {author} {\bibfnamefont {K.}~\bibnamefont {West}}, \bibinfo {author}
  {\bibfnamefont {L.}~\bibnamefont {Pfeiffer}}, \bibinfo {author}
  {\bibfnamefont {E.}~\bibnamefont {Del~Valle}}, \bibinfo {author}
  {\bibfnamefont {D.}~\bibnamefont {Sanvitto}},\ and\ \bibinfo {author}
  {\bibfnamefont {F.~P.}\ \bibnamefont {Laussy}},\ }\bibfield  {title}
  {\bibinfo {title} {The colored {{Hanbury Brown}}{\textendash}{{Twiss}}
  effect},\ }\href {https://doi.org/10.1038/srep37980} {\bibfield  {journal}
  {\bibinfo  {journal} {Scientific Reports}\ }\textbf {\bibinfo {volume} {6}},\
  \bibinfo {pages} {37980} (\bibinfo {year} {2016})}\BibitemShut {NoStop}%
\bibitem [{\citenamefont {Shammah}\ \emph {et~al.}(2018)\citenamefont
  {Shammah}, \citenamefont {Ahmed}, \citenamefont {Lambert}, \citenamefont
  {De~Liberato},\ and\ \citenamefont {Nori}}]{ShammahOpenQuantum2018}%
  \BibitemOpen
  \bibfield  {author} {\bibinfo {author} {\bibfnamefont {N.}~\bibnamefont
  {Shammah}}, \bibinfo {author} {\bibfnamefont {S.}~\bibnamefont {Ahmed}},
  \bibinfo {author} {\bibfnamefont {N.}~\bibnamefont {Lambert}}, \bibinfo
  {author} {\bibfnamefont {S.}~\bibnamefont {De~Liberato}},\ and\ \bibinfo
  {author} {\bibfnamefont {F.}~\bibnamefont {Nori}},\ }\bibfield  {title}
  {\bibinfo {title} {Open quantum systems with local and collective incoherent
  processes: {{Efficient}} numerical simulations using permutational
  invariance},\ }\href {https://doi.org/10.1103/PhysRevA.98.063815} {\bibfield
  {journal} {\bibinfo  {journal} {Physical Review A}\ }\textbf {\bibinfo
  {volume} {98}},\ \bibinfo {pages} {063815} (\bibinfo {year}
  {2018})}\BibitemShut {NoStop}%
\bibitem [{\citenamefont {Mlynek}\ \emph {et~al.}(2014)\citenamefont {Mlynek},
  \citenamefont {Abdumalikov}, \citenamefont {Eichler},\ and\ \citenamefont
  {Wallraff}}]{MlynekObservationDicke2014}%
  \BibitemOpen
  \bibfield  {author} {\bibinfo {author} {\bibfnamefont {J.~A.}\ \bibnamefont
  {Mlynek}}, \bibinfo {author} {\bibfnamefont {A.~A.}\ \bibnamefont
  {Abdumalikov}}, \bibinfo {author} {\bibfnamefont {C.}~\bibnamefont
  {Eichler}},\ and\ \bibinfo {author} {\bibfnamefont {A.}~\bibnamefont
  {Wallraff}},\ }\bibfield  {title} {\bibinfo {title} {Observation of {{Dicke}}
  superradiance for two artificial atoms in a cavity with high decay rate},\
  }\href {https://doi.org/10.1038/ncomms6186} {\bibfield  {journal} {\bibinfo
  {journal} {Nature Communications}\ }\textbf {\bibinfo {volume} {5}},\
  \bibinfo {pages} {5186} (\bibinfo {year} {2014})}\BibitemShut {NoStop}%
\bibitem [{\citenamefont {Krummheuer}\ \emph {et~al.}(2002)\citenamefont
  {Krummheuer}, \citenamefont {Axt},\ and\ \citenamefont
  {Kuhn}}]{KrummheuerTheoryPure2002}%
  \BibitemOpen
  \bibfield  {author} {\bibinfo {author} {\bibfnamefont {B.}~\bibnamefont
  {Krummheuer}}, \bibinfo {author} {\bibfnamefont {V.~M.}\ \bibnamefont
  {Axt}},\ and\ \bibinfo {author} {\bibfnamefont {T.}~\bibnamefont {Kuhn}},\
  }\bibfield  {title} {\bibinfo {title} {Theory of pure dephasing and the
  resulting absorption line shape in semiconductor quantum dots},\ }\href
  {https://doi.org/10.1103/PhysRevB.65.195313} {\bibfield  {journal} {\bibinfo
  {journal} {Physical Review B}\ }\textbf {\bibinfo {volume} {65}},\ \bibinfo
  {pages} {195313} (\bibinfo {year} {2002})}\BibitemShut {NoStop}%
\bibitem [{\citenamefont {{del Pino}}\ \emph {et~al.}(2015)\citenamefont {{del
  Pino}}, \citenamefont {Feist},\ and\ \citenamefont
  {{Garcia-Vidal}}}]{delPinoQuantumTheory2015}%
  \BibitemOpen
  \bibfield  {author} {\bibinfo {author} {\bibfnamefont {J.}~\bibnamefont {{del
  Pino}}}, \bibinfo {author} {\bibfnamefont {J.}~\bibnamefont {Feist}},\ and\
  \bibinfo {author} {\bibfnamefont {F.~J.}\ \bibnamefont {{Garcia-Vidal}}},\
  }\bibfield  {title} {\bibinfo {title} {Quantum theory of collective strong
  coupling of molecular vibrations with a microcavity mode},\ }\href
  {https://doi.org/10.1088/1367-2630/17/5/053040} {\bibfield  {journal}
  {\bibinfo  {journal} {New Journal of Physics}\ }\textbf {\bibinfo {volume}
  {17}},\ \bibinfo {pages} {053040} (\bibinfo {year} {2015})}\BibitemShut
  {NoStop}%
\bibitem [{\citenamefont {Reitz}\ \emph {et~al.}(2020)\citenamefont {Reitz},
  \citenamefont {Sommer}, \citenamefont {Gurlek}, \citenamefont {Sandoghdar},
  \citenamefont {{Martin-Cano}},\ and\ \citenamefont
  {Genes}}]{ReitzMoleculephotonInteractions2020}%
  \BibitemOpen
  \bibfield  {author} {\bibinfo {author} {\bibfnamefont {M.}~\bibnamefont
  {Reitz}}, \bibinfo {author} {\bibfnamefont {C.}~\bibnamefont {Sommer}},
  \bibinfo {author} {\bibfnamefont {B.}~\bibnamefont {Gurlek}}, \bibinfo
  {author} {\bibfnamefont {V.}~\bibnamefont {Sandoghdar}}, \bibinfo {author}
  {\bibfnamefont {D.}~\bibnamefont {{Martin-Cano}}},\ and\ \bibinfo {author}
  {\bibfnamefont {C.}~\bibnamefont {Genes}},\ }\bibfield  {title} {\bibinfo
  {title} {Molecule-photon interactions in phononic environments},\ }\href
  {https://doi.org/10.1103/PhysRevResearch.2.033270} {\bibfield  {journal}
  {\bibinfo  {journal} {Physical Review Research}\ }\textbf {\bibinfo {volume}
  {2}},\ \bibinfo {pages} {033270} (\bibinfo {year} {2020})}\BibitemShut
  {NoStop}%
\bibitem [{\citenamefont {Prasanna~Venkatesh}\ \emph
  {et~al.}(2018)\citenamefont {Prasanna~Venkatesh}, \citenamefont {Juan},\ and\
  \citenamefont {{Romero-Isart}}}]{PrasannaVenkateshCooperativeEffects2018}%
  \BibitemOpen
  \bibfield  {author} {\bibinfo {author} {\bibfnamefont {B.}~\bibnamefont
  {Prasanna~Venkatesh}}, \bibinfo {author} {\bibfnamefont {M.~L.}\ \bibnamefont
  {Juan}},\ and\ \bibinfo {author} {\bibfnamefont {O.}~\bibnamefont
  {{Romero-Isart}}},\ }\bibfield  {title} {\bibinfo {title} {Cooperative
  {{Effects}} in {{Closely Packed Quantum Emitters}} with {{Collective
  Dephasing}}},\ }\href {https://doi.org/10.1103/PhysRevLett.120.033602}
  {\bibfield  {journal} {\bibinfo  {journal} {Physical Review Letters}\
  }\textbf {\bibinfo {volume} {120}},\ \bibinfo {pages} {033602} (\bibinfo
  {year} {2018})}\BibitemShut {NoStop}%
\bibitem [{\citenamefont {Wang}\ \emph {et~al.}(2015)\citenamefont {Wang},
  \citenamefont {Ji}, \citenamefont {Li},\ and\ \citenamefont
  {Zhou}}]{WangDissipationDecoherence2015}%
  \BibitemOpen
  \bibfield  {author} {\bibinfo {author} {\bibfnamefont {Z.~H.}\ \bibnamefont
  {Wang}}, \bibinfo {author} {\bibfnamefont {Y.~J.}\ \bibnamefont {Ji}},
  \bibinfo {author} {\bibfnamefont {Y.}~\bibnamefont {Li}},\ and\ \bibinfo
  {author} {\bibfnamefont {D.~L.}\ \bibnamefont {Zhou}},\ }\bibfield  {title}
  {\bibinfo {title} {Dissipation and decoherence induced by collective
  dephasing in a coupled-qubit system with a common bath},\ }\href
  {https://doi.org/10.1103/PhysRevA.91.013838} {\bibfield  {journal} {\bibinfo
  {journal} {Physical Review A}\ }\textbf {\bibinfo {volume} {91}},\ \bibinfo
  {pages} {013838} (\bibinfo {year} {2015})}\BibitemShut {NoStop}%
\bibitem [{\citenamefont {Tomm}\ \emph {et~al.}(2021)\citenamefont {Tomm},
  \citenamefont {Javadi}, \citenamefont {Antoniadis}, \citenamefont {Najer},
  \citenamefont {L{\"o}bl}, \citenamefont {Korsch}, \citenamefont {Schott},
  \citenamefont {Valentin}, \citenamefont {Wieck}, \citenamefont {Ludwig},\
  and\ \citenamefont {Warburton}}]{TommBrightFast2021}%
  \BibitemOpen
  \bibfield  {author} {\bibinfo {author} {\bibfnamefont {N.}~\bibnamefont
  {Tomm}}, \bibinfo {author} {\bibfnamefont {A.}~\bibnamefont {Javadi}},
  \bibinfo {author} {\bibfnamefont {N.~O.}\ \bibnamefont {Antoniadis}},
  \bibinfo {author} {\bibfnamefont {D.}~\bibnamefont {Najer}}, \bibinfo
  {author} {\bibfnamefont {M.~C.}\ \bibnamefont {L{\"o}bl}}, \bibinfo {author}
  {\bibfnamefont {A.~R.}\ \bibnamefont {Korsch}}, \bibinfo {author}
  {\bibfnamefont {R.}~\bibnamefont {Schott}}, \bibinfo {author} {\bibfnamefont
  {S.~R.}\ \bibnamefont {Valentin}}, \bibinfo {author} {\bibfnamefont {A.~D.}\
  \bibnamefont {Wieck}}, \bibinfo {author} {\bibfnamefont {A.}~\bibnamefont
  {Ludwig}},\ and\ \bibinfo {author} {\bibfnamefont {R.~J.}\ \bibnamefont
  {Warburton}},\ }\bibfield  {title} {\bibinfo {title} {A bright and fast
  source of coherent single photons},\ }\href
  {https://doi.org/10.1038/s41565-020-00831-x} {\bibfield  {journal} {\bibinfo
  {journal} {Nature Nanotechnology}\ }\textbf {\bibinfo {volume} {16}},\
  \bibinfo {pages} {399} (\bibinfo {year} {2021})}\BibitemShut {NoStop}%
\bibitem [{\citenamefont {Wang}\ \emph {et~al.}(2017)\citenamefont {Wang},
  \citenamefont {Kelkar}, \citenamefont {{Martin-Cano}}, \citenamefont
  {Utikal}, \citenamefont {G{\"o}tzinger},\ and\ \citenamefont
  {Sandoghdar}}]{WangCoherentCoupling2017}%
  \BibitemOpen
  \bibfield  {author} {\bibinfo {author} {\bibfnamefont {D.}~\bibnamefont
  {Wang}}, \bibinfo {author} {\bibfnamefont {H.}~\bibnamefont {Kelkar}},
  \bibinfo {author} {\bibfnamefont {D.}~\bibnamefont {{Martin-Cano}}}, \bibinfo
  {author} {\bibfnamefont {T.}~\bibnamefont {Utikal}}, \bibinfo {author}
  {\bibfnamefont {S.}~\bibnamefont {G{\"o}tzinger}},\ and\ \bibinfo {author}
  {\bibfnamefont {V.}~\bibnamefont {Sandoghdar}},\ }\bibfield  {title}
  {\bibinfo {title} {Coherent {{Coupling}} of a {{Single Molecule}} to a
  {{Scanning Fabry-Perot Microcavity}}},\ }\href
  {https://doi.org/10.1103/PhysRevX.7.021014} {\bibfield  {journal} {\bibinfo
  {journal} {Physical Review X}\ }\textbf {\bibinfo {volume} {7}},\ \bibinfo
  {pages} {021014} (\bibinfo {year} {2017})}\BibitemShut {NoStop}%
\bibitem [{\citenamefont {Wang}\ \emph {et~al.}(2019)\citenamefont {Wang},
  \citenamefont {Kelkar}, \citenamefont {{Martin-Cano}}, \citenamefont
  {Rattenbacher}, \citenamefont {Shkarin}, \citenamefont {Utikal},
  \citenamefont {G{\"o}tzinger},\ and\ \citenamefont
  {Sandoghdar}}]{WangTurningMolecule2019}%
  \BibitemOpen
  \bibfield  {author} {\bibinfo {author} {\bibfnamefont {D.}~\bibnamefont
  {Wang}}, \bibinfo {author} {\bibfnamefont {H.}~\bibnamefont {Kelkar}},
  \bibinfo {author} {\bibfnamefont {D.}~\bibnamefont {{Martin-Cano}}}, \bibinfo
  {author} {\bibfnamefont {D.}~\bibnamefont {Rattenbacher}}, \bibinfo {author}
  {\bibfnamefont {A.}~\bibnamefont {Shkarin}}, \bibinfo {author} {\bibfnamefont
  {T.}~\bibnamefont {Utikal}}, \bibinfo {author} {\bibfnamefont
  {S.}~\bibnamefont {G{\"o}tzinger}},\ and\ \bibinfo {author} {\bibfnamefont
  {V.}~\bibnamefont {Sandoghdar}},\ }\bibfield  {title} {\bibinfo {title}
  {Turning a molecule into a coherent two-level quantum system},\ }\href
  {https://doi.org/10.1038/s41567-019-0436-5} {\bibfield  {journal} {\bibinfo
  {journal} {Nature Physics}\ }\textbf {\bibinfo {volume} {15}},\ \bibinfo
  {pages} {483} (\bibinfo {year} {2019})}\BibitemShut {NoStop}%
\bibitem [{\citenamefont {Nguyen}\ \emph {et~al.}(2019)\citenamefont {Nguyen},
  \citenamefont {Sukachev}, \citenamefont {Bhaskar}, \citenamefont {Machielse},
  \citenamefont {Levonian}, \citenamefont {Knall}, \citenamefont {Stroganov},
  \citenamefont {Riedinger}, \citenamefont {Park}, \citenamefont {Lon{\v
  c}ar},\ and\ \citenamefont {Lukin}}]{NguyenQuantumNetwork2019}%
  \BibitemOpen
  \bibfield  {author} {\bibinfo {author} {\bibfnamefont {C.~T.}\ \bibnamefont
  {Nguyen}}, \bibinfo {author} {\bibfnamefont {D.~D.}\ \bibnamefont
  {Sukachev}}, \bibinfo {author} {\bibfnamefont {M.~K.}\ \bibnamefont
  {Bhaskar}}, \bibinfo {author} {\bibfnamefont {B.}~\bibnamefont {Machielse}},
  \bibinfo {author} {\bibfnamefont {D.~S.}\ \bibnamefont {Levonian}}, \bibinfo
  {author} {\bibfnamefont {E.~N.}\ \bibnamefont {Knall}}, \bibinfo {author}
  {\bibfnamefont {P.}~\bibnamefont {Stroganov}}, \bibinfo {author}
  {\bibfnamefont {R.}~\bibnamefont {Riedinger}}, \bibinfo {author}
  {\bibfnamefont {H.}~\bibnamefont {Park}}, \bibinfo {author} {\bibfnamefont
  {M.}~\bibnamefont {Lon{\v c}ar}},\ and\ \bibinfo {author} {\bibfnamefont
  {M.~D.}\ \bibnamefont {Lukin}},\ }\bibfield  {title} {\bibinfo {title}
  {Quantum {{Network Nodes Based}} on {{Diamond Qubits}} with an {{Efficient
  Nanophotonic Interface}}},\ }\href
  {https://doi.org/10.1103/PhysRevLett.123.183602} {\bibfield  {journal}
  {\bibinfo  {journal} {Physical Review Letters}\ }\textbf {\bibinfo {volume}
  {123}},\ \bibinfo {pages} {183602} (\bibinfo {year} {2019})}\BibitemShut
  {NoStop}%
\bibitem [{\citenamefont {Akselrod}\ \emph {et~al.}(2014)\citenamefont
  {Akselrod}, \citenamefont {Argyropoulos}, \citenamefont {Hoang},
  \citenamefont {Cirac{\`i}}, \citenamefont {Fang}, \citenamefont {Huang},
  \citenamefont {Smith},\ and\ \citenamefont
  {Mikkelsen}}]{AkselrodProbingMechanisms2014}%
  \BibitemOpen
  \bibfield  {author} {\bibinfo {author} {\bibfnamefont {G.~M.}\ \bibnamefont
  {Akselrod}}, \bibinfo {author} {\bibfnamefont {C.}~\bibnamefont
  {Argyropoulos}}, \bibinfo {author} {\bibfnamefont {T.~B.}\ \bibnamefont
  {Hoang}}, \bibinfo {author} {\bibfnamefont {C.}~\bibnamefont {Cirac{\`i}}},
  \bibinfo {author} {\bibfnamefont {C.}~\bibnamefont {Fang}}, \bibinfo {author}
  {\bibfnamefont {J.}~\bibnamefont {Huang}}, \bibinfo {author} {\bibfnamefont
  {D.~R.}\ \bibnamefont {Smith}},\ and\ \bibinfo {author} {\bibfnamefont
  {M.~H.}\ \bibnamefont {Mikkelsen}},\ }\bibfield  {title} {\bibinfo {title}
  {Probing the mechanisms of large {{Purcell}} enhancement in plasmonic
  nanoantennas},\ }\href {https://doi.org/10.1038/nphoton.2014.228} {\bibfield
  {journal} {\bibinfo  {journal} {Nature Photonics}\ }\textbf {\bibinfo
  {volume} {8}},\ \bibinfo {pages} {835} (\bibinfo {year} {2014})}\BibitemShut
  {NoStop}%
\bibitem [{\citenamefont {Chikkaraddy}\ \emph {et~al.}(2016)\citenamefont
  {Chikkaraddy}, \citenamefont {{de Nijs}}, \citenamefont {Benz}, \citenamefont
  {Barrow}, \citenamefont {Scherman}, \citenamefont {Rosta}, \citenamefont
  {Demetriadou}, \citenamefont {Fox}, \citenamefont {Hess},\ and\ \citenamefont
  {Baumberg}}]{ChikkaraddySinglemoleculeStrong2016}%
  \BibitemOpen
  \bibfield  {author} {\bibinfo {author} {\bibfnamefont {R.}~\bibnamefont
  {Chikkaraddy}}, \bibinfo {author} {\bibfnamefont {B.}~\bibnamefont {{de
  Nijs}}}, \bibinfo {author} {\bibfnamefont {F.}~\bibnamefont {Benz}}, \bibinfo
  {author} {\bibfnamefont {S.~J.}\ \bibnamefont {Barrow}}, \bibinfo {author}
  {\bibfnamefont {O.~A.}\ \bibnamefont {Scherman}}, \bibinfo {author}
  {\bibfnamefont {E.}~\bibnamefont {Rosta}}, \bibinfo {author} {\bibfnamefont
  {A.}~\bibnamefont {Demetriadou}}, \bibinfo {author} {\bibfnamefont
  {P.}~\bibnamefont {Fox}}, \bibinfo {author} {\bibfnamefont {O.}~\bibnamefont
  {Hess}},\ and\ \bibinfo {author} {\bibfnamefont {J.~J.}\ \bibnamefont
  {Baumberg}},\ }\bibfield  {title} {\bibinfo {title} {Single-molecule strong
  coupling at room temperature in plasmonic nanocavities},\ }\href
  {https://doi.org/10.1038/nature17974} {\bibfield  {journal} {\bibinfo
  {journal} {Nature}\ }\textbf {\bibinfo {volume} {535}},\ \bibinfo {pages}
  {127} (\bibinfo {year} {2016})}\BibitemShut {NoStop}%
\bibitem [{\citenamefont {Baumberg}\ \emph {et~al.}(2019)\citenamefont
  {Baumberg}, \citenamefont {Aizpurua}, \citenamefont {Mikkelsen},\ and\
  \citenamefont {Smith}}]{BaumbergExtremeNanophotonics2019}%
  \BibitemOpen
  \bibfield  {author} {\bibinfo {author} {\bibfnamefont {J.~J.}\ \bibnamefont
  {Baumberg}}, \bibinfo {author} {\bibfnamefont {J.}~\bibnamefont {Aizpurua}},
  \bibinfo {author} {\bibfnamefont {M.~H.}\ \bibnamefont {Mikkelsen}},\ and\
  \bibinfo {author} {\bibfnamefont {D.~R.}\ \bibnamefont {Smith}},\ }\bibfield
  {title} {\bibinfo {title} {Extreme nanophotonics from ultrathin metallic
  gaps},\ }\href {https://doi.org/10.1038/s41563-019-0290-y} {\bibfield
  {journal} {\bibinfo  {journal} {Nature Materials}\ }\textbf {\bibinfo
  {volume} {18}},\ \bibinfo {pages} {668} (\bibinfo {year} {2019})}\BibitemShut
  {NoStop}%
\bibitem [{\citenamefont {Molesky}\ \emph {et~al.}(2018)\citenamefont
  {Molesky}, \citenamefont {Lin}, \citenamefont {Piggott}, \citenamefont {Jin},
  \citenamefont {Vuckovi{\'c}},\ and\ \citenamefont
  {Rodriguez}}]{MoleskyInverseDesign2018}%
  \BibitemOpen
  \bibfield  {author} {\bibinfo {author} {\bibfnamefont {S.}~\bibnamefont
  {Molesky}}, \bibinfo {author} {\bibfnamefont {Z.}~\bibnamefont {Lin}},
  \bibinfo {author} {\bibfnamefont {A.~Y.}\ \bibnamefont {Piggott}}, \bibinfo
  {author} {\bibfnamefont {W.}~\bibnamefont {Jin}}, \bibinfo {author}
  {\bibfnamefont {J.}~\bibnamefont {Vuckovi{\'c}}},\ and\ \bibinfo {author}
  {\bibfnamefont {A.~W.}\ \bibnamefont {Rodriguez}},\ }\bibfield  {title}
  {\bibinfo {title} {Inverse design in nanophotonics},\ }\href
  {https://doi.org/10.1038/s41566-018-0246-9} {\bibfield  {journal} {\bibinfo
  {journal} {Nature Photonics}\ }\textbf {\bibinfo {volume} {12}},\ \bibinfo
  {pages} {659} (\bibinfo {year} {2018})}\BibitemShut {NoStop}%
\bibitem [{\citenamefont
  {{Navarrete-Benlloch}}(2022)}]{Navarrete-BenllochIntroductionQuantum2022}%
  \BibitemOpen
  \bibfield  {author} {\bibinfo {author} {\bibfnamefont {C.}~\bibnamefont
  {{Navarrete-Benlloch}}},\ }\href@noop {} {\bibinfo {title} {Introduction to
  {{Quantum Optics}}}} (\bibinfo {year} {2022}),\ \Eprint
  {https://arxiv.org/abs/2203.13206} {arxiv:2203.13206} \BibitemShut {NoStop}%
\end{thebibliography}%

\let\addcontentsline\oldaddcontentsline

\end{document}